\documentclass[fleqn,10pt]{wlscirep}
\usepackage{longtable}
\usepackage{array} 
\usepackage{xcolor,amsmath}
\usepackage{float}
\usepackage{amsfonts}
\usepackage{dsfont}
\usepackage{yfonts}
\usepackage{tabularx}
\usepackage{titletoc}
\usepackage{booktabs}
\usepackage{subfigure}
\usepackage{comment}
\usepackage[switch]{lineno}
\usepackage[mathcal]{eucal}
\usepackage[many]{tcolorbox}
\usepackage{multirow}
\usepackage{bibunits}
\usepackage{makecell}
\usepackage{bbding}
\usepackage{algorithm}
\usepackage[noend]{algpseudocode}

\usepackage{epstopdf}
\usepackage{graphicx}
\usepackage{bbm}

\newcommand{\blue}[1]{{#1}}

\usepackage{soul}
\soulregister\cite7 
\soulregister\cite7 
\soulregister\citet7 
\soulregister\ref7 
\soulregister\pageref7 
\soulregister\it7 
\title{\LARGE AtomDisc: An Atom-level Tokenizer that Boosts Molecular LLMs and Reveals Structure–Property Associations}

\author[1]{Mingxu Zhang}
\author[2]{Dazhong Shen}
\author[1*]{Ying Sun}

\affil[1]{The Thrust of Artificial Intelligence, The Hong Kong University of Science and Technology (Guangzhou), Guangzhou, China.}
\affil[2]{The College  of Computer Science and Technology, The Nanjing University of Aeronautics and Astronautics, Nanjing, China}
\affil[*]{Correspondence: yings@hkust-gz.edu.cn}
\begin{abstract}

Advances in large language models (LLMs) are accelerating discovery in molecular science. However, adapting molecular information to the serialized, token-based processing of LLMs remains a key challenge. Compared to other representations, molecular graphs explicitly encode atomic connectivity and local topological environments, which are key determinants of atomic behavior and molecular properties. Despite recent efforts to tokenize overall molecular topology, there still lacks effective fine-grained tokenization of local atomic environments, which are critical for determining sophisticated chemical properties and reactivity. To address these issues, we introduce AtomDisc, a novel framework that quantizes atom-level local environments into structure-aware tokens embedded directly in LLM’s token space. Our experiments show that AtomDisc, in a data-driven way, can distinguish chemically meaningful structural features that reveal structure–property associations. Equipping LLMs with AtomDisc tokens injects an interpretable inductive bias that delivers state-of-the-art performance on property prediction and molecular generation. Our methodology and findings can pave the way for constructing more powerful molecular LLMs aimed at mechanistic insight and complex chemical reasoning.
\end{abstract}

\makeatletter
\newcommand{\MainTotalPages}{%
  \@ifundefined{r@MainLastPage}{\pageref{LastPage}}{\pageref{MainLastPage}}%
}
\makeatother

\begin{document}
\maketitle
\rfoot{\small\sffamily\bfseries\thepage/\MainTotalPages}

\section{Introduction}
Large Language Models (LLMs) have transformed diverse domains by embedding domain-specific knowledge and exhibiting advanced reasoning~\cite{wu2025towards}. In chemistry and molecular sciences, LLMs demonstrate substantial scientific expertise~\cite{liu2025quantitative} and show strong promise for tasks such as molecular property prediction~\cite{ross2022large}, reaction outcome forecasting~\cite{zhang2025large}, and de novo drug design~\cite{poongavanam2024molecular}.

Since LLMs primarily process serialized textual data, encoding molecular information into a format they can effectively process is essential. Molecules can be represented in different modalities, each highlighting specific structural features: strings (such as SMILES~\cite{weininger1988smiles} or SELFIES~\cite{selfies}, compacting atomic connectivity), graphs~\cite{fang2023knowledge} (depicting molecular topology), and 3D conformations~\cite{zhou2023uni} (showing spatial arrangement). Recent research has treated molecular strings as textual inputs~\cite{fang2023mol} for LLMs, aligning with both LLM infrastructure and chemical content in pre-training corpora. However, despite being character-based, molecular strings encode information differently from natural language, making it difficult for LLMs to infer complex molecular structures from these compact representations. For example, an LLM may understand multiple atoms or bonds as a single token (Supplementary Table \blue{S3}), severing the direct link between tokens and specific atomic units.

Unlike other modalities, molecular graphs directly store raw atomic connectivity, capturing a molecule’s constitutional information. Within a molecule, different atoms and their neighboring structures form complex local contexts, establishing stable, environment-independent characteristics that profoundly influence key physicochemical properties, such as boiling point, melting point, and molar refractivity~\cite{sorgun2025vertex}. By providing the most complete blueprint of detailed topology, molecular graphs have become the most widely used modality for capturing the intrinsic relationship between molecular structure and properties~\cite{fang2023knowledge, liu2022pretraining, ross2022large}.

However, adapting molecular graph topology to the serialized, token-based processing of LLMs remains a key challenge. Some pioneering studies have transformed the entire molecule graph into continuous embeddings~\cite{cao2023instructmol, liu2023molca, li2023blip}, which are then discretized into tokens for LLMs~\cite{zhang2024unimot}. However, these molecule-level tokens encode the molecule as a single holistic entity, inevitably obscuring fine-grained local atomic environment information—such as individual atom and bond types, atomic degrees, electronic characteristics, and polarity features. Such local details are closely related to reactive centers and reaction feasibility~\cite{wang2023retrosynthesis}, as well as properties like melting and boiling points~\cite{sorgun2025vertex}, and have been shown to be beneficial for generalizable molecular modeling~\cite{townsend2020representation}. Therefore, preserving these details is essential for enhancing LLMs’ molecular understanding. However, to our knowledge, there still lacks an effective tokenization method that retains such detailed local topological information.

In this work, we present AtomDisc, an atom-level discrete tokenizer for molecular language models. AtomDisc discretizes local atomic environments into structure-aware tokens via a learnable vector-quantized codebook~\cite{vq}, transforming molecular graphs into atomic token sequences aligned with the LLM embedding space. These tokens, interleaved with SMILES and instructions, enable LLMs to incorporate detailed structural information without changing the input format or architecture.

\begin{figure}[t]
    \centering
    \includegraphics[width=\linewidth]{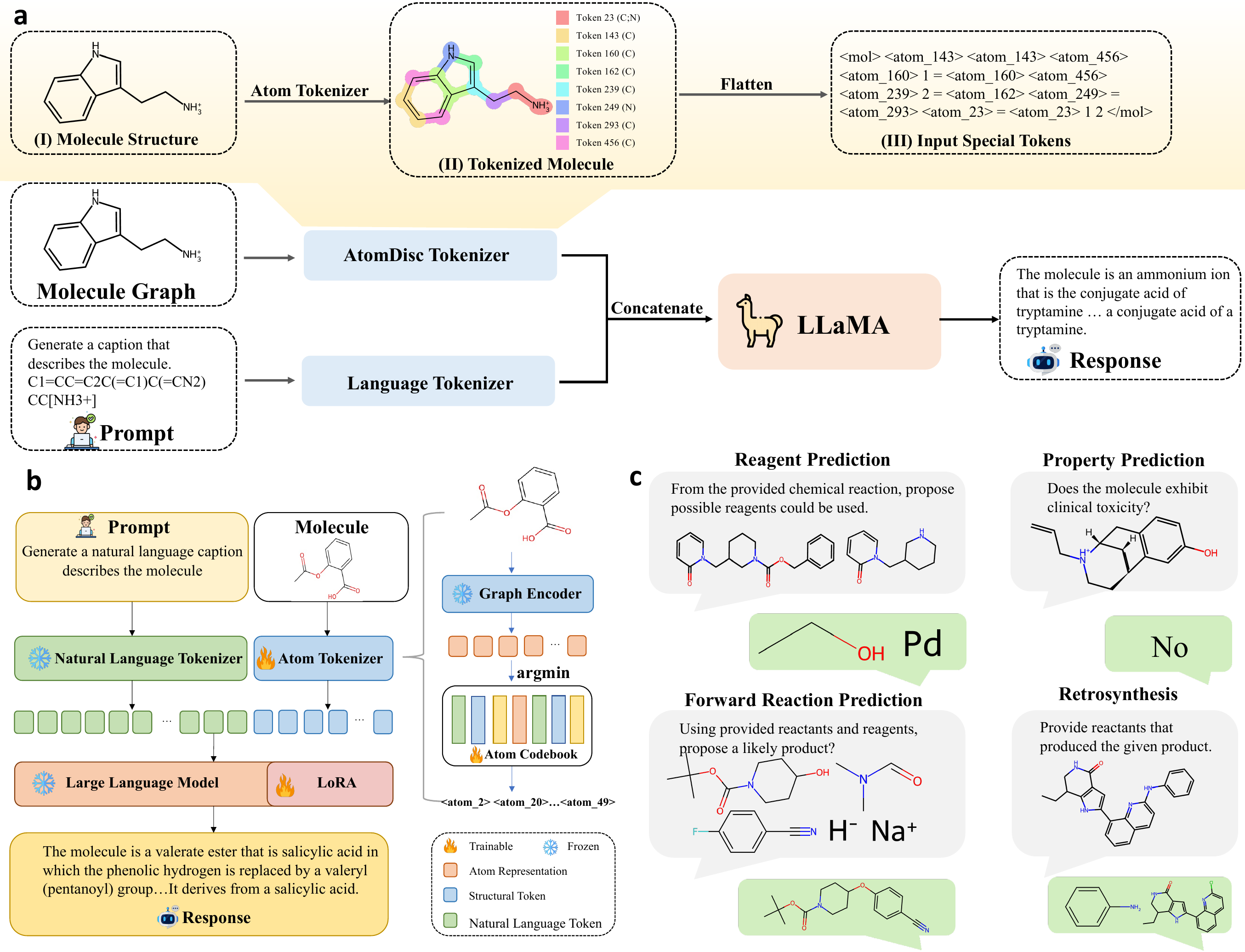}
\caption{\textbf{Overview of our approach.} \textbf{(a)} A molecule graph is first processed by the AtomDisc Tokenizer, which assigns context-specific codes to atoms, converting each into a token that reflects its unique chemical environment. These structure tokens are concatenated with language tokens and input to the LLM. 
\textbf{(b)} AtomDisc-LLM workflow: molecular graphs are embedded, discretized into atom-level tokens, and combined with language tokens for unified modeling.
\textbf{(c)} Example downstream tasks, including reagent prediction, property prediction, forward reaction prediction, and retrosynthesis.}
    \label{fig:1}
\end{figure}

Our analysis demonstrates that AtomDisc maps atom-level structures to chemically meaningful tokens without explicit prior knowledge, automatically capturing chemical environments beyond predefined features—such as similarities across different functional groups and diversity within the same group arising from local context variations. These tokens can thus distinctly differentiate atomic descriptors and molecular property patterns. Incorporating AtomDisc’s structural tokens into LLMs improves molecular comprehension and achieves SOTA performance on tasks including various property prediction and generative tasks (e.g., forward reaction prediction, retrosynthesis, and reagent identification). For property prediction, AtomDisc strengthens functional-group grounding and enforces property-discriminative motif embeddings; for generative tasks, it enables reactive substructure localization and structure-centric instruction routing, thereby enhancing the LLM’s understanding of reaction mechanisms. Meanwhile, AtomDisc improves interpretability by bridging the model’s outputs with human-understandable chemical concepts. Our findings open the door to versatile atom-level tokenization that seamlessly integrates interpretable, fine-grained molecular structural information into LLMs. This flexibility promises to extend LLM advances to molecular science, driving innovation in materials, drug discovery, and catalysis. Our approach and findings equip LLMs with atomic-scale chemical intuition and enable deep reasoning based on subtle atomic structural information. This facilitates more accurate predictions, rational design, and interpretable insights, laying a strong foundation for harnessing the full potential of large language models in chemistry and accelerating discovery and innovation in materials science, drug development, and catalysis.

\section{Results}
\subsection{The Overview of AtomDisc}

\begin{figure}[t]
    \centering
    \includegraphics[width=\linewidth]{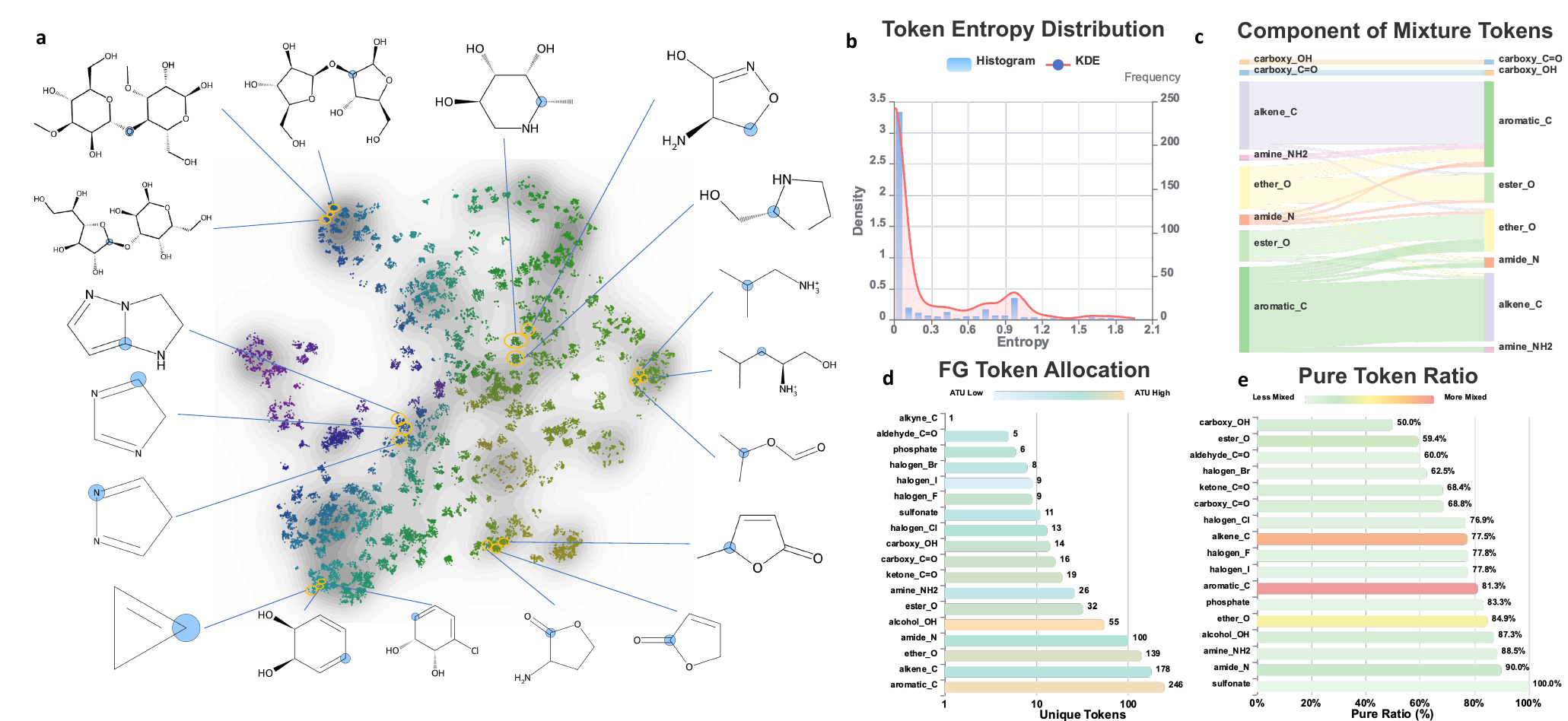}
    \caption{\textbf{Atom-level tokenization analysis and integration with LLMs.}
\textbf{(a)} 2D t-SNE visualization of atom-level representations before quantization; points within the same circle form a token cluster.
\textbf{(b)} Token entropy distribution: KDE plots of token density (left) and frequency (right).
\textbf{(c)} Mixture-token composition: Sankey diagram showing functional group nodes and edges representing shared mixture tokens (edge thickness indicates frequency).
\textbf{(d)} Functional-group token allocation: number of unique tokens per group; color encodes Average Token Utilization (ATU, $\text{ATU} = \tfrac{\text{Total Usage}}{\text{Unique Tokens}}$), with darker shades for higher ATU.
\textbf{(e)} Pure token ratio for each functional group (fraction of tokens exclusive to that group); color intensity reflects the number of associated mixture tokens.
}

    \label{fig2_Tokenizer}
\end{figure}

Figure~\ref{fig:1} provides an overview of our method, illustrating how AtomDisc integrates atom-level structural information into LLMs for downstream molecular applications. As shown in Fig.~\ref{fig:1}(a), given a task description and molecular structure, AtomDisc maps each atom’s local chemical environment to a code in a shared codebook. The resulting atomic tokens are combined with natural language tokens—such as the task description and molecular string—into a unified input sequence for the LLM, which then generates task-specific responses. Specifically, as shown in Fig.~\ref{fig:1}(b), molecular graphs are first embedded by a pretrained GNN encoder, which propagates local structural information to each atom via message passing. These embeddings are discretized into atom-level tokens using a vector-quantized codebook, ensuring each atom’s fine-grained, context-specific environment is captured in a single token. Within the LLM, atom and language tokens are jointly processed in a unified embedding space and interact through cross-attention. LoRA fine-tuning is then applied to help the LLM effectively incorporate the new tokens. As illustrated in Fig.~\ref{fig:1}(c), this unified framework enables a range of molecular tasks simply by modifying the task description, including reagent prediction, property prediction, forward reaction prediction, and retrosynthesis. Notably, AtomDisc’s tokenizer encodes molecular structures in an unsupervised, data-driven fashion, providing informative key-structural features while enhancing interpretability. Attention pattern analysis can link outputs to structure tokens representing meaningful substructures, explaining model decisions and potentially revealing new structure-property associations for scientific discovery.

\subsection{Atom-Level Tokenization Analysis}
Without explicit supervision, AtomDisc maps topological patterns around atoms into discrete tokens that can distinguish their chemical diversities and associations, which is shown in Fig.~\ref{fig2_Tokenizer}. 

\paragraph{AtomDisc maps atom-level structures to chemically meaningful tokens.}
Fig.~\ref{fig2_Tokenizer} (a) shows an overview of the atom-level embeddings during tokenization. The embeddings form clusters corresponding to chemically related motifs (e.g., oxygen-containing groups), with closer clusters reflecting greater chemical similarity. Analyzing the correspondence between tokens and functional groups reveals that most tokens exhibit low entropy in their functional group distribution (Fig.~\ref{fig2_Tokenizer}(b)), typically mapping to a single group (referred to as ``pure tokens''). This demonstrates that AtomDisc autonomously identifies chemically meaningful substructures, enabling LLMs to interpret their roles and apply chemical rules in downstream tasks. Notably, such a data-driven method captures subtle chemical environments beyond the predefined function groups, which we introduce in the subsequent paragraphs.

\paragraph{AtomDisc captures context-driven chemical similarities across different functional groups.}  
Fig.~\ref{fig2_Tokenizer}(b) also shows that while most tokens are group-specific, a subset exhibits higher entropy, indicating AtomDisc assigns certain atoms from distinct functional groups to the same ``mixture token.” This reveals that atoms from different groups can share chemically similar environments in particular contexts. Notably, the most common mixtures are alkene carbons with aromatic carbons, and ester oxygens with ether oxygens (Fig.~\ref{fig2_Tokenizer}(c)). Indeed, alkene and aromatic carbons share delocalized $\pi$-electronic characteristics, while ester and ether oxygens both function as electron-donating groups in oxygen-rich motifs. This indicates that AtomDisc effectively encode cross-group similarities by their local chemical context beyond conventional functional group boundaries. Focusing on token 20, a representative mixture token covering both aromatic and alkene carbons in conjugated systems, we observe that their chemical descriptor (e.g., Mulliken charge and PSA) distributions are distinct for pure tokens but nearly overlap when encoded by mixture token 20 (Supplementary Fig.~S4(b)). For instance, Mulliken charge distributions show separate peaks for pure tokens but merge into an intermediate peak for mixture token 20, indicating blended electronic character. In terms of polar surface area (PSA), pure-token alkene carbons have a higher peak at low PSA, while aromatic carbons show two similar peaks; mixture token 20 yields a unified, overlapping PSA distribution, distinct from either pure type. This suggests that embedding these carbons in conjugated frameworks with polar substituents homogenizes their solvation-accessible surface area. Indeed, mixture token 20 commonly appears in conjugated rings adjacent to polar substituents (Supplementary Fig.~S4(c)), such as hydroxyl or amino groups, which elevate and even out PSA values across both carbon types. This structural context also explains the intermediate charge and occupancy: extended conjugation delocalizes $\pi$-electron density, while nearby OH/NH groups introduce opposing inductive and resonance effects, pushing Mulliken charge and $\pi$-occupancy toward mid-range, resulting in overlapping distributions for token 20. Analysis of other mixture tokens (Supplementary Table~\blue{S14–15} and Fig.~\blue{S12–13}) shows similar trends, capturing convergent property distributions in various contexts. Detailed methods and statistics are in Section~\ref{sec:case_setup} and Supplementary Table~\blue{S13}.

\paragraph{AtomDisc captures context-driven chemical diversity for the same functional group.}  
For example, hydroxyl (OH) can be assigned to either token 319 or 338, each reflecting distinct structural features (Supplementary Fig.~S4(d)). Token 338 mainly corresponds to chain-like, solvent-exposed hydroxyls with high hydrogen-bond accessibility, while token 319 is typically attached to cyclic or ether rings, where steric hindrance limits solvent accessibility and increases polarization. Consequently, these tokens show distinct distributions in atomic descriptors (Supplementary Fig.~S4(a), Table~S12): token 338 has sharp, concentrated peaks in PSA and Mulliken charge, whereas token 319 exhibits broader distributions with lower PSA and higher polarization. Their $\pi$-electron occupancy distributions largely overlap, suggesting similar conjugation. 
Besides, we conduct a significance test on 45 property distribution pairs (three token pairs per group, five groups, three properties; Supplementary Fig.~S14-S28, Tables~S16–S30) and found 29 statistically significant differences, with each functional group having at least one significant token pair. This confirms that AtomDisc’s tokenization reliably captures chemically meaningful atom features, benefiting molecular understanding for LLM. For instance, swapping hydroxyl tokens alters prediction of aqueous solubility (LogS) —which are highly sensitive to hydrogen-bonding, polarity, and solvent accessibility—with token 338 increasing and token 319 decreasing the predicted value (Supplementary Fig.~S4(e), Table~S11). Moreover, different functional groups vary in token allocation—including token count, Average Token Utilization (ATU) (Fig.~\ref{fig2_Tokenizer}(d)), and the ratio of pure tokens (Fig.~\ref{fig2_Tokenizer}(e))—indicating that AtomDisc adaptively assigns tokens based on structural associations, thus distinguishing diverse chemical environments with few tokens. For example, alcohol–OH and aromatic carbons have similar ATU, but aromatic carbons are assigned far more tokens, reflecting greater local diversity. Alcohol–OH variability is dominated by a smaller set of continuous axes (hydrogen-bond acidity/basicity and polarity, modulated by steric accessibility) well captured by $pK_\mathrm{BHX}$ and related scales~\cite{OH_Case}, whereas aromatic carbons have context-dependent reactivity shaped by substituent electronics, positional effects, and conjugation topology—e.g., arenium-ion chemistry in superacids—resulting in a wider range of environments~\cite{aromatic_C_example}.

\subsection{Molecular Property Prediction}
\begin{table}[t]
\centering
\resizebox{\textwidth}{!}{
\begin{tabular}{lcccccccc}
\toprule
\textbf{Method}      & \textbf{BBBP↑} & \textbf{Tox21↑} & \textbf{ToxCast↑} & \textbf{Sider↑} & \textbf{ClinTox↑} & \textbf{HIV↑} & \textbf{BACE↑} & \textbf{Avg↑} \\
\midrule
KV-PLM~\cite{zeng2022deep}               & 72.0±0.9 &70.0±0.5 &55.0±1.7& 59.8±1.5& 89.2±2.7 &71.8±1.4& 78.5±2.7           & 70.9         \\
InstructMol~\cite{cao2023instructmol}          & 70.0           & 74.7            & 64.3            & 57.8            & 91.5            & 68.9           & 82.3           & 72.2         \\
UniMoT~\cite{zhang2024unimot} & 71.4           & 76.4            & 65.8            & 59.8            & 92.9            & 78.5           & 83.7           & 75.5         \\
MoMu~\cite{su2022molecular}      & 70.5$\pm$2.0 & 75.6$\pm$0.3 & 63.4$\pm$0.5 & 60.5$\pm$0.9 & 79.9$\pm$4.1 & 75.9$\pm$0.8 & 76.7$\pm$2.1 & 71.8         \\
MolLM~\cite{tang2024mollm}     & 75.7$\pm$1.7 &80.0$\pm$1.7 &68.2$\pm$0.4& \textbf{71.0$\pm$0.8} &91.1$\pm$1.0&  80.2$\pm$0.7 &84.1$\pm$0.9&78.6         \\
TokenMol~\cite{wang2025token}             & 93.4$\pm$0.1        & 82.9$\pm$0.5            & \textbf{74.6$\pm$1.2}& {64.4$\pm$2.0}& 92.7$\pm$2.1            & –               & \underline{89.6$\pm$1.5}& \underline{82.9}         \\
MoLFormer-XL~\cite{ross2022large}         & \underline{93.7}& \underline{84.7}& –                & {69.0}& \underline{94.8}& –               & 88.2           & -\\
AtomDisc               & \textbf{95.2$\pm$0.7}  & \textbf{85.6$\pm$0.5}   & \underline{73.9$\pm$0.9}   & \underline{70.5$\pm$1.1}   & \textbf{96.4$\pm$1.6}   & \textbf{81.4$\pm$0.3}  & \textbf{89.9$\pm$0.9}  & \textbf{84.7}\\
\bottomrule
\end{tabular}
}
\caption{
\textbf{ROC-AUC (\%) on MoleculeNet classification benchmarks. }
For each column, the best is \textbf{bolded} and the runner-up is \underline{underlined}.  
Our results are reported as the mean~$\pm$~standard deviation over five random seeds.  
Baseline numbers are taken directly from the results reported in the original papers.
}
\label{tab:molnet_classification_full}
\end{table}

We analyze the effectiveness of {AtomDisc} on diversified molecular property prediction tasks across the full suite of MoleculeNet~\cite{wu2018moleculenet}  datasets. The detailed experimental setup can be found in  Section~\ref{sec:classification_setup}.

\paragraph{AtomDisc boosts LLMs’ molecular property understanding, achieving SOTA prediction performance.}  As shown in Table~\ref{tab:molnet_classification_full}, {AtomDisc} achieves the strongest overall performance on MoleculeNet among LM-based methods. Crucially, this improvement does not come from scaling up data or model size, but from injecting explicit inductive biases through atom-level structural tokens. Unlike approaches that rely solely on SMILES strings (e.g., MoLFormer-XL~\cite{ross2022large}), global molecular embeddings (e.g., InstructMol~\cite{cao2023instructmol}), external continuous representations (e.g., UniMoT~\cite{zhang2024unimot}), or 3D modality features (e.g., TokenMol~\cite{wang2025token}), AtomDisc encodes (i) local structural priors, forcing the model to reason over each atom’s chemical environment, (ii) discrete structural tokens that integrate directly into the LLM’s native token space, allowing joint modeling with natural language tokens, and (iii) rich 2D topology information such as local connectivity patterns and atom-level physicochemical attributes (e.g., charge, polarity). These inductive biases provide richer and more chemically grounded supervision, enabling AtomDisc to surpass all the existing models. Especially, it surpasses MoLFormer-XL, which adopted far larger pretraining dataset. We also show in the Supplementary Table~\blue{S5} that AtomDisc's overall performance also surpasses leading \emph{Non-LM} methods such as KANO~\cite{fang2023knowledge}. We perform an ablation to isolate AtomDisc’s contribution: removing its structural tokens consistently reduces AUC across datasets (Fig.~3(a)), confirming their necessity.

\begin{figure}[t]
    \centering
    \includegraphics[width=\linewidth]{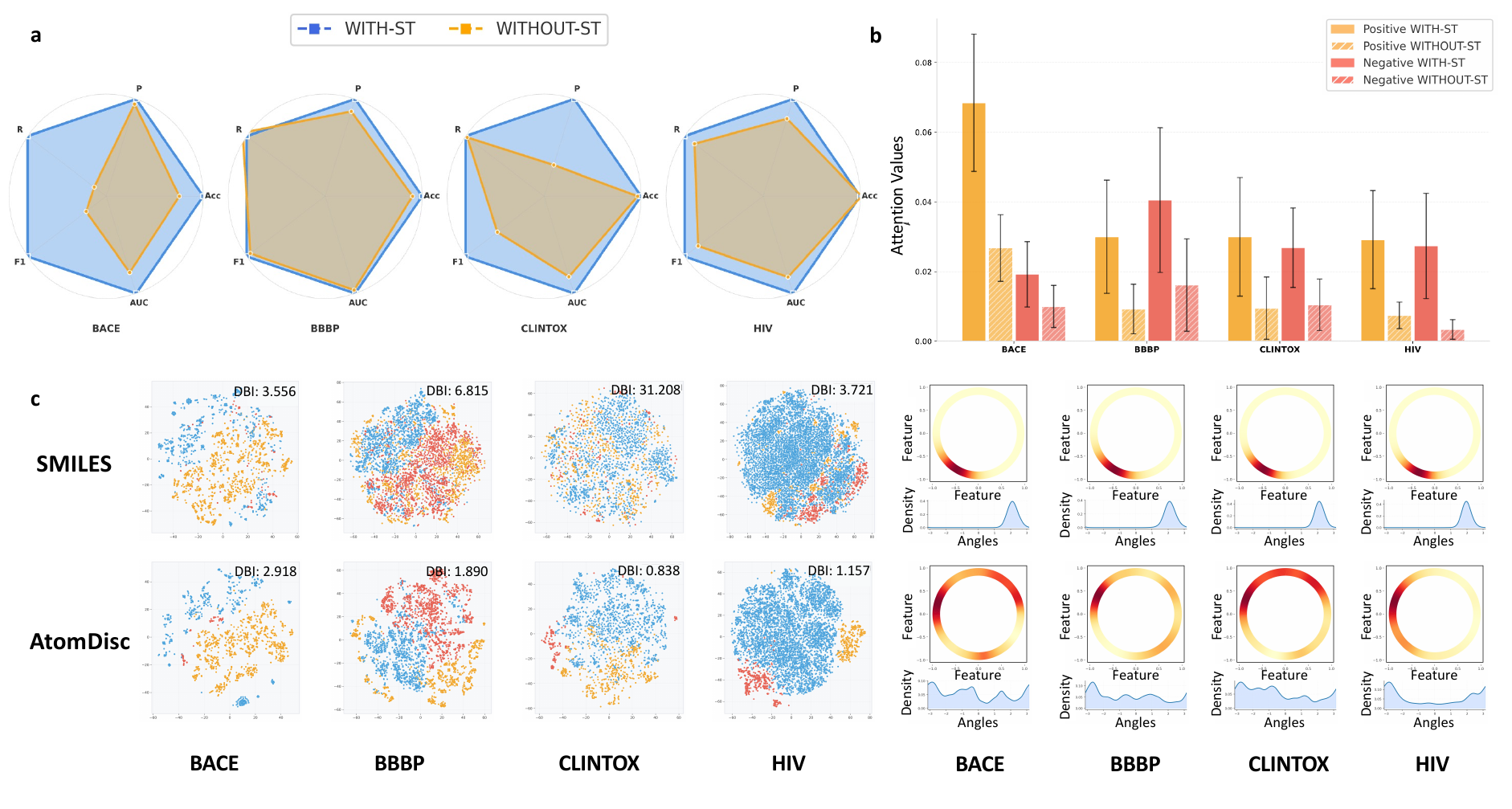}
    \caption{{Visualizations for property prediction.}
    \textbf{(a)} {Ablation study}: comparison of performance with vs. without ST.  
    \textbf{(b)} {Attention allocation}: attention values toward positive/negative functional groups in models trained with vs. without ST.
    \textbf{(c)} {Representation analysis}: for each functional group and its $r{=}2$ neighborhood, motif embeddings (mean-pooled from SMILES or AtomDisc tokens) are projected to $\mathbb{R}^2$ via t-SNE and colored by effect class (red: negative, yellow: positive, blue: neutral). Cluster compactness is quantified by the Davies–Bouldin (DB) Index. The ring plot shows the Gaussian KDE over angles $\theta=\arctan2(y,x)$ after normalized t-SNE coordinates $(x,y)$ to unit-circle; brighter regions denote higher density.
    }
    \label{fig:3_property_prediction_analysis}
\end{figure}

\paragraph{AtomDisc strengthens LLMs’ property-related functional-group grounding.}  
To examine how AtomDisc helps LLM understand structural features for property prediction, we categorize functional groups into three effect levels based on the empirical probability difference of molecules with and without $f$ showing that property: strong positive, strong negative, and weak or neutral effects, the details are shown in Section~\ref{sec:classification_setup} and Supplementary Fig.~S2. Accordingly, we analyze the LLM's attention on the related tokens (including both SMILES and AtomDisc tokens) for atoms in functional group with different levels of effects, with and without the AtomDisc structural token (ST), as shown in Fig.~\ref{fig:3_property_prediction_analysis}(b). It can be observed that armed with AtomDisc, our molecule LLM allocates significantly more attention to functional groups with strong positive or negative effects on the target property when making predictions, while the baseline without ST distributes attention more diffusely. This targeted focus demonstrates that ST guides the model toward functionally decisive chemical motifs, thereby strengthening both the accuracy and interpretability of property predictions.

\paragraph{AtomDisc enforces isotropic, property-discriminative motif embeddings.}

Beyond attention allocation, we also notice that our AtomDisc yields substantially more separable and chemically faithful representations than SMILES-only baselines. As shown in Fig.~\ref{fig:3_property_prediction_analysis}(c), AtomDisc embeddings form visibly tighter, better-separated clusters than the SMILES counterparts: positive (yellow) and negative (red) motifs aggregate into compact islands with clearer boundaries and far less mixing with neutral (blue), whereas SMILES embeddings remain diffuse and intermingled. The pattern holds across BACE, BBBP, ClinTox, and HIV, and is corroborated by lower Davies–Bouldin indices (BACE: 3.56→2.92; BBBP: 6.82→1.89; ClinTox: 31.21→1.16; HIV: 3.72→0.84). Moreover, the embedding distributions also show higher isotropies. In 2D Gaussian-KDE plots, the intensity spread around the full circle rather than concentrated in a single sector, implying balanced dispersion. Correspondingly, the angular KDE curves display low-amplitude, multi-peak profiles rather than one sharp lobe, suggesting local modes that reflect diversified chemical contexts. Such a uniform–yet–multimodal organization separates context-specific semantics more cleanly than a unimodal, directionally biased embedding, improving discriminability across contexts and thereby enhancing property-prediction performance.

Meanwhile, AtomDisc generalizes to regression tasks and outperforms existing methods on continuous molecular property prediction (Supplementary Table~\blue{S6}).

\subsection{Molecule Generation}
\label{sec:exp_molecule_generation}
\begin{figure}[htbp]
    \centering
    \includegraphics[width=\linewidth]{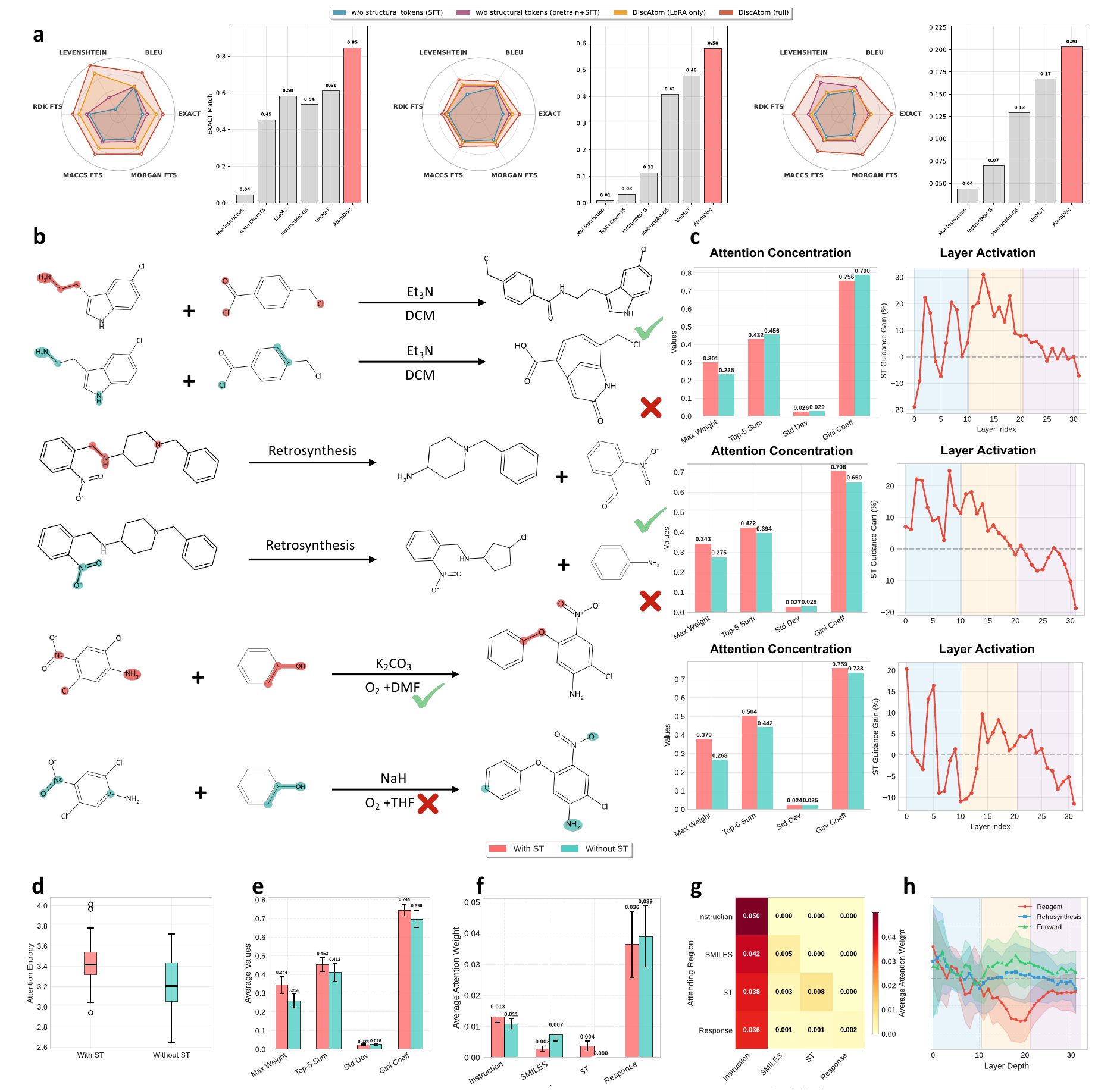}
    \caption{\textbf{Analysis of molecular generation tasks on Mol-Instructions datasets.}  
    \textbf{(a)} {Quantitative evaluation:} benchmark performance with baseline and ablation (with vs.\ without structural tokens, ST) comparisons; radial plots show multiple metrics, bar charts show dataset-level scores.  
    \textbf{(b)} {Case studies:} representative prediction examples from models with and without ST, highlighting top-3 attended atoms (red = with ST, blue = without ST). 
    \textbf{(c)} {Attention dynamics:} task-specific attention concentration and layer-wise activation; metrics include max weight, top-5 cumulative weight, standard deviation, and Gini coefficient. 
    \textbf{(d–h)} Dataset-level statistics: aggregate forward-prediction analysis covering \textbf{(d)} attention entropy distribution, \textbf{(e)} attention concentration,\textbf{(f)} regional attention, \textbf{(g)} cross-region bidirectional attention (color = weight intensity), and \textbf{(h)} layer-activation comparison across task. Mean values and standard-deviation bands are shown.}
    \label{fig:4_reaction_prediction}
\end{figure}

\begin{table}[htbp]
\centering
\footnotesize
\setlength{\tabcolsep}{3pt} 
\begin{tabular}{l lccccccc}
\toprule
\textbf{Task} & \textbf{Model} & \textbf{EXACT↑} & \textbf{BLEU↑} & \textbf{LEVENSHTEIN↓} & \textbf{RDK FTS↑} & \textbf{MACCS FTS↑} & \textbf{MORGAN FTS↑} & \textbf{VALIDITY↑} \\
\midrule
\multicolumn{9}{l}{\textit{Reagent Prediction}} \\
\midrule
& Alpaca$\dagger$~\cite{alpaca}                 & 0.000 & 0.026 & 29.037 & 0.029 & 0.016 & 0.001 & {0.186} \\
& Baize$\dagger$~\cite{xu2023baize} & 0.000 & 0.051 & 30.628 & 0.022 & 0.018 & 0.004 & 0.099 \\
& ChatGLM$\dagger$~\cite{zeng2023glm130bopenbilingualpretrained}                 & 0.000 & 0.019 & 29.169 & 0.017 & 0.006 & 0.002 & 0.074 \\
& LLaMA $\dagger$~\cite{touvron2023llamaopenefficientfoundation}                   & 0.000 & 0.003 & 28.040 & 0.037 & 0.001 & 0.001 & 0.001 \\
& Vicuna$\dagger$~\cite{vicuna2023} & 0.000 & 0.010 & 27.948 & 0.038 & 0.002 & 0.001 & 0.007 \\
& Mol-Instructions~\cite{fang2023mol}         & 0.044 & 0.224 & 23.167 & 0.237 & 0.364 & 0.213 & 1.000 \\
& InstructMol-G~\cite{cao2023instructmol}           & 0.070 & \textbf{0.890} & 24.732 & 0.469 & \textbf{0.691} & 0.426 & 1.000 \\
& InstructMol-GS~\cite{cao2023instructmol}          & 0.129 & 0.610 & 19.664 & 0.444 & 0.539 & 0.400 & 1.000 \\
& UniMoT~\cite{zhang2024unimot}                  & \underline{0.167} & \underline{0.728} & \textbf{14.588} & \underline{0.549} & 0.621 & \underline{0.507} & \underline{1.000}\\
& AtomDisc                  & \textbf{0.203} & 0.623 & \underline{15.362} & \textbf{0.583} & \underline{0.672} & \textbf{0.559} & \textbf{1.000} \\
\midrule
\multicolumn{9}{l}{\textit{Forward Prediction}} \\
\midrule
& Alpaca$\dagger$~\cite{alpaca}                  & 0.000 & 0.065 & 41.989 & 0.004 & 0.024 & 0.008 & 0.138 \\
& Baize$\dagger$~\cite{xu2019powerfulgraphneuralnetworks}                  & 0.000 & 0.044 & 41.500 & 0.004 & 0.025 & 0.009 & 0.097 \\
& ChatGLM$\dagger$~\cite{zeng2023glm130bopenbilingualpretrained}                 & 0.000 & 0.183 & 40.008 & 0.050 & 0.100 & 0.044 & 0.108 \\
& LLaMA $\dagger$~\cite{touvron2023llamaopenefficientfoundation}                  & 0.000 & 0.020 & 42.002 & 0.001 & 0.002 & 0.001 & 0.039 \\
& Vicuna$\dagger$~\cite{vicuna2023}                 & 0.000 & 0.057 & 41.690 & 0.007 & 0.016 & 0.006 & 0.059 \\
& Galactica-6.7B~\cite{taylor2022galactica}          & 0.000 & 0.468 & 35.021 & 0.156 & 0.257 & 0.097 & 0.946 \\
& Mol-Instructions~\cite{fang2023mol}     & 0.045 & 0.654 & 27.262 & 0.313 & 0.509 & 0.262 & 1.000 \\
& MolecularTransformer~\cite{schwaller2019molecular}    & 0.000 & 0.476 & 45.979 & 0.761 & 0.673 & 0.540 & 1.000 \\
& Text+ChemT5~\cite{Text+ChemT5}             & 0.454 & 0.602 & 26.545 & 0.729 & 0.773 & 0.700 & 0.851 \\
& LLaMo~\cite{park2024llamo}                   & 0.584 & 0.894 & 6.162  & 0.857 & 0.918 & 0.841 & 0.938 \\
& InstructMol-G~\cite{cao2023instructmol}           & 0.153 & 0.906 & 20.155 & 0.519 & 0.717 & 0.457 & 1.000 \\
& InstructMol-GS~\cite{cao2023instructmol}          & 0.536 & 0.967 & 10.851 & 0.776 & 0.878 & 0.741 & 1.000 \\
& UniMoT~\cite{zhang2024unimot}                 & \underline{0.611} & \textbf{0.980} & \underline{8.297} & \underline{0.836} & \underline{0.911} & \underline{0.807} & \underline{1.000} \\
& AtomDisc                  & \textbf{0.846} & \underline{0.971} & \textbf{2.366} & \textbf{0.949} & \textbf{0.966} & \textbf{0.940} & \textbf{1.000} \\
\midrule
\multicolumn{9}{l}{\textit{Retrosynthesis}} \\
\midrule
& Alpaca$\dagger$~\cite{alpaca}                  & 0.000 & 0.063 & 46.915 & 0.005 & 0.023 & 0.007 & 0.160 \\
& Baize$\dagger$~\cite{xu2019powerfulgraphneuralnetworks}                  & 0.000 & 0.095 & 44.714 & 0.025 & 0.050 & 0.023 & 0.112 \\
& ChatGLM$\dagger$~\cite{zeng2023glm130bopenbilingualpretrained}                 & 0.000 & 0.117 & 48.365 & 0.056 & 0.075 & 0.043 & 0.046 \\
& LLaMA $\dagger$~\cite{touvron2023llamaopenefficientfoundation}                  & 0.000 & 0.036 & 46.844 & 0.018 & 0.029 & 0.017 & 0.010 \\
& Vicuna$\dagger$~\cite{vicuna2023}                 & 0.000 & 0.057 & 46.877 & 0.025 & 0.030 & 0.021 & 0.017 \\
& Mol-Instructions~\cite{fang2023mol}         & 0.009 & 0.705 & 31.227 & 0.283 & 0.487 & 0.230 & 1.000 \\
& Text+ChemT5~\cite{Text+ChemT5}             & 0.033 & 0.314 & 88.672 & 0.457 & 0.469 & 0.350 & 0.632 \\
& InstructMol-G~\cite{cao2023instructmol}           & 0.114 & 0.586 & 21.271 & 0.422 & 0.523 & 0.285 & 1.000 \\
& InstructMol-GS~\cite{cao2023instructmol}          & 0.407 & \underline{0.941} & 13.967 & 0.753 & 0.852 & 0.714 & \underline{1.000} \\
& LLaMo~\cite{park2024llamo}                   & 0.341 & 0.830 & 12.263 & 0.793 & 0.868 & 0.750 & 0.954 \\
& UniMoT~\cite{zhang2024unimot}                 & \underline{0.478} & \textbf{0.974} & \underline{11.634} & \underline{0.810} & \underline{0.909} & \underline{0.771} & 1.000 \\
& AtomDisc                  & \textbf{0.580} & {0.907} & \textbf{8.340} & \textbf{0.880} & \textbf{0.916} & \textbf{0.830} & \textbf{1.000} \\
\bottomrule
\end{tabular}
\caption{
Performance on three molecular generation tasks: reagent prediction, forward reaction prediction, and retrosynthesis. Results are reported for exact match (EXACT), BLEU score (BLEU), Levenshtein distance (LEVENSHTEIN; lower is better), Tanimoto similarities computed over RDKit, MACCS, and Morgan fingerprints (RDK FTS, MACCS FTS, MORGAN FTS), and chemical validity (VALIDITY). In each column, the top-performing model is \textbf{bolded} and the runner-up is \underline{underlined}. $\dagger$ means few-shot ICL results from Mol-Instructions~\cite{fang2023mol}.
}
\label{tab:combined_tasks}
\end{table}

We evaluate the effectiveness of {AtomDisc} on generative molecular tasks, including reagent prediction, forward reaction prediction, and retrosynthesis. Building on the experimental setup outlined in Section~\ref{sec:generation_setup}, we assess our model using the Mol-Instructions dataset~\cite{fang2023mol}. These tasks provide a challenging testbed for understanding how well AtomDisc can generate molecules with realistic and syntactically valid structures. The results are shown in Fig.~\ref{fig:4_reaction_prediction} and Table~\ref{tab:combined_tasks}.

\paragraph{AtomDisc strengthens grasp of reaction mechanisms, driving SOTA molecular generation.} As shown in Fig.~\ref{fig:4_reaction_prediction}(a), incorporating AtomDisc's special token consistently increases LLM performance on molecule generation across different adaptation paradigms—fine-tuning alone or combined re-pretraining and fine-tuning (detailed statistics in Supplementary Table~\blue{S7}). As a result, our LLM armed with AtomDisc substantially outperforms both LM-based and domain-specific baselines (Table~\ref{tab:combined_tasks}), especially on exact match scores (Fig.~\ref{fig:4_reaction_prediction}(a)), which strictly assess if the model generates the completely (rather than partially) correct target, ensuring real-world chemical usability. 
Indeed, different from existing methods that only rely on SMILES (Mol-Instructions~\cite{fang2023mol}, MolecularTransformer~\cite{schwaller2019molecular}, Text+ChemT5~\cite{Text+ChemT5}) or global molecular embeddings (UniMoT~\cite{zhang2024unimot}, InstructMol~\cite{cao2023instructmol}), AtomDisc injects atom-level inductive bias—local topological information such as atomic electronic states and bond connectivity. This equips the LLM with a deeper mechanistic understanding of reactions, robustly improving generation quality. Notably, AtomDisc generalizes well and can generate chemically plausible alternatives beyond rote memorization, as shown in Supplementary Table~\blue{S9}–\blue{S10}.

\paragraph{AtomDisc enables LLMs’ reaction-related substructure localization.}
As illustrated in Fig.~\ref{fig:4_reaction_prediction}(b), incorporating Special Tokens (ST) of AtomDisc, the model’s attention concentrates on atoms that are meaningful for the reaction. Specifically, (1) in the forward prediction case, the model with ST assigns peak attention to the nucleophilic amine N of the first reactant and the acyl-chloride carbonyl C (and its leaving-group Cl) of the second reactant—i.e., the canonical nucleophile–electrophile pair for amide formation—yielding the correct product. However, the model without ST disperses attention over ring Cl and generic aromatic carbons and fails. (2) In retrosynthesis case, the model with ST assigns peak attention to the benzylic C–N bond linking the nitro-aryl unit and the amine, identifying the chemically plausible disconnection into a secondary amine and a nitrobenzaldehyde precursor. However, the model without ST instead fixates on nitro O/N atoms, which are spectators for this cut, and proposes incorrect precursors. (3) In reagent prediction for an SNAr-type aryl ether formation, the model with ST directs attention jointly to the phenolic O, the activated aryl C–Cl (para/ortho to –NO$_2$), and the nitro group that facilitates addition–elimination, leading to conditions that include a carbonate base and a polar aprotic medium. However, the model without ST shifts attention to nonreactive ring sites and nitro oxygens, producing implausible reagents and failure. 
Indeed, we observe a general increase in attention concentration metrics and a greater attention entropy after incorporating ST (Fig.~\ref{fig:4_reaction_prediction}(c), Fig.~\ref{fig:4_reaction_prediction}(d), Fig.~\ref{fig:4_reaction_prediction}(e)). This means ST helps alleviate diffuse, non-causal focus and anchors the LLM on the compact set of reaction-critical structures—nucleophile/electrophile centers, disconnection atoms, and activation motifs—needed for generation.
Notably, ST also enforces the multi-site focus required by particular reactions. For example, in the forward-prediction case, the elementary step involves two spatially separated reactive centers (e.g., a nucleophilic N and an acyl chloride). ST explicitly marks both centers and their leaving-group context, so the model deliberately splits attention across these few sites. Proper allocation should assign high attention to multiple atoms. In contrast, the model without ST shows a lower top-5 sum and Gini coefficient, implying over-concentration on a single—often incorrect—site.
Also, we find that ST increases activations in early layers (0–6), indicating heightened sensitivity to reactive centers and their local context during feature extraction. This can enhance model performance by providing stronger structural cues for subsequent reasoning stages.

\paragraph{AtomDisc facilitates structure-centric instruction information routing in molecular generation.} We partition the prompt into four regions—Instruction, SMILES, Structural Tokens(ST), and Response (the “\#\#\# Response” anchor; prompt examples in \blue{Supplementary Information C.1} )—to quantify how structural tokens reshape information flow. In the Fig.~\ref{fig:4_reaction_prediction}(f), adding ST shifts attention mass away from the Response anchor and toward the Instruction and molecular content (SMILES+ST). This indicates the model conditions more on the task specification and explicit structure. 
Fig.~\ref{fig:4_reaction_prediction}(g) further reveals that other regions devote little attention to ST, whereas ST actively attend to Instruction and SMILES. This implies our AtomDisc ST operate as \emph{communication hubs} that aggregate information from the prompt and relay it into the generative pathway. With ST, the model is better conditioned on structurally relevant cues. 
Finally, Fig.~\ref{fig:4_reaction_prediction}(h) shows task-dependent depth usage with a common pattern of elevated gains in the early layers when ST are included—suggesting that structural cues are injected and organized at shallow depths, stabilizing subsequent reasoning. For forward prediction, gains persist across most layers, plausibly because composing the product requires coordinating multiple reactive centers throughout the stack. In retrosynthesis, gains peak early and then taper: ST rapidly fix the candidate bond disconnection sites; once these sites are set, later layers mainly recover the corresponding precursors by using learned priors (SMILES grammar and common transformation patterns), so activation gain adds less. In reagent prediction, gains are mainly realized in the early stage, while later layers show little or negative gain. This pattern is consistent with the idea that the reagent class can be inferred once reactive roles are recognized from ST, which aligns with established understanding in chemical research. Taken together, in addition to facilitating the localization of important chemical structures, ST also acts as a high-signal router for feature extraction from different signals that promotes instruction understanding centered around molecular structures and guides task-dependent optimization of the modeling process.


\section{Discussion}


In this work, we introduced AtomDisc, an atom-level tokenizer that bridges the divide between linear SMILES notations and the rich, graph-based nature of molecular structures. Our experiments demonstrate that AtomDisc effectively bridges molecular structure and language understanding, enabling large language models (LLMs) to reason over chemical systems in a more mechanistic, interpretable, and data-efficient manner. Across diverse benchmarks—from molecular property prediction to generative tasks such as reagent prediction, forward reaction prediction, and retrosynthesis—AtomDisc consistently improves both quantitative performance and qualitative interpretability. At the molecular property level, AtomDisc enhances LLMs’ grounding in chemically meaningful substructures, leading to sharper functional-group discrimination and isotropic, well-separated motif embeddings. These embeddings capture context-specific reactivity and physicochemical variations, suggesting that the atom-level inductive bias introduced by AtomDisc allows the model to internalize fine-grained chemical principles. At the generative level, the introduction of AtomDisc’s structural tokens provides the LLM with explicit cues about reactive centers and their electronic environments. This yields a stronger grasp of reaction mechanisms, better attention localization on causal atomic sites, and improved reaction condition inference. The observed shift of information flow toward instruction and structural regions further indicates that AtomDisc facilitates structure-centric information routing, enabling chemically aware reasoning paths within the LLM architecture.

Beyond raw performance, the broader impact of AtomDisc lies in its potential to reshape how LLMs contribute to scientific discovery. At the task level, AtomDisc provides explicit interpretability: predictions can be directly traced to reactive centers, functional groups, and even context-dependent token assignments, enabling chemists to understand not only what the model predicts but also why. By uncovering implicit structure–property associations—such as differentiating hydroxyl environments and linking them to solubility—AtomDisc demonstrates the capacity to rediscover fundamental chemical principles without explicit supervision. Moreover, its emergent “mixture tokens” capture shared or transitional chemical contexts, allowing a richer and more continuous understanding of chemical space than rigid human-defined categories permit. Taken together, these capabilities position AtomDisc-empowered LLMs not merely as task-specific predictors but as general scientific assistants capable of guiding hypothesis generation, revealing hidden mechanistic insights, and accelerating discovery pipelines in drug design, catalysis, and materials science. Ultimately, AtomDisc establishes a foundation for a new paradigm in which AI and chemists collaborate through a shared, interpretable atomic language, advancing chemistry toward interpretable and data-driven innovation. Future work will focus on developing a unified model that integrates both 2D and 3D molecular representations, bridging topological and spatial information within a single framework. Such integration could enable AtomDisc-based LLMs to reason over conformations, reaction pathways, and stereochemical effects in a unified chemical space.

\section{Methods}
\label{sec:methods}
To augment SMILES representations with explicit structural context and establish a unified token space for both chemical structure and natural language semantics, we introduce \textbf{AtomDisc}. Our framework discretizes the local chemical environment of each atom in a molecule and represents it as a special token. This is enabled by the \textbf{AtomDisc Tokenizer} (\blue{Section}~\ref{sec:architecture}), which transforms continuous, atom-level structural features into discrete tokens that are seamlessly integrated into the autoregressive input sequence of a large language model (LLM).

We employ a four-stage training pipeline: (1) training an atomic codebook to discretize structural representations (\blue{Section}~\ref{sec:training_atom_codebook}); (2) learning a projection network to align structure token embeddings with the LLM's latent space (\blue{Section}~\ref{sec:training_projector}); (3) pretraining the base language model on multiple tasks using the extended vocabulary (\blue{Section}~\ref{sec:training_base}); and (4) fine-tuning the model on downstream molecular tasks (\blue{Section}~\ref{sec:training_downstream}). This staged approach allows AtomDisc to achieve state-of-the-art performance across a range of molecular understanding and generation benchmarks. The training procedures, hyper-parameters, and datasets details are provided in Supplementary Table~S1 and S2.

\subsection{AtomDisc Tokenizer Architecture}
\label{sec:architecture}

As shown in Fig.~\ref{fig:1}(b), the AtomDisc tokenizer operates in three main stages to convert a given molecule into discrete tokens interpretable by the LLM. First, the molecular graph is embedded using a pretrained graph encoder. Next, each atom embedding is mapped to the closest codeword $k$ in the vector-quantized codebook via an argmin operation. Finally, the corresponding index is used to generate a structural token of the form \texttt{<atom\_k>}.

\paragraph{Molecule Encoder.}
To encode molecular structure, we leverage the GIN-based graph encoder from MoleculeSTM~\cite{liu2023multi}. This encoder is pretrained on the GEOM~\cite{axelrod2022geom} dataset and further refined on molecule-text pairs from PubChem~\cite{pubchem}. While MoleculeSTM typically aggregates atom representations into a single global embedding for the entire molecule, we utilize the atom-wise outputs directly. For a molecular graph constructed from a SMILES string, the encoder produces a set of atom-level embeddings, $\mathbf{H} = \{ \mathbf{h}_i \in \mathbb{R}^{d_{\text{enc}}} \}_{i=1}^{N}$, where $N$ is the number of atoms and $\mathbf{h}_i$ is the $d_{\text{enc}}$-dimensional feature vector for atom $i$. These fine-grained embeddings serve as the input for our atom-level tokenization.

\paragraph{Atom-level Discretization via Vector Quantization.}
To convert the continuous atom embeddings into interpretable, discrete tokens, we apply vector quantization (VQ). Given the embedding $\mathbf{h}_i$ for atom $i$, we assign it to the nearest vector in a learnable codebook $\mathcal{E} = \{\mathbf{e}_1, \dots, \mathbf{e}_K\}$, where each codeword $\mathbf{e}_k \in \mathbb{R}^{d_{\text{enc}}}$ and $K$ is the codebook size. The assignment is determined by finding the index $z_i$ that minimizes the Euclidean distance:
$$z_i = \arg\min_{k \in \{1, \dots, K\}} \|\mathbf{h}_i - \mathbf{e}_k\|_2.$$
The resulting integer index $z_i$ functions as a discrete token representing the local chemical environment of atom $i$. Each index is then mapped to a unique symbolic token, denoted \texttt{<atom\_$z_i$>}, for integration into the LLM's input sequence.

\paragraph{Embedding Alignment and Vocabulary Augmentation.}
To integrate the structural tokens into the LLM, we align the codebook embeddings with the LLM's latent space. Each codebook vector $\mathbf{e}_k$ is projected by a learnable multilayer perceptron (MLP), $f_{\theta}$, into the LLM's embedding dimension, $d_{\text{llm}}$:
$$\tilde{\mathbf{e}}_k = f_{\theta}(\mathbf{e}_k), \quad \text{where } \tilde{\mathbf{e}}_k \in \mathbb{R}^{d_{\text{llm}}}.$$
The LLM's vocabulary is then augmented with these $K$ new structural tokens, and its embedding matrix is extended by $K$ corresponding rows, each initialized with the projected vector $\tilde{\mathbf{e}}_k$. In practice, we reserve special tokens such as \texttt{<atom\_k>} for each structural code, as well as \texttt{<mol>} and \texttt{</mol>} to explicitly delimit the molecular structure span in the input sequence. This allows the LLM to process structural and textual inputs within a unified sequence, using a shared modeling interface and standard autoregressive objectives. A detailed comparison of our structural tokenization scheme with traditional SMILES tokenization is provided in Supplementary Table \blue{S3}.

\subsection{Stage-1: Training an Atomic Codebook}
\label{sec:training_atom_codebook}
We train a VQ module to learn a codebook for discretizing the continuous atom-level embeddings. Using a frozen MoleculeSTM encoder, we generate atom embeddings for nearly 220,000 molecules with 7,000,000 atoms sampled from PubChem~\cite{pubchem}. The VQ module is trained to map each atom embedding $\mathbf{h}_i$ to its closest codeword $\mathbf{e}_{z_i}$ while ensuring that the quantized representation remains informative. The module, which consists of the codebook $\mathcal{E}$ and a reconstruction decoder $D$, is optimized using a loss function with two components:
$$
\mathcal{L}_{\text{VQ}} = \underbrace{\| \mathbf{h}_i - D(\mathbf{e}_{z_i}) \|^2_2}_{\text{reconstruction loss}} + \beta\underbrace{\| \mathbf{h}_i - \text{sg}(\mathbf{e}_{z_i}) \|^2_2}_{\text{commitment loss}},
$$
where $\text{sg}(\cdot)$ is the stop-gradient operator and $\beta$ is a scaling hyperparameter that balances the trade-off between reconstruction accuracy and the stability of the encoder–codebook interaction.. The reconstruction loss trains the decoder $D$ to accurately reconstruct the original atom embedding from the quantized vector. The codebook loss updates the codeword $\mathbf{e}_{z_i}$ to move closer to the encoder output $\mathbf{h}_i$, effectively training the codebook. The training details are provided in \blue{Supplementary Information A.3}.

\subsection{Stage-2: Training Embeddings for Atomic Tokens}
\label{sec:training_projector}

In this stage, we adopt a \textbf{“runtime projection, bake after convergence”} strategy to integrate structural codewords into the LLM’s embedding space. Concretely, we freeze all parameters of both the molecule encoder and the LLM, and train only a lightweight MLP projector $f_\theta: \mathbb{R}^{d_{\text{enc}}} \to \mathbb{R}^{d_{\text{llm}}}$. At runtime, each learned structural codeword $\mathbf{e}_k$ is mapped through the projector to a projected vector $\tilde{\mathbf{e}}_k = f_\theta(\mathbf{e}_k)$, which is dynamically returned during embedding lookup for the corresponding structural token \texttt{<atom\_$k$>}. This setup ensures that, during training, the autoregressive language modeling loss:
$$\mathcal{L}_{\text{LM}} = -\sum_{t=1}^{T} \log p(x_t \mid x_{<t}),$$
—applied to sequences containing both SMILES characters and structural tokens—updates only the parameters of $f_\theta$.

After the projector converges, we \emph{bake} the learned vectors $\tilde{\mathbf{e}}_k$ directly into the LLM’s embedding matrix by overwriting the rows corresponding to \texttt{<atom\_$k$>}. The projector is then discarded, so that inference proceeds without additional computational overhead. Detailed training settings are provided in \blue{Supplementary Information A.4}.
\subsection{Stage-3: Multi-task Pretraining of the Base Model}
\label{sec:training_base}
We adopt LLaMA-2-7B~\cite{llama2} as the base language model. The model’s vocabulary and embedding matrix are augmented with the structural tokens and their corresponding embeddings learned in Stage-2. To enable efficient adaptation, we freeze the LLM backbone and introduce lightweight LoRA adapters into the attention layers. The model is then optimized with the standard autoregressive loss, $\mathcal{L}_{\text{LM}}$, on a multi-task instruction-tuning corpus covering multiple downstream tasks. Through this process, the model learns to effectively exploit the structural tokens across a broad range of chemical reasoning tasks. Detailed training configurations are provided in \blue{Supplementary Information A.5}.

\subsection{Stage-4: Supervised Fine-Tuning on Downstream Tasks}
\label{sec:training_downstream}
In the final stage, we perform supervised fine-tuning (SFT) on specific downstream tasks. Building on the structure-aware base model from Stage-3, we continue to keep the GNN encoder and the LLM backbone frozen, while further updating the LoRA adapters and, optionally, the structure token embeddings. Fine-tuning is performed on task-specific instruction datasets, where each example (e.g., for forward reaction prediction) is formulated as an instruction-response pair. For each input SMILES string, structural tokens are generated on-the-fly using the frozen GNN and VQ modules, enabling the construction of structure-aware prompts. Evaluation details and calculation of metrics for each task are provided in \blue{Supplementary Information B.2}.

\subsection{Baseline}

For property prediction tasks, we compare AtomDisc against a range of strong baselines, including both non-LM-based and LM-based methods. The non-LM-based models comprise MoleculeSTM~\cite{liu2023multi}, UniMol~\cite{zhou2023uni}, KPGT~\cite{li2022kpgt}, and KANO~\cite{fang2023knowledge}, among others, while the LM-based approaches include global 2D representation based methods like UniMoT~\cite{zhang2024unimot} and InstructionMol~\cite{cao2023instructmol}, SMILES based method like MoLFormer-XL~\cite{ross2022large}, 3D structural information based method Token-Mol~\cite{wang2025token}, and related models. All baseline results are reported from the respective original publications for comparability.

For generation tasks, we compare AtomDisc with a broad set of large language models and molecular generation baselines, including general-purpose LLMs (e.g., Alpaca~\cite{alpaca}, ChatGLM~\cite{zeng2023glm130bopenbilingualpretrained}, LLaMA~\cite{touvron2023llamaopenefficientfoundation}, Vicuna~\cite{vicuna2023}), molecularly fine-tuned LLMs (e.g., Mol-Instructions~\cite{fang2023mol}, and state-of-the-art specialized models such as UniMoT~\cite{zhang2024unimot}, InstructionMol~\cite{cao2023instructmol}, and LLaMo~\cite{park2024llamo}. For all few-shot ICL baselines, we directly report published results from Mol-Instructions~\cite{fang2023mol}.


\subsection{Experimental Setup for Atomic Descriptors}
\label{sec:case_setup}
To illustrate how AtomDisc provides chemically interpretable insights beyond classical functional group definitions, we conduct two representative case studies that highlight both the specificity and ambiguity captured by our atom-level tokenization. Specifically, we evaluated three atomic properties that collectively capture electronic, steric, and chemical environment effects: Mulliken charge (reflecting local electronic polarization and reactivity), local $\pi$-electron occupancy (probing aromaticity and conjugation), and local polar surface area (PSA) (a well-established descriptor in drug discovery, correlated with aqueous solubility and membrane permeability\cite{poongavanam2024molecular}). Mulliken charge and $\pi$-electron occupancy were computed via DFT and orbital projection using PySCF, while PSA was calculated as the sum of polar surface area within a 4-bond environment using RDKit. 

\subsection{Experimental Setup for Molecular Property Classification}
\label{sec:classification_setup}
We evaluate AtomDisc on a suite of molecular property classification benchmarks from MoleculeNet~\cite{wu2018moleculenet}. Dataset descriptions and task definitions are provided in \blue{Supplementary Information B.1}, with evaluation metrics detailed in \blue{Supplementary Information B.2}. For all classification tasks, we report the area under the receiver operating characteristic curve (ROC-AUC, \%) as the primary metric with five random seeds.
Following the official recommendation, each dataset uses scaffold splitting into training, validation, and test sets in an 8:1:1 ratio. All results are obtained using our stage-2 AtomDisc model, where the 300-dimensional embeddings of structural tokens are projected into the LLM’s native embedding space and integrated into its vocabulary, without any additional multi-task pretraining. On top of this stage-2 backbone, we apply lightweight LoRA fine-tuning for each downstream classification task. This setup isolates the contribution of AtomDisc itself—separating it from potential confounding effects of large-scale pretraining—while still enabling efficient adaptation across diverse benchmarks. Each property prediction task is cast as a generative classification problem: the model is prompted with a molecule’s SMILES augmented by structural tokens, and tasked to autoregressively generate a single output token—\texttt{yes} or \texttt{no}—as the predicted label. ROC-AUC is then computed from the logits corresponding to these two outputs.
For multi-task classification tasks such as SIDER, we leverage prompt engineering to specify the property of interest, enabling a single model to perform multiple prediction tasks within a unified architecture. In cases where a particular subtask label for a molecule is missing, we discard only that subtask instance while retaining the molecule’s labels for all other subtasks. Illustrative prompt templates are provided in Supplementary Table~S4.
To systematically examine how AtomDisc influences the model’s attention to chemically salient functional groups (FGs), we first identify—separately for each dataset—which FGs are associated with the target outcome. For every FG $f$, we compute a label-based effect $E(f)$ as the empirical probability difference: 
$$E(f)=\hat{P}(y{=}1\mid f)-\hat{P}(y{=}1\mid \neg f),$$
where $\hat{P}(\cdot)$ denotes empirical probabilities are estimated from counts: 
$$\hat{P}(y{=}1\mid f)\;=\;\frac{n_{+}(f)}{n(f)},\
\hat{P}(y{=}1\mid \neg f)\;=\;\frac{n_{+}(\neg f)}{n(\neg f)}.$$ 
Here, $n(f)$ is the number of molecules containing FG $f$ and $n_{+}(f)$ the number of those labeled $y{=}1$; $n(\neg f)$ and $n_{+}(\neg f)$ are defined analogously for molecules that do not contain $f$. We then assign a categorical polarity to each FG using a fixed threshold $\tau=0.05$: Positive if $E(f)\ge\tau$, Negative if $E(f)\le -\tau$, and Neutral otherwise. The complete per-dataset inventories and assigned FG categories are reported in \blue{Supplementary Information, B.3.1}.

\subsection{Experimental Setup for Molecular Generation Tasks}
\label{sec:generation_setup}
We evaluate AtomDisc on three challenging molecule generation benchmarks: reagent prediction, forward reaction prediction, and retrosynthesis from the Mol-Instructions datasets~\cite{fang2023mol}. For each task, models are assessed using a comprehensive suite of metrics, including exact match (EXACT), BLEU score, Levenshtein distance, fingerprint-based Tanimoto similarities (RDK, MACCS, MORGAN), and chemical validity. Task definitions and detailed evaluation protocols are described in \blue{Supplementary Information B.1-B.2}, and representative instruction prompts and model output cases are provided in \blue{Supplementary Information C.1}. AtomDisc augments all SMILES inputs with discretized structure tokens, enabling unified sequence modeling across tasks. All results are obtained using the pretrained AtomDisc model, followed by supervised fine-tuning with LoRA adapters.

\section*{Data Availability}
All datasets used in this study are publicly available:

\begin{itemize}
  \item \textbf{Mol-Instructions}: \href{https://huggingface.co/datasets/zjunlp/Mol-Instructions}{Hugging Face — zjunlp/Mol-Instructions} (CC BY 4.0).
  \item \textbf{MoleculeNet} benchmarks: \href{https://moleculenet.org}{moleculenet.org} and \href{https://deepchem.readthedocs.io/en/latest/moleculenet.html}{DeepChem MoleculeNet docs}.  
  The curated distribution in DeepChem follows the MIT License; individual source datasets retain their original terms.
  \item \textbf{ChEBI-20-MM}: \href{https://huggingface.co/datasets/liupf/ChEBI-20-MM}{Hugging Face — liupf/ChEBI-20-MM} (MIT).
  \item \textbf{ChEBI-Augmented (this work)}: \href{https://huggingface.co/datasets/anonymous041/ChEBI-Augmented}{Hugging Face — anonymous041/ChEBI-Augmented} (Apache-2.0).
  \item \textbf{PubChem} Dataset: \href{https://ftp.ncbi.nlm.nih.gov/pubchem/Compound/Extras/}{PubChem Compound FTP — CID–SMILES mapping file (\texttt{CID-SMILES.gz})}.  
  This dataset is maintained by the National Center for Biotechnology Information (NCBI) as part of the PubChem Compound Database and is publicly available under the \href{https://creativecommons.org/publicdomain/mark/1.0/}{Public Domain (CC0)} dedication.
\end{itemize}
\section*{Code Availability}
The source code for AtomDisc is available at \url{https://anonymous.4open.science/r/AtomDisc-F197}. 
Please refer to the repository for installation and usage instructions. 
The code is released under the license specified in the repository. 
This work builds upon LLaMA-2-7B; users must accept the LLaMA 2 Community License to obtain the base weights.




\section*{Inclusion \& Ethics Statement}
We confirm that this study adheres to all applicable ethical regulations. 

\bibliography{ref}

\newpage

\makeatletter
\begingroup
  \count@=\value{page}\advance\count@ by -1 
  \protected@write\@auxout{}{\string\newlabel{MainLastPage}{{}{\number\count@}}}%
\endgroup
\makeatother

\appendix

\rfoot{\small\sffamily\bfseries\thepage/\pageref{LastPage}}

\clearpage
\begin{bibunit}[naturemag] 

\setcounter{page}{1} 
\setcounter{figure}{0}
\setcounter{table}{0}
\renewcommand{\thefigure}{S\arabic{figure}} 
\renewcommand{\thetable}{S\arabic{table}}   
\makeatletter
\renewcommand{\theHfigure}{S\arabic{figure}} 
\renewcommand{\theHtable}{S\arabic{table}}   
\makeatother
\setcounter{secnumdepth}{3}
\titlecontents{section}
  [3.2em]{\small\bfseries}{\contentslabel{1.6em}}{}%
  {\titlerule*[0.5pc]{.}\contentspage}[\vspace{4pt}]

\titlecontents{subsection}
  [5em]{\small}{\contentslabel{2.3em}}{}%
  {\titlerule*[0.5pc]{.}\contentspage}[\vspace{2pt}]

\titlecontents{subsubsection}
  [7em]{\small}{\contentslabel{3.2em}}{}%
  {\titlerule*[0.5pc]{.}\contentspage}[\vspace{2pt}]

\startcontents[app] 
\newcommand{\printappendixtoc}{%
  \section*{Appendix Contents}%
  \setcounter{tocdepth}{3}
  \printcontents[app]{l}{1}{}%
  \clearpage
}
\printappendixtoc
\section{Training Details}

\subsection{Hyper-parameters}
\label{sec:hyper-parameters}

\begin{table}[ht]
\centering
\caption{Detailed training hyperparameters across the three stages.}
\label{tab:training-params}
\begin{tabular}{lccc}
\toprule
\textbf{Configuration} & \textbf{Training Codebook} & \textbf{Training Projector} & \textbf{Training LoRA} \\
\midrule
Molecule Encoder       & GIN (MoleculeSTM)   & GIN (MoleculeSTM)    & GIN (MoleculeSTM) \\
LLM Base               & -                   & LLaMA-2-7B            & LLaMA-2-7B \\
Codebook Size          & 512                 & 512                   & 512 \\
Molecule Emb. Dim.     & 300                 & 300                   & 300 \\
Projection Layer       & -                   & MLP                   & MLP \\
Number of Special Tokens    & -              & 514            & 514  \\
Batch Size             & 64                  & 16                    & 4 \\
Epochs                 & 10                  & 5                    & 10 \\
Optimizer              & AdamW               & AdamW                 & AdamW \\
Learning Rate          & 1e-4                & 2e-5                  & 2e-5 \\
Weight Decay           & -                & 1e-2                  & 1e-2 \\
Warmup Ratio           & -                 & 0.1                   & 0.1 \\
Seed                   & 42                  & 42                    & 42\\
LR Scheduler           & Cosine              & Cosine                & Cosine \\
Precision              & float32            & bfloat16              & bfloat16 \\
LoRA Config            & -                   & -                     & $r=8,\ \alpha=32,\ \text{dropout}=0.05$ \\
Gradient Accum. Steps  & -                   & 6                     & 6 \\
Max Prompt Length      & -                   & 2048                  & 2048 \\
Max Response Length    & -                   & 512                   & 512 \\
\bottomrule
\end{tabular}
\end{table}
\newpage

\subsection{Datasets}
\label{sec:datasets}

To train our molecular instruction-tuned models, we constructed a large and diverse dataset by mixing data from Mol-Instructions sources, following a multi-task setup to enhance instruction-following capability.

\paragraph{ChEBI-20 Augmented Dataset.}
We first leveraged the ChEBI database, which contains curated chemical entities and their hierarchical relationships. Starting from the official ChEBI graph database\footnote{\url{https://ftp.ebi.ac.uk/pub/databases/chebi/}}, we extracted both the molecular descriptions and the ontology hierarchy, such as “is-a” and other parent-child relationships. Each entry was converted into natural language statements describing the molecule and its context within the ChEBI hierarchy (e.g., “This molecule is a steroid and belongs to the class of…”). For every valid entry, we retained the SMILES string and associated textual description. We merged these new samples with the original ChEBI-20~\cite{molt5} dataset, deduplicated the entries, and verified SMILES validity using RDKit. After cleaning, the final dataset contained approximately 100,000 high-quality (text, SMILES) pairs. We constructed both “mol2text” (SMILES-to-description) and “text2mol” (description-to-SMILES) formats for multi-directional instruction tuning.

\paragraph{Mol-Instructions Dataset}
The Mol-Instructions collection provides diverse molecular generative tasks. We specifically utilized the training split of this resource, including three sub-tasks: (1) forward reaction prediction, (2) retrosynthesis prediction, and (3) reagent prediction. Each sub-task comprises about 120,000 samples, yielding a total of approximately 370,000 training pairs in this component. Each entry provides task-specific prompts and target SMILES for molecular generation, allowing us to train the model in both “instruction-to-molecule” and “molecule-to-instruction” directions.

By integrating these resources, our training set contains over 570,000 instruction–molecule pairs, supporting robust learning for diverse molecular tasks.

\begin{table}[htbp]
    \centering
    \caption{Summary of datasets used for molecular instruction tuning.}
    \label{tab:dataset_summary}
    \begin{tabular}{lll}
        \toprule
        Dataset & Number of Samples & Task / Description \\
        \midrule
        ChEBI-20 Augmented & $\sim$$100,000 \times 2$ & Hierarchical description $\leftrightarrow$ SMILES \\
        Mol-Instructions: Forward Reaction Prediction & $\sim$120,000 & Reactants $\to$ Product \\
        Mol-Instructions: Retrosynthesis Prediction & $\sim$120,000 & Product $\to$ Reactants \\
        Mol-Instructions: Reagent Prediction & $\sim$120,000 & Reagents $\to$ Product \\
        \bottomrule
    \end{tabular}
\end{table}

\newpage
\subsection{Training Details of Codebook}
\label{sec:training_codebook_details}
AtomDisc provides a unified framework for incorporating interpretable, atom-level structural information into large language models (LLMs). Given an input molecule, a frozen graph encoder extracts atom-wise representations that capture each atom’s unique chemical environment. These continuous atom embeddings are then discretized using a learned codebook, producing a sequence of interpretable structure tokens—each corresponding to an individual atom in the molecule. 

To enable seamless integration with language models, these structure tokens are embedded directly into the input token sequence, interleaved with natural language instructions or SMILES tokens. The resulting unified sequence is then processed by an LLM (e.g., LLaMA-2) equipped with parameter-efficient fine-tuning (LoRA), allowing the model to jointly reason over both linguistic and structural information at the atomic scale. By grounding molecular representation at the atom level and maintaining interpretability throughout the modeling pipeline, AtomDisc not only delivers strong performance on generative and predictive tasks, but also enables direct attribution of model outputs to specific chemical substructures. This approach opens up new avenues for scientific discovery by revealing how large language models internalize and leverage detailed molecular features.

We train our atom-level structure tokenizer using a vector-quantized autoencoder (VQ-VAE) on a curated subset of approximate 220,000 molecules from PubChem~\cite{pubchem}, comprising a total of 7.8 million atoms. Each SMILES string is first converted into a molecular graph via RDKit~\cite{rdkit}, and then encoded into a 300-dimensional atom embedding using a frozen, pretrained MoleculeSTM~\cite{liu2023multi} graph encoder.

To discretize these continuous atom representations, we employ a VQ module with a learnable codebook of size 512. The reason why we choose 512 as the codebook size is given in \blue{Supplementary Information B.3.6}. Each atom embedding $\mathbf{h}_i \in \mathbb{R}^{300}$ is assigned to its nearest codeword $\mathbf{e}_{z_i}$ by minimizing the Euclidean distance, while a decoder $f_{\text{dec}}$ is trained to reconstruct the original embedding. The model is optimized with the standard VQ-VAE loss:
$$
\mathcal{L}_{\text{VQ}} = \| \mathbf{h}_i - f_{\text{dec}}(\mathbf{e}_{z_i}) \|^2 + \beta \left( \| \mathbf{h}_i - \operatorname{sg}[\mathbf{e}_{z_i}] \|^2 \right),
$$
where $\beta$ balances the reconstruction and commitment losses.

To mitigate codebook collapse and encourage diverse code utilization, we initialize the codebook via $k$-means clustering over a randomly sampled set of 500,000 atom embeddings. This initialization ensures that over 90\% of codewords are utilized within the first epoch. The VQ-VAE is then trained for 10 epochs to convergence using the entire dataset. As a result, we obtain a discretizer that maps each atom in a molecule to a symbolic token \texttt{<atom\_$z_i$>} representing its unique structural context.

This discrete tokenizer enables fine-grained differentiation between atoms of the same element type in varying chemical environments. For instance, two carbon atoms with different hybridizations or neighboring groups can be mapped to distinct codewords. Each code thus encapsulates a reusable, structure-aware atom-level pattern, capturing local chemical semantics beyond simple atom type—forming the foundation for symbolic structural augmentation in downstream molecular modeling.

\begin{table}[htbp]
\centering
\renewcommand{\arraystretch}{1.1}
\scriptsize
\begin{tabularx}{\textwidth}{l|X|X}
\toprule
\textbf{Molecule} & \textbf{SMILES Token Sequence} & \textbf{AtomDisc Token Sequence} \\
\midrule
CCI &
\texttt{CC\quad I} &
\texttt{\detokenize{<atom_406>}\quad\detokenize{<atom_406>}\quad\detokenize{<atom_406>}} \\
\midrule
O=C(O)C1=CC=CC=C1Br &
\texttt{O\quad =\quad C\quad (\quad O\quad )\quad C\quad 1\quad =\quad CC\quad =\quad CC\quad =\quad C\quad 1\quad Br}
&
\texttt{\detokenize{<atom_433>}\quad =\quad \detokenize{<atom_140>}\quad (\quad \detokenize{<atom_449>}\quad )\quad \detokenize{<atom_344>}\quad 1\quad =\quad \detokenize{<atom_353>}\quad \detokenize{<atom_353>}\quad =\quad \detokenize{<atom_353>}\quad \detokenize{<atom_344>}\quad =\quad \detokenize{<atom_344>}\quad 1\quad \detokenize{<atom_344>}}
\\
\midrule
CN(C)C=O &
\texttt{CN\quad (\quad C\quad )\quad C\quad =\quad O} &
\texttt{\detokenize{<atom_369>}\quad \detokenize{<atom_369>}\quad (\quad \detokenize{<atom_369>}\quad )\quad \detokenize{<atom_263>}\quad =\quad \detokenize{<atom_263>}}
\\
\midrule
O=C([O-1])O &
\texttt{O\quad =\quad C\quad ([\quad O\quad -\quad 1\quad ]\quad )\quad O} &
\texttt{\detokenize{<atom_433>}\quad =\quad \detokenize{<atom_262>}\quad (\quad \detokenize{<atom_262>}\quad )\quad \detokenize{<atom_262>}}
\\
\bottomrule
\end{tabularx}
\caption{Comparison between conventional SMILES tokenization and AtomDisc's structure-aware tokenization for selected molecules. AtomDisc tokens reflect atom-level distinctions with specialized identifiers, offering richer structural granularity than SMILES.}
\label{tab:smiles-vs-AtomDisc}
\end{table}

To better illustrate the differences between traditional SMILES tokenization and our proposed AtomDisc structural tokenization, we provide several representative examples in Table~\ref{tab:smiles-vs-AtomDisc}. Each row displays a molecule’s SMILES along with the corresponding token sequence generated by standard SMILES parsing and by the AtomDisc tokenizer. These examples highlight that AtomDisc assigns specialized structure-aware tokens based on atomic environments, offering richer and more context-sensitive representations than conventional SMILES-based tokens. This detailed tokenization process underpins AtomDisc’s superior performance in downstream molecular tasks.

\begin{figure}
    \centering
    \includegraphics[width=\linewidth]{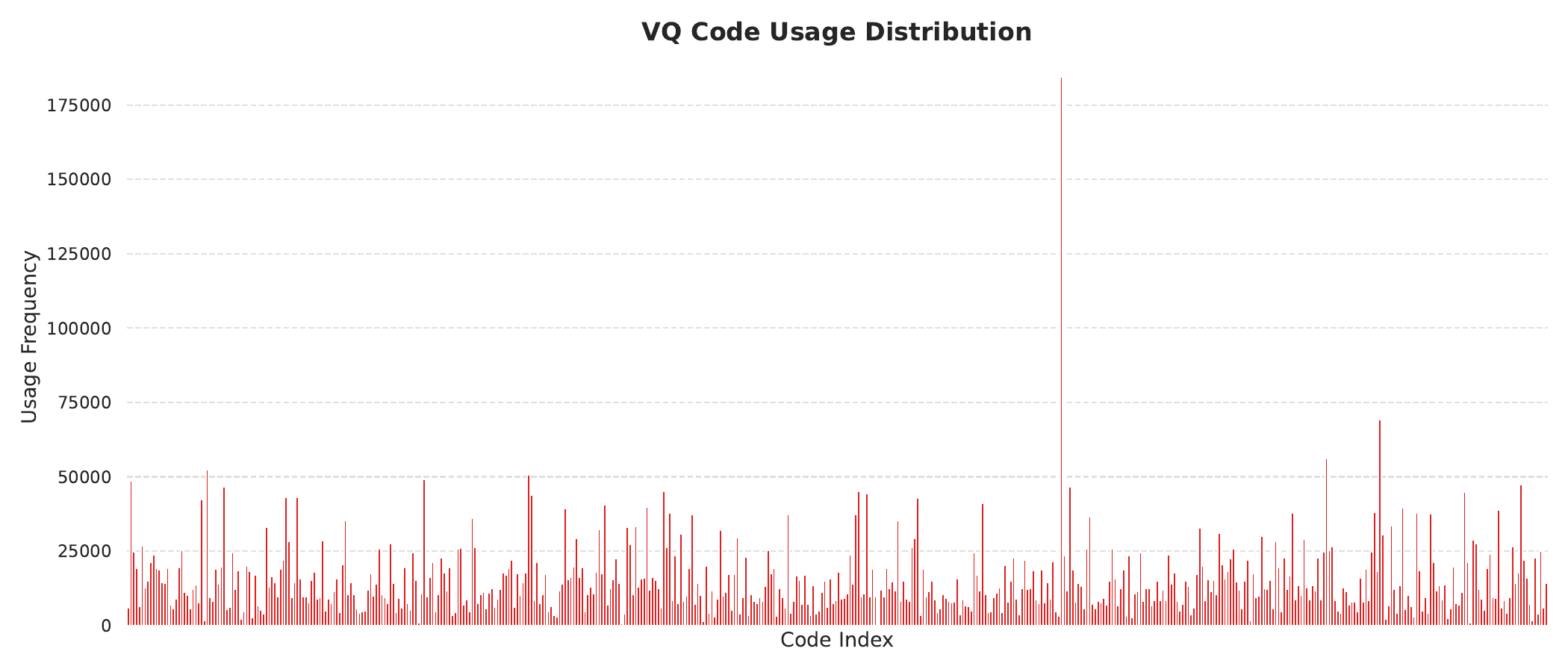}
    \caption{The vast majority of codewords are actively utilized, indicating that the VQ discretizer successfully captures a diverse range of atomic environments. This high code diversity demonstrates that our VQ tokenizer can represent many distinct structural and chemical contexts, rather than collapsing onto a few trivial patterns. Such diverse code usage reflects the model’s capacity to distinguish atoms with different chemical properties, even among atoms of the same element type.}
    \label{fig:code_usage_distribution}
\end{figure}

After training, the utilization of each codeword in the learned codebook is visualized in Figure~\ref{fig:code_usage_distribution}. Nearly all 512 codes are actively used across the dataset, confirming that the VQ discretizer captures a broad diversity of atomic environments and avoids codebook collapse. This demonstrates that our atom tokenizer can differentiate and encode a wide spectrum of structural contexts—supporting fine-grained, chemically meaningful token assignments for downstream molecular modeling.

\newpage
\subsection{Training Details of Projector}
\label{sec:training_projector_details}

After obtaining a discrete 512-entry atom codebook, we learn a lightweight \emph{projector} that embeds each structural token into the same 4\,096-dimensional semantic space used by the LLaMA-2 backbone.  Concretely, each codeword $\mathbf{e}_{k}\!\in\!\mathbb{R}^{300}$ is passed through a two-layer MLP and an RMSNorm layer,
$$
\tilde{\mathbf{e}}_{k}= \operatorname{RMSNorm}\!\bigl(\mathbf{W}_2\,\sigma(\mathbf{W}_1\mathbf{e}_{k})\bigr)\,\cdot s,
$$
where $\sigma$ denotes \textsc{GELU} and the scalar $s$ matches the expected root-mean-square norm of native LLaMA embeddings.  
This scale factor is critical: without it the projected vectors exhibit larger magnitudes, which in turn distort the attention logits and degrade generation quality.

During training we freeze the GNN, atom discretizer, and the entire LLaMA model; only the projector weights are updated.  
We extend the tokenizer vocabulary with the symbols
$$
\texttt{\small<atom\_0>},\;\dots,\;\texttt{\small<atom\_{511}>}
$$
as well as \texttt{\small<mol>} and \texttt{\small</mol>}, and treat every special token as an ordinary input token.  
For each training sample we first convert the SMILES portion into its sequence of structural codes, substitute the corresponding \texttt{\small<atom\_k>} tokens, and feed the resulting mixed sequence into the frozen LLaMA.  
The model’s language-modeling loss $\mathcal{L}_{\text{LM}}$ is then back-propagated \emph{only} through the projector, allowing it to learn an embedding alignment that is fully supervised by downstream, instruction-style objectives. Once training converges, the final projected codebook $\{\tilde{\mathbf{e}}_{k}\}$ is \emph{baked} into LLaMA’s embedding matrix, after which the projector layer itself can be discarded at inference time.  
The resulting model can natively ingest structural tokens and plain text in a single autoregressive stream, enabling seamless structure-aware generation and understanding.

\newpage
\subsection{Training Details of Pre-Training}
\label{sec:training_pretraining_details}

In the pre-training stage, we utilize the mixed dataset described in \blue{Supplementary Information A.2}, including both the ChEBI-20 augmented dataset and Mol-Instructions collections, covering a wide range of molecular understanding and generation tasks. Each sample is processed by first generating atom-level structure tokens from the frozen GNN encoder and vector quantizer, followed by mapping these discrete codes into the LLM embedding space using the fixed projector trained in Stage-2. For each SMILES sequence, the corresponding sequence of special structure tokens (\texttt{<atom\_k>}) is concatenated, forming a unified input that augments the original molecule representation with explicit structural context.

During pre-training, the backbone LLaMA-2-7B language model is employed, with only the LoRA adapters updated; all other components, including the GNN encoder, VQ tokenizer, and projector, are kept frozen. We train the model for 10 epochs using the standard autoregressive language modeling objective. Each mini-batch contains a mixture of molecule-to-text, text-to-molecule, and various instruction-based generative tasks, supporting both SMILES and structure-token-augmented sequences. Every prompt and instruction follows the unified sequence-to-sequence format, with typical examples provided in the case studies (see \blue{Supplementary Information C.1}). This multi-task pre-training paradigm equips our model with robust molecular reasoning, structural understanding, and generalizable instruction-following capabilities, laying the foundation for strong downstream performance.

\newpage
\section{Experimental Details}
\subsection{Tasks Specification and Datasets}
\label{sec:tasks_datasets}
\subsubsection{Molecular Property Prediction Tasks}

\paragraph{Classification Tasks.} 
For molecular property classification, we adopt several benchmark datasets from the MoleculeNet collection~\cite{wu2018moleculenet}, which are widely used in evaluating the performance of molecular representation models. Our focus lies on seven binary or multi-label classification tasks: BBBP, Tox21, ToxCast, SIDER, ClinTox, HIV, and BACE. These datasets cover a broad spectrum of biomedical and physicochemical properties, including blood-brain barrier permeability (BBBP), various toxicological endpoints (Tox21, ToxCast, SIDER, ClinTox), antiviral activity (HIV), and inhibition of human $\beta$-secretase 1 (BACE).

Each molecule is represented by its SMILES string, and the corresponding structural tokens derived via our AtomDisc tokenizer are used to enhance representation learning. We evaluate classification performance using the ROC-AUC metric, following standard scaffold splits and preprocessing procedures provided by MoleculeNet.

To adapt multi-label classification datasets (e.g., Tox21, ToxCast, SIDER) for autoregressive language models, we convert each label prediction into an independent instance with its own instruction. Specifically, for a given molecule, each binary subtask (e.g., "Does the molecule cause Cardiac disorders?") is treated as a separate single-label classification example. This design aligns with the instruction-following paradigm of LLMs and enables fine-grained supervision at the subtask level. At inference time, we predict each subtask individually and report the average ROC-AUC across all subtasks as the final score for the dataset. Table~\ref{tab:classification_instruction_example} presents examples of such reformulated instructions for the SIDER dataset.

\begin{table}[htbp]
\centering
\begin{tabular}{p{3.5cm}p{8.5cm}p{1cm}}
\toprule
\textbf{SMILES} & \textbf{Instruction} & \textbf{Label} \\
\midrule
C(CNCCNCCNCCN)N & Does the molecule cause Hepatobiliary disorders? & no \\
C(CNCCNCCNCCN)N & Does the molecule cause Investigations? & yes \\
C(CNCCNCCNCCN)N & Does the molecule cause Product issues? & no \\
C(CNCCNCCNCCN)N & Does the molecule cause Eye disorders? & no \\
C(CNCCNCCNCCN)N & Does the molecule cause Product issues? & no \\
C(CNCCNCCNCCN)N & Does the molecule cause Gastrointestinal disorders? & yes \\
\bottomrule
\end{tabular}
\caption{Examples of decomposed multi-label classification instances from the SIDER dataset. Each subtask is expressed as an individual instruction with its corresponding molecular SMILES and binary label.}
\label{tab:classification_instruction_example}
\end{table}

\paragraph{Regression Tasks.}  
For molecular property regression, we utilize a combination of quantum chemistry and ADMET-related datasets. Specifically, from the QM9~\cite{wu2018moleculenet} dataset, we focus on predicting three electronic properties: the energies of the Highest Occupied Molecular Orbital (HOMO), the Lowest Unoccupied Molecular Orbital (LUMO), and their corresponding gap (HOMO–LUMO gap), which are crucial indicators of molecular reactivity and stability. 

Molecular inputs are encoded as SMILES strings and augmented with structure tokens generated by our AtomDisc tokenizer. For property prediction, we use the hidden state of the last input token—corresponding to the model state just before generating the first output token—as the sequence-level representation. A regression head is attached to this representation, and both the regression head and LoRA parameters are optimized jointly in an end-to-end manner using a mean squared error (MSE) loss. For evaluation, performance is reported with either Mean Absolute Error (MAE) or Root Mean Square Error (RMSE), following the standard protocol of each benchmark dataset.

\subsubsection{Molecule Captioning Task}

We adopt the ChEBI-20 dataset as the benchmark for molecular captioning. This dataset is derived from the ChEBI (Chemical Entities of Biological Interest) database, a manually curated repository of small molecular entities relevant to biology. In ChEBI-20, each sample consists of a SMILES representation paired with a human-written textual description. This task requires the model to generate fluent and semantically accurate textual descriptions for input molecular structures, bridging chemical representation and natural language. Evaluation metrics include BLEU (2, 4), ROUGE (1, 2, L), and METEOR, which measure lexical and semantic alignment between the generated captions and ground-truth references.

\subsubsection{Molecule Generation Tasks.}

We evaluate the model on three core molecular generation tasks: forward reaction prediction, retrosynthesis, description guided molecule generation, and reagent prediction. In the forward reaction prediction task, the model is given the reactants and is required to generate the product. The retrosynthesis task reverses this direction, where the product molecule is provided, and the model predicts the set of plausible reactants. The reagent prediction task involves inferring potential reagents given the reactants and the target product. These tasks test the model’s ability to reason over molecular transformations and synthetic logic. These tasks are sourced from the Mol-Instructions datasets~\cite{fang2023mol}, a curated instruction-tuning corpus specifically constructed for large language models in chemistry. 

\subsubsection{Task Descriptions}

\paragraph{BBBP.} This task assesses whether a compound can cross the blood-brain barrier (BBB), a crucial consideration for central nervous system (CNS) drugs. Molecules are classified into BBB-permeable or non-permeable categories based on experimental permeability data.

\paragraph{Tox21.} A multi-label classification task involving 12 distinct toxicity pathways. These include nuclear receptor signaling and stress response pathways, reflecting the compound’s potential toxic effects across multiple biological processes.

\paragraph{ToxCast.} This task covers a broader panel of in vitro toxicity assays than Tox21, measuring compound bioactivity across hundreds of biochemical and cellular targets. It enables high-throughput evaluation of toxicity at scale.

\paragraph{SIDER.} The Side Effect Resource dataset includes marketed drugs annotated with known adverse drug reactions. The goal is to predict side effect profiles across 27 categories, making this a multi-label classification task.

\paragraph{ClinTox.} This binary classification task distinguishes drugs approved by the FDA from those that failed clinical trials due to toxicity. It provides a stringent benchmark for modeling toxicity-related risk.

\paragraph{HIV.} The HIV dataset involves binary classification of compounds based on their ability to inhibit HIV replication in a human cell assay. It evaluates antiviral activity and molecular efficacy.

\paragraph{BACE.} This task aims to predict inhibitors of human $\beta$-secretase 1 (BACE-1), an enzyme linked to Alzheimer’s disease. Compounds are labeled based on their biochemical assay outcomes.

\paragraph{QM9.} From the QM9 quantum chemistry dataset, we focus on predicting three scalar properties: the energies of the Highest Occupied Molecular Orbital (HOMO), the Lowest Unoccupied Molecular Orbital (LUMO), and their difference (the HOMO–LUMO gap). These properties are central to understanding molecular reactivity, stability, and electronic behavior.








\paragraph{Forward Reaction Prediction.} In this task, the model is given the reactants and tasked with predicting the major product of the reaction. This simulates the process of forward synthesis, requiring the model to understand chemical reactivity and apply transformation rules to produce plausible outputs.

\paragraph{Retrosynthesis.} Retrosynthesis involves predicting plausible reactant sets given a target product molecule. It mimics how chemists plan synthetic routes by reasoning backwards from the desired compound. This task is more open-ended than forward prediction, as multiple reactant combinations may yield the same product.

\paragraph{Reagent Prediction.} This task requires the model to infer appropriate reagents (e.g., catalysts, solvents) given the main reactants and the product. Reagents are essential for driving and optimizing chemical reactions, making this task critical for practical synthesis planning.

\newpage
\subsection{Evaluation Metrics}
\label{sec:metrics}
\subsubsection{Metrics in Property Prediction Tasks}

\paragraph{AUROC.}
For classification tasks, we report the Area Under the Receiver Operating Characteristic Curve (AUROC) as the primary evaluation metric. AUROC measures the ability of the model to rank positive instances higher than negative ones, with a value of 1 indicating perfect classification and 0.5 indicating random guessing. Given prediction logits $s_i$ and ground truth labels $y_i \in \{0,1\}$, AUROC is computed as:
$$
\text{AUROC} = \frac{1}{n_+ n_-} \sum_{i:y_i=1} \sum_{j:y_j=0} \mathbb{1}(s_i > s_j)
$$
In our implementation, for each molecule, we extract the logits of the \texttt{yes} and \texttt{no} tokens from the model output and use the difference between them as the prediction score $s_i$.

\paragraph{RMSE.}
For regression tasks, we report the Root Mean Squared Error (RMSE) defined as:
$$
\text{RMSE} = \sqrt{\frac{1}{n} \sum_{i=1}^n (\hat{y}_i - y_i)^2}
$$
where $\hat{y}_i$ is the predicted value and $y_i$ is the ground truth label.

\paragraph{MAE.}
We also report the Mean Absolute Error (MAE), which is defined as:
$$
\text{MAE} = \frac{1}{n} \sum_{i=1}^n |\hat{y}_i - y_i|
$$

\subsubsection{Metrics in Molecule Generation Tasks}

\paragraph{Exact Match.}
Exact match measures the percentage of generated SMILES strings that are exactly identical to the reference SMILES.

\paragraph{BLEU.}
We use BLEU to assess lexical similarity between generated and reference SMILES.

\paragraph{Levenshtein Distance.}
This metric captures the minimal number of edit operations (insertions, deletions, substitutions) needed to transform the generated SMILES into the reference.

\paragraph{FTS.}
Following previous work, we compute Fingerprint Tanimoto Similarity (FTS)~\cite{bajusz2015tanimoto} between the generated and reference molecules using three types of molecular fingerprints: RDK, MACCS, and Morgan. FTS is defined as:
$$
\text{FTS}(x, y) = \frac{|F(x) \cap F(y)|}{|F(x) \cup F(y)|}
$$
where $F(\cdot)$ denotes the set of fingerprint bits for a molecule.

\paragraph{Validity.}
To evaluate chemical validity, we use RDKit~\cite{rdkit} to check whether the generated SMILES strings can be successfully parsed into chemically valid molecules. Specifically, we call \text{Chem.MolFromSmiles(smi, sanitize=True)} and require that the returned molecule object is non-empty and passes valence checks. Invalid SMILES are discarded in the final performance calculation.

\newpage
\subsection{More Experimental Results}
\label{sec:more_results}

\subsubsection{Classification Tasks}

As shown in Table~\ref{tab:molnet_classification_full_appendix}, AtomDisc achieves state-of-the-art performance across all MoleculeNet classification bench- marks, consistently outperforming both LM-based and non-LM-based baselines. For example, on key datasets such as BBBP, Tox21, and ClinTox, AtomDisc attains ROC-AUCs of 95.2\%, 85.6\%, and 96.4\%, respectively—substantially exceeding the best prior models. In particular, compared to MoLFormer-XL, which was previously the leading LM-based approach, AtomDisc achieves notable improvements of +0.9\% and +1.6\% ROC-AUC on Tox21 and ClinTox, respectively. Furthermore, AtomDisc sets a new benchmark for average ROC-AUC (84.7\%) across all tasks, clearly advancing beyond the best-performing non-LM-based baseline KANO (84.3\%). These results highlight AtomDisc’s ability to seamlessly integrate discrete atom-level structural information into molecular language modeling, thereby enhancing both predictive accuracy and generalization. 

\begin{table}[htbp]
\centering
\resizebox{\textwidth}{!}{
\begin{tabular}{lcccccccc}
\toprule
\textbf{Method}      & \textbf{BBBP↑} & \textbf{Tox21↑} & \textbf{ToxCast↑} & \textbf{Sider↑} & \textbf{ClinTox↑} & \textbf{HIV↑} & \textbf{BACE↑} & \textbf{Avg↑} \\
\midrule
\multicolumn{9}{l}{\textit{Non-LM-based Methods}} \\
\midrule
AttrMask~\cite{hu2019strategies}             & 66.5$\pm$2.5 &77.9$\pm$0.4 &65.1$\pm$0.3& 63.9$\pm$0.9& 73.7$\pm$2.8& 77.1$\pm$1.2& 80.3$\pm$0.9 &73.2         \\
InfoGraph~\cite{sun2019infograph}            & 69.2$\pm$0.8 & 73.0$\pm$0.7 & 62.0$\pm$0.3 & 59.2$\pm$0.2 & 75.1$\pm$5.0 & 74.5$\pm$1.8 & 73.9$\pm$2.5 & 70.1         \\
MolCLR~\cite{wang2022molecular}               & 73.3$\pm$1.0 & 74.1$\pm$5.3 & 65.9$\pm$2.1 & 61.2$\pm$3.6 & 89.8$\pm$2.7 &    77.4$\pm$0.6  & 82.8$\pm$0.7  &   74.9       \\
GraphMVP~\cite{liu2022pretraining}             & 72.4$\pm$1.6 &  75.9$\pm$0.5 & 63.1$\pm$0.4 &  63.9$\pm$1.2 & 79.1$\pm$2.8 & 77.0$\pm$1.2 & 81.2$\pm$0.9         & 73.2         \\
MoleculeSTM~\cite{liu2023multi} & 70.0$\pm$0.5 & 76.9$\pm$0.5 & 65.1$\pm$0.40 & 61.0$\pm$1.1 & 92.5$\pm$1.1 & 76.9$\pm$1.8 & \underline{80.8$\pm$1.3} & 74.6         \\
UniMol~\cite{zhou2023uni}               & 72.9$\pm$0.6   & 79.6$\pm$0.5   & 69.6$\pm$0.1   & \textbf{65.9$\pm$1.3}   & 91.9$\pm$1.8 & 80.8$\pm$0.3  & 85.7$\pm$0.2& 78.3\\
KPGT~\cite{li2022kpgt}                 & \underline{90.8$\pm$1.0}& \textbf{84.8$\pm$1.3}  & \textbf{74.6$\pm$0.2}   & 64.9$\pm$0.9            & \textbf{94.6$\pm$2.2}   & -  & \textbf{86.8$\pm$1.1}  & \underline{82.8}\\
KANO~\cite{fang2023knowledge}                 & \textbf{96.0$\pm$1.6}  & \underline{83.7$\pm$1.3}& \underline{73.2$\pm$1.6}& \underline{65.2$\pm$0.8} & \underline{94.4$\pm$0.3}& \textbf{83.7$\pm$2.3}& \underline{85.1$\pm$2.2}         & \textbf{84.3}\\
\midrule
\multicolumn{9}{l}{\textit{LM-based Methods}} \\
\midrule
KV-PLM~\cite{zeng2022deep}               & 72.0±0.9 &70.0±0.5 &55.0±1.7& 59.8±1.5& 89.2±2.7 &71.8±1.4& 78.5±2.7           & 70.9         \\
InstructMol~\cite{cao2023instructmol}          & 70.0           & 74.7            & 64.3            & 57.8            & 91.5            & 68.9           & 82.3           & 72.2         \\
UniMoT~\cite{zhang2024unimot} & 71.4           & 76.4            & 65.8            & 59.8            & 92.9            & 78.5           & 83.7           & 75.5         \\
MoMu~\cite{su2022molecular}      & 70.5$\pm$2.0 & 75.6$\pm$0.3 & 63.4$\pm$0.5 & 60.5$\pm$0.9 & 79.9$\pm$4.1 & 75.9$\pm$0.8 & 76.7$\pm$2.1 & 71.8         \\
MolLM~\cite{tang2024mollm}     & 75.7$\pm$1.7 &80.0$\pm$1.7 &68.2$\pm$0.4& \textbf{71.0$\pm$0.8} &91.1$\pm$1.0&  80.2$\pm$0.7 &84.1$\pm$0.9&78.6         \\
TokenMol~\cite{wang2025token}             & 93.4$\pm$0.1        & 82.9$\pm$0.5            & \textbf{74.6$\pm$1.2}& {64.4$\pm$2.0}& 92.7$\pm$2.1            & –               & \underline{89.6$\pm$1.5}& \underline{82.9}         \\
MoLFormer-XL~\cite{ross2022large}         & \underline{93.7}& \underline{84.7}& –                & {69.0}& \underline{94.8}& –               & 88.2           & -\\
AtomDisc               & \textbf{95.2$\pm$0.7}  & \textbf{85.6$\pm$0.5}   & \underline{73.9$\pm$0.9}   & \underline{70.5$\pm$1.1}   & \textbf{96.4$\pm$1.6}   & \textbf{81.4$\pm$0.3}  & \textbf{89.9$\pm$0.9}  & \textbf{84.7}\\
\bottomrule
\end{tabular}
}
\caption{
ROC-AUC (\%) on MoleculeNet classification benchmarks.  
\textit{Non-LM-based} (top) and \textit{LM-based} (bottom) methods are each evaluated separately: for each column, the best in its group is \textbf{bolded} and the runner-up is \underline{underlined}.  
Our results are reported as the mean~$\pm$~standard deviation over five random seeds.  
Baseline numbers are taken directly from the results reported in the original papers.
}
\label{tab:molnet_classification_full_appendix}

\end{table}

\begin{figure}[htbp]
    \centering
    \includegraphics[width=\linewidth]{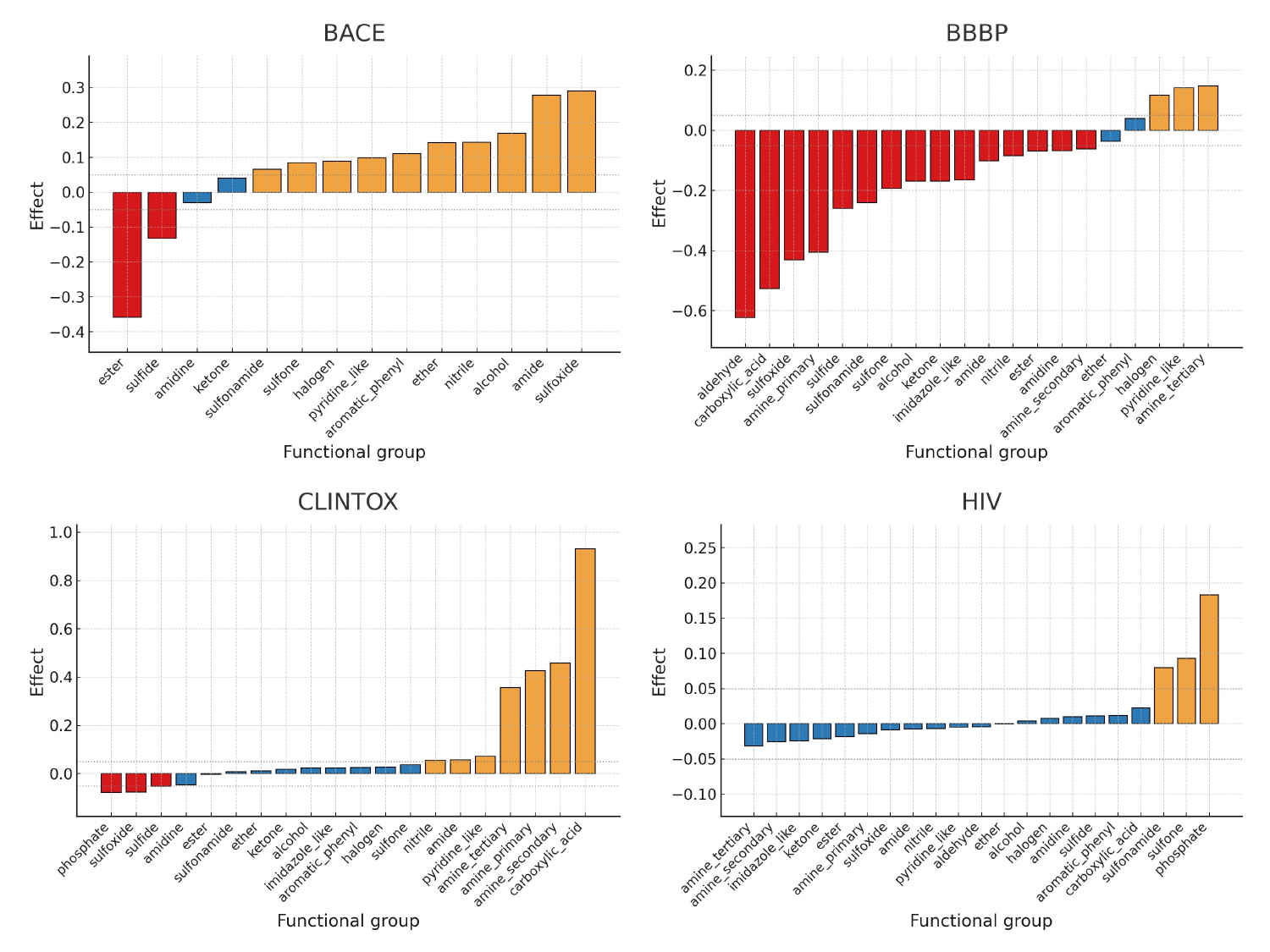}
    \caption{Functional-group (FG) effects across datasets (BACE, BBBP, ClinTox, HIV). Each bar corresponds to one FG; bar height is the empirical label effect $E(f)=\hat{P}(y{=}1\mid f)-\hat{P}(y{=}1\mid \neg f)$. Bars are sorted by $E(f)$ within each dataset. Colors denote the polarity class: positive (yellow), negative (red), and neutral (blue), using a fixed threshold $\tau=0.05$ (positive if $E(f)\ge \tau$, negative if $E(f)\le -\tau$, neutral otherwise).}
    \label{fig:effect}
\end{figure}

To identify chemically salient groups per dataset, we compute for each FG f the effect:
$$E(f)=\hat{P}(y{=}1\mid f)-\hat{P}(y{=}1\mid \neg f),\quad
\hat{P}(y{=}1\mid f)=\frac{n_{+}(f)}{n(f)},\
\hat{P}(y{=}1\mid \neg f)=\frac{n_{+}(\neg f)}{n(\neg f)} ,$$
where $n(f)$ and $n_{+}(f)$ are, respectively, the number of molecules containing f and the number of those labeled $y{=}1$ (analogously for $\neg f$). We then categorize each FG using $\tau=0.05$: positive if $E(f)\ge \tau$, negative if $E(f)\le -\tau$, and neutral otherwise. The figure visualizes these per-dataset landscapes, providing the reference inventory of FG polarities used in our downstream attention and representation analyses.
\newpage
\subsubsection{Regression Tasks}
We further evaluate AtomDisc on a series of standard molecular property regression tasks. Table~\ref{tab:frontier_orbitals} reports results for quantum chemical property prediction on QM9, focusing on the HOMO, LUMO, and $\Delta\epsilon$ (HOMO-LUMO gap) endpoints. AtomDisc achieves the lowest mean absolute errors across all three metrics, setting a new state-of-the-art (0.0033 for HOMO, 0.0032 for LUMO, and 0.0042 for the gap). These results represent significant improvements over prior LLM-based and specialized baselines (e.g., UniMoT, InstructMol), highlighting AtomDisc’s effectiveness in quantum chemical regression tasks.
\begin{table}[htbp]
\centering
\begin{tabular}{lcccc}
\toprule
\textbf{Model}                         & \textbf{HOMO↓}          & \textbf{LUMO↓}          & \textbf{$\Delta\epsilon$↓} & \textbf{AVG↓}            \\
\midrule
Alpaca (Llama-7B)                     & –                       & –                       & –                          & 322.109                  \\
Baize (Llama-7B)                      & –                       & –                       & –                          & 261.343                  \\
Llama-2-7B                            & 0.7367                  & 0.8641                  & 0.5152                     & 0.7510                   \\
Vicuna-13B                            & 0.7135                  & 3.6807                  & 1.5407                     & 1.9783                   \\
Mol-Instructions (Llama-7B)           & 0.0210                  & 0.0210                  & 0.0203                     & 0.0210                   \\
InstructMol (Vicuna-7B)               & 0.0048                  & 0.0050                  & 0.0061                     & 0.0050                   \\
UniMoT (Llama-2-7B)                   & \underline{0.0042}      & \underline{0.0047}      & \underline{0.0055}         & 0.0049                   \\
UniMol                                & –                       & –                       & –                          & \underline{0.0047}       \\
AtomDisc (Llama-2-7B)                   & \textbf{0.0033}         & \textbf{0.0032}         & \textbf{0.0042}            & \textbf{0.0035}          \\
\bottomrule
\end{tabular}
\caption{
Comparison of predicted molecular frontier orbital properties in QM9 dataset. Values are mean absolute errors (lower is better). For each column, the best result per metric is \textbf{bolded} and the second-best is \underline{underlined}.
}
\label{tab:frontier_orbitals}
\end{table}

\newpage
\subsubsection{Molecule Generation Tasks}

As summarized in Table~2, AtomDisc achieves top performance across all three generation tasks. For reagent prediction, AtomDisc attains the highest exact match (0.203), validity (1.000), and fingerprint similarity, surpassing UniMoT and InstructMol-GS. For forward reaction prediction, AtomDisc delivers the best exact match (0.846), the lowest Levenshtein distance (2.366), and near-perfect validity, outperforming prior SOTA models. For retrosynthesis, AtomDisc leads in exact match (0.580), fingerprint similarity, and validity, consistently outperforming all baselines. Notably, while AtomDisc’s BLEU scores are in some cases slightly lower than those of a few baselines (such as UniMoT), this is not due to an inability to generate correct answers, but rather reflects the model’s strong generalization ability: AtomDisc is able to propose chemically plausible and valid alternatives rather than simply memorizing training set answers. This enhanced diversity and chemical reasoning are illustrated by representative cases in \blue{Supplementary Information C.2}. Collectively, these results demonstrate that integrating interpretable atom-level structure tokens directly into the sequence modeling process enhances both accuracy and chemical validity across diverse molecular generation tasks, while simultaneously promoting generalization beyond rote recall.

\begin{figure}[htbp]
    \centering
    \includegraphics[width=\linewidth]{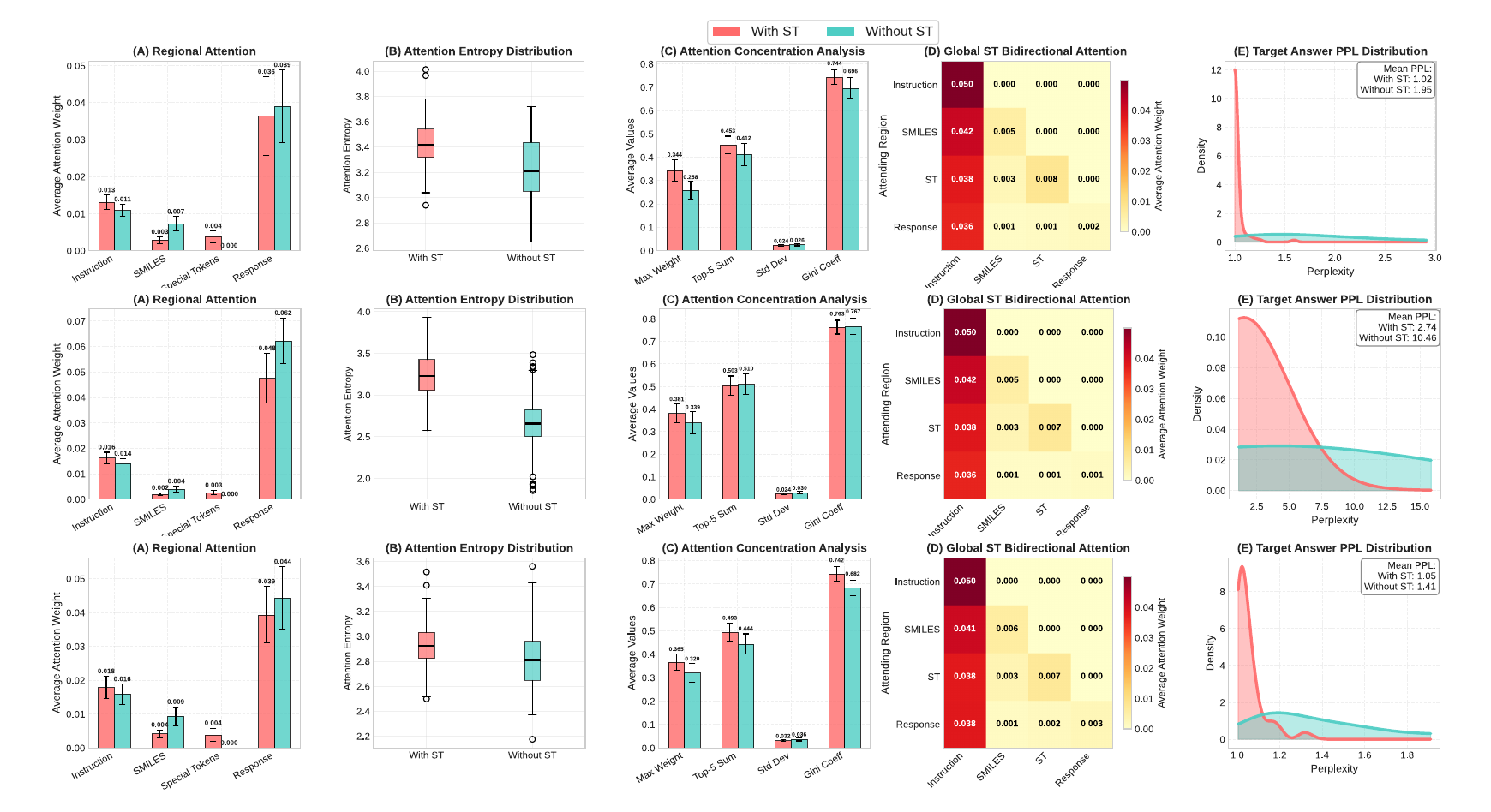}
    \caption{Aggregate analysis on the Mol-Instructions datasets, covering functional-region attention allocation, entropy of attention weights, attention concentration, cross-region bidirectional attention, perplexity distribution, and shared multi-task activations. Together, these results show that ST enhances not only predictive accuracy but also interpretability, yielding structured and chemically coherent representations.}
    \label{fig:generation_whole_datasets}
\end{figure}

As shown in Figure~4 and Figure~S3, AtomDisc exhibits both consistent and task-specific behaviors across the three molecular generation tasks, with forward reaction prediction shown at the top, reagent prediction in the middle, and retrosynthesis at the bottom. A shared pattern is that the inclusion of structural tokens (ST) increases attention entropy and specificity, encouraging a more balanced allocation of attention and reinforcing the role of ST as communication hubs between molecular and textual regions. In contrast, Fig.~\ref{fig:4_reaction_prediction}(h) reveals that the layer activation profiles differ markedly across tasks: forward reaction prediction elicits strong activation across nearly all layers, reflecting the need to integrate both local environments and long-range transformations; retrosynthesis shows heightened activity in the early layers (1–8) and again in the mid-to-deep layers (12–20), consistent with the interplay of local rearrangements and global retrosynthetic planning; and reagent prediction relies mainly on the earliest layers, with activation rapidly diminishing thereafter, suggesting that shallow contextual cues are sufficient for identifying suitable reagents. These divergent profiles imply that AtomDisc adapts its representational depth to the reasoning demands of each task—ranging from broad integration (forward prediction), to dual-scale reasoning (retrosynthesis), to localized context extraction (reagent prediction). This highlights AtomDisc’s capacity to provide a unifying inductive bias while tailoring its internal computation to the complexity and information requirements of distinct generation challenges.
\newpage
\newpage
\subsubsection{Case Study}

\begin{figure}[htbp]
    \centering
    \includegraphics[width=\linewidth]{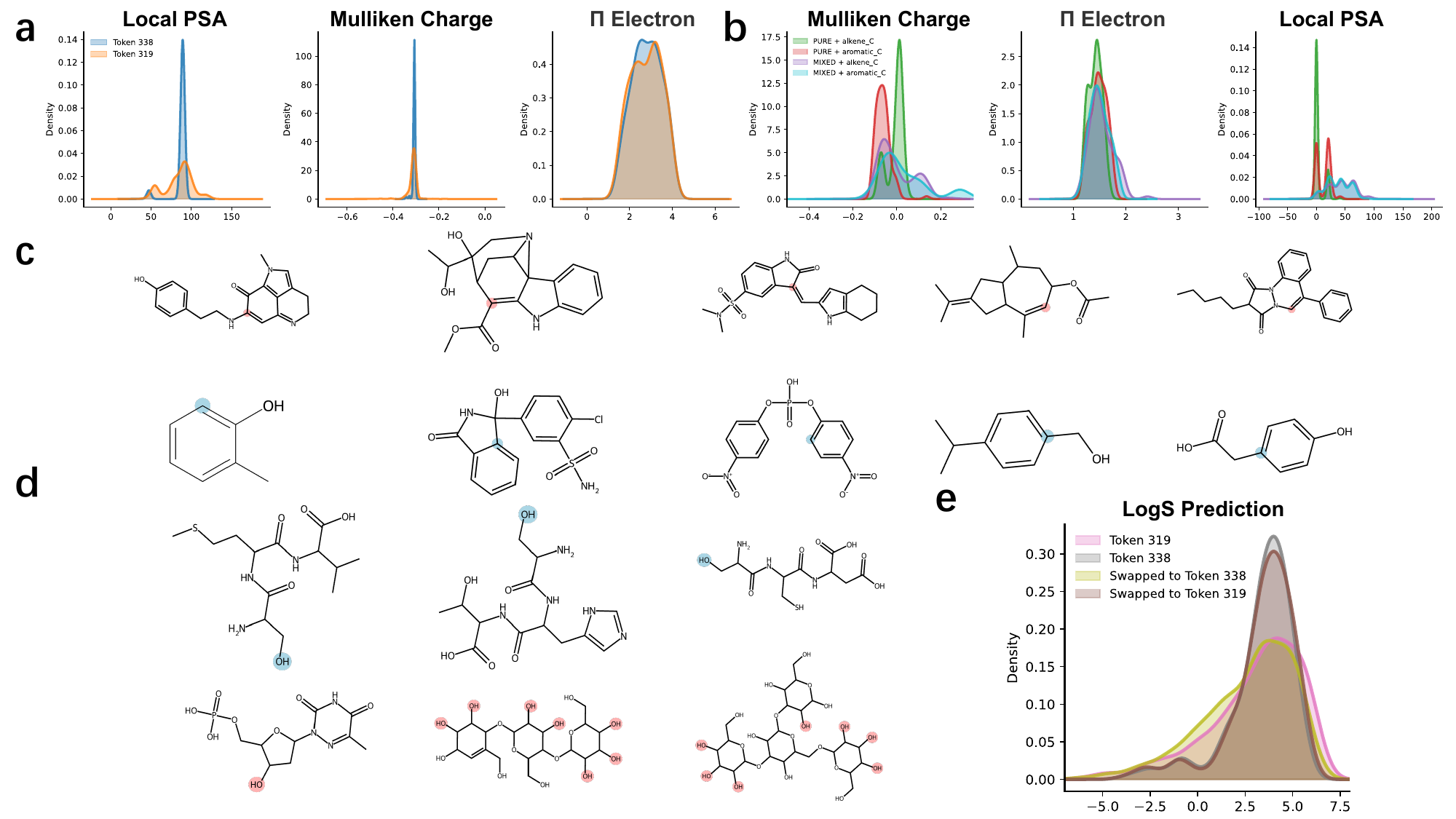}
    \caption[Functional group token analysis and downstream property prediction.]{
    Functional group token analysis and downstream property prediction. 
    \textbf{(a)} Kernel density estimates (KDEs) of atomic property distributions for hydroxyl groups (OH), comparing token 319 vs.~338; subpanels show local polar surface area (PSA), Mulliken charge, and $\pi$-electron occupancy. 
    \textbf{(b)} Kernel density estimates of atomic property distributions for mixture token 20, spanning both aromatic and alkene carbons; subpanels show Mulliken charge, $\pi$-electron occupancy, and local PSA. 
    \textbf{(c)} Representative molecules containing mixture token 20. 
    \textbf{(d)} Representative examples of hydroxyl groups assigned to token 319 (bottom) versus token 338 (top). 
    \textbf{(e)} Distribution of predicted aqueous solubility (LogS) values for test molecules when exchanging the two OH tokens (319 vs.~338) in the model input.}
    \label{fig:5_case_combine}
\end{figure}

\newpage
\subsubsection{Ablation Study on AtomDisc}

To assess the impact of atom-level structure tokens, we conduct ablation studies on all three molecular generation tasks (reagent prediction, forward reaction prediction, and retrosynthesis), using the same evaluation metrics and protocols as in the main experiments.

We compare AtomDisc with three variants: (1) removing structure tokens only during downstream fine-tuning (\textit{w/o structural tokens (SFT)}), (2) removing structure tokens for both pretraining and fine-tuning (\textit{w/o structural tokens (pretraining+SFT)}), and (3) training LoRA adapters on a plain LLM without any structure tokens (\textit{AtomDisc (LoRA only)}).

All ablation variants show substantial drops in exact match, fingerprint similarity, and validity across tasks, especially when structure tokens are omitted throughout pretraining and fine-tuning or when only LoRA is used. These results confirm that interpretable atom-level structure tokens are essential for enabling strong generalization and chemical reasoning in molecular LLMs. The full AtomDisc model achieves the best results in all metrics, underscoring the necessity of explicit structural augmentation.
\begin{table}[H]
\centering
\footnotesize
\setlength{\tabcolsep}{3pt}
\begin{tabular}{l lccccccc}
\toprule
\textbf{Task} & \textbf{Model} & \textbf{EXACT↑} & \textbf{BLEU↑} & \textbf{LEVENSHTEIN↓} & \textbf{RDK FTS↑} & \textbf{MACCS FTS↑} & \textbf{MORGAN FTS↑} & \textbf{VALIDITY↑} \\
\midrule
\multicolumn{9}{l}{\textit{Reagent Prediction}} \\
\midrule
& w/o structural tokens (SFT)      &   0.114   &  0.496    &  21.853    &  0.421    &   0.522   &   0.397   &   1.000   \\
& w/o structural tokens (pretrain+SFT)  &     0.148 &    0.538  &  17.649    &   0.472   &   0.563   &   0.446   &  1.000    \\
& AtomDisc (LoRA only)  &   0.154  &   0.512   &    20.999  &   0.453   & 0.553     &  0.431    &   1.000         \\                
& \textbf{AtomDisc (full)}              & \textbf{0.203} & \textbf{0.623} & \textbf{15.362} & \textbf{0.583} & \textbf{0.672} & \textbf{0.559} & \textbf{1.000} \\
\midrule
\multicolumn{9}{l}{\textit{Forward Prediction}} \\
\midrule
& w/o structural tokens (SFT)       & 0.657   &  0.929    &  5.357    &  0.868    &    0.904  &    0.842  & 1.000         \\
& w/o structural tokens (pretrain+SFT)  &    0.710  &  0.941  & 4.579 &   0.905   &   0.937   &  0.883   &  1.000         \\
& AtomDisc (LoRA only) &  0.794  &  0.936    &  2.919    &   0.932   &   0.953   &  0.917    &  1.000          \\ 
& \textbf{AtomDisc (full)}              & \textbf{0.846} & \textbf{0.971} & \textbf{2.366}  & \textbf{0.949} & \textbf{0.966} & \textbf{0.940} & \textbf{1.000} \\
\midrule
\multicolumn{9}{l}{\textit{Retrosynthesis}} \\
\midrule
& w/o structural tokens (SFT)   &    0.448  &  0.870    &  11.489    & 0.834  & 0.883    &   0.772   &   1.000        \\
& w/o structural tokens (pretrain+SFT)  &    0.499  &   0.889   &     9.715 &   0.866   &   0.903   &   0.808   &  1.000    \\
& AtomDisc (LoRA only)     &     0.525 &   0.894   &  9.485    &   0.861   &   0.902   &   0.806   &    1.000  \\  
& \textbf{AtomDisc (full)}              & \textbf{0.580} & \textbf{0.907} & \textbf{8.340}  & \textbf{0.880} & \textbf{0.916} & \textbf{0.830} & \textbf{1.000} \\
\bottomrule
\end{tabular}
\caption{
Ablation study for AtomDisc on three molecular generation tasks. “w/o structural tokens (SFT)” removes atom-level tokens during fine-tuning and inference; “w/o structural tokens (pretraining+SFT)” removes them throughout pretraining and fine-tuning; “AtomDisc (LoRA only)” applies LoRA on a baseline LLM. Last row is our full method. Metrics as in Table~\ref{tab:combined_tasks}.
}
\label{tab:ablation_generation}
\end{table}
\newpage
\subsubsection{Ablation Study on Codebook Size}
\label{sec:ablate_codebook_size}

To better explain why we chose a codebook size of 512, we conducted an ablation study. Let $T$ be the discrete atom token and $G$ the functional–group (FG) label assigned by SMARTS. We measure global interpretability by the mutual information
$$
I(T; G) = \sum_{t, g} p(t, g) \log \frac{p(t, g)}{p(t) \, p(g)},
$$
estimated over all atoms. From an information-theoretic perspective, $I(T;G)$ quantifies the reduction in uncertainty about the functional group identity ($G$) given the token ($T$). A higher value signifies a stronger statistical dependency; it means that observing a specific token provides significant information about which functional group is present, implying a more reliable and selective token-FG association. For each token, we also compute a normalized Shannon entropy:
$$
H_{\text{norm}}(t) =
\frac{-\sum_{g} p(g \mid t) \log p(g \mid t)}
{\log |\mathcal{G}|},
$$
Lower $H_{\mathrm{norm}}$ means a token concentrates its probability mass on fewer FGs and is thus easier to interpret. Finally, per‐token MI contributions reveal how much each token adds to $I(T;G)$; comparing pure vs. mixture tokens quantifies whether ambiguity concentrates in a small subset of codes.

To analyze the effect of codebook size on token–FG disentanglement and interpretability, we vary the VQ codebook size $K \in \{256, 512, 1024\}$. We report the global MI ($I_{\text{global}}$), the number of distinct tokens used (Code Usage) and the corresponding usage efficiency, the mixture token ratio, the average per‐token MI for pure vs. mixture tokens, and the mean of $H_{\text{norm}}$ across tokens.

\begin{table}[htbp]
\centering
\caption{Ablation study on VQ codebook size, comparing key interpretability metrics. \textbf{Usage Ratio} measures vocabulary efficiency. \textbf{Mixture Ratio} indicates the proportion of tokens associated with multiple functional groups (FGs). \textbf{Pure/Mixed Avg. MI} shows the average per-token mutual information for tokens linked to single (pure) vs. multiple (mixed) FGs. \textbf{Global MI} ($I_{\text{global}}$) quantifies overall explanatory power, while \textbf{Avg $H_{\text{norm}}$} measures token sharpness (lower is better).}
\label{tab:codebook_ablation}
\renewcommand{\arraystretch}{1.2}
\newcolumntype{C}{>{\centering\arraybackslash}X} 
\begin{tabularx}{\linewidth}{lCCCCCC}
\toprule
\textbf{Size} & \textbf{Usage Ratio} & \textbf{Mixture Ratio} & \textbf{Pure Avg. MI} & \textbf{Mixed Avg. MI} & \textbf{Global MI} & \textbf{Avg $\boldsymbol{H_{\text{norm}}}$} \\
\midrule
256  & 0.945 & 0.376 & 2.31e$-$3 & 2.00e$-$3 & 0.530 & 0.062 \\
512  & 0.912 & 0.313 & 1.28e$-$3 & 1.18e$-$3 & 0.582 & 0.054 \\
1024 & 0.865 & 0.252 & 6.74e$-$4  & 7.22e$-$4  & 0.608 & 0.043 \\
\bottomrule
\end{tabularx}
\end{table}

The results show that larger codebooks increase explanatory power, but with diminishing returns. The global mutual information, $I_{\text{global}}$, rises from 0.530 ($K=256$) to 0.582 ($K=512$), a 9.7\% increase, and then to 0.608 ($K=1024$), a smaller 4.6\% gain over the 512-codebook. This suggests that larger codebooks capture finer FG distinctions, though the benefit tapers beyond $K=512$. As the codebook size $K$ grows, tokens become sharper and more specialized, which is beneficial for interpretability. The average normalized entropy, $H_{\text{norm}}$, drops from 0.062 to 0.054 and finally to 0.043. Concurrently, ambiguity becomes more concentrated, with the mixture token ratio falling from 37.6\% to 31.3\% and then to 25.2\%, indicating improved disentanglement.

However, this specialization comes at a cost. The average per-token MI declines as $K$ increases, a fragmentation effect that makes individual long-tail tokens less informative. Furthermore, code utilization drops from 94.5\% to 91.2\% and then to 86.5\%, signaling a growing number of rarely used codes that waste vocabulary capacity. Critically, a larger vocabulary also increases the computational cost and model parameters for downstream tasks, such as training a projector layer. Thus, an excessively large codebook is not only inefficient but also introduces a significant computational burden.

Based on these trade-offs, a codebook size of $K=512$ offers a practical balance. It achieves a significant 9.7\% MI gain over $K=256$ and realizes most of the attainable disentanglement, reflected in marked improvements in entropy and mixture ratio. It maintains high utilization (91.2\%) without creating an excessive tail of rare tokens, and it substantially reduces the mixture share relative to the 256-codebook (a 17\% relative drop). This choice avoids the diminished returns in interpretability and efficiency seen when doubling the vocabulary to 1024 for only a modest 4.6\% additional MI gain. In practice, $K=512$ balances global explanatory power with local token readability while keeping training overhead and prompt length manageable.

\newpage
\subsection{Atomic Property Computation}
\label{sec:atomic_property_computation}

To systematically analyze and interpret AtomDisc’s learned token assignments, we computed three key atomic-level properties: Mulliken charge, local $\pi$-electron occupancy, and local polar surface area (PSA). These complementary descriptors provide quantitative insight into the electronic and structural environments captured by the tokenizer.

\paragraph{Mulliken Charge.}
Mulliken atomic charges were calculated using single-point quantum chemical calculations performed with the PySCF package. Geometries were extracted directly from the dataset, and computations were carried out at the Hartree–Fock (HF) or density functional theory (DFT) level with the STO-3G basis set. For each atom, the Mulliken population analysis provides a measure of local electronic polarization, which is widely used to characterize reactivity, intermolecular interactions, and the influence of substituents. To confirm robustness, we repeated the analysis with a heavier basis set in PySCF, with consistent results Supplementary Table~\ref{tab:pi_elec_light_vs_heavy_redux}.

\paragraph{Local $\pi$-Electron Occupancy.}
To quantify the extent of conjugation and aromaticity in each atomic environment, we computed local $\pi$-electron occupancy by projecting molecular orbitals onto atomic $p_z$ orbitals. This was implemented using PySCF quantum calculations, where the $p_z$ contribution for each atom was summed over occupied molecular orbitals. This descriptor highlights variations in delocalized electron density, which are relevant for understanding chemical stability and reactivity in unsaturated systems.

\paragraph{Local Polar Surface Area (PSA).}
Local PSA was calculated as the sum of polar surface area contributions for a given atom and all atoms within a 4-bond neighborhood. This was performed using the RDKit cheminformatics toolkit, following established descriptor definitions. PSA is a widely used metric in drug discovery and medicinal chemistry, known to correlate with aqueous solubility, membrane permeability, and overall bioavailability.

All property calculations were automated and performed on representative subsets of atoms from each token or functional group class. These quantitative descriptors provide the basis for our statistical analyses of token-property relationships.

\newpage
\section{Case Study}
In this section, we present a series of case studies to illustrate the versatility and interpretability of our proposed framework. First, we showcase several representative instruction examples (\ref{sec:case_study_ins_example}), demonstrating how AtomDisc handles diverse molecular queries and user prompts in a unified fashion. Next, we provide concrete examples of model generalization in challenging tasks such as retrosynthesis (\ref{sec:Generalization_retro}) and reagent prediction (\ref{sec:Generalization_reagent}), highlighting AtomDisc’s ability to generate chemically plausible solutions beyond the training data. Finally, we present data-driven scientific discovery cases (\ref{sec:scientific_discovery}), where the interpretable atom-level tokenization of AtomDisc uncovers novel structure–property relationships and chemical insights inaccessible to traditional approaches. And more cases are provided in~\ref{appendix:more_cases}.
\subsection{Instructions Examples}
\label{sec:case_study_ins_example}
To illustrate the versatility and practical utility of AtomDisc, we provide several representative instruction–response examples covering key tasks: forward reaction prediction, molecular property classification, retrosynthesis, reagent prediction, and molecule captioning. In each case, the model receives an instruction prompt alongside molecular inputs, with special atom-level structure tokens (\texttt{<atom\_$k$>}) inserted into the input sequence to encode local chemical environments. The language model then generates a textual or SMILES-format response, as appropriate for the task.

Each example below is presented in a unified format: the prompt (including natural language instruction, input molecules, and their corresponding structure tokens), the model output, and the reference answer (when available) for direct comparison. These cases demonstrate AtomDisc’s ability to process complex, chemistry-specific instructions, interpret atom-level context, and deliver accurate predictions or chemically meaningful descriptions across a wide range of scenarios.
\begin{tcolorbox}[
    title=Case Example: Forward Reaction Prediction,
    colback=gray!5!white,
    colframe=black!80!black,
    fonttitle=\bfseries,
    breakable,
    enhanced
]

\textbf{Prompt:}

\begin{tcolorbox}[
    colback=white,
    colframe=gray!60!black,
    boxrule=0.4pt,
    left=2pt, right=2pt, top=2pt, bottom=2pt,
    fontupper=\ttfamily\small,
    breakable
]
\#\#\# Instruction: 

Using the provided reactants and reagents, can you propose a likely product?

\#\#\# Input: 

Molecules (SMILES):  

CC(C)(C)OC(=O)N1CCC(O)CC1  
N\#CC1=CC=C(F)C=C1  
CN(C)C=O  
[H-1]  
[Na+1]  

Structural representations:  

<mol> <atom\_370> <atom\_370> ( <atom\_370> ) ( <atom\_370> ) <atom\_225> <atom\_225> ( = <atom\_225> ) <atom\_392> 1 <atom\_392> <atom\_183> <atom\_100> ( <atom\_491> ) <atom\_183> <atom\_392> 1 </mol>  

<mol> <atom\_220> \# <atom\_220> <atom\_455> 1 = <atom\_125> <atom\_434> = <atom\_188> ( <atom\_410> ) <atom\_434> = <atom\_125> 1 </mol>  

<mol> <atom\_369> <atom\_369> ( <atom\_369> ) <atom\_263> = <atom\_263> </mol>  

<mol> <atom\_406> </mol>  

<mol> <atom\_442> </mol>

\#\#\# Response: 
\end{tcolorbox}

\vspace{1mm}
\textbf{Model Output:}

\begin{tcolorbox}[colback=white, colframe=yellow!60!black, fontupper=\ttfamily\small]
CC(C)(C)OC(=O)N1CCC(OC2=CC=C(C\#N)C=C2)CC1
\end{tcolorbox}

\vspace{1mm}
\textbf{Reference Answer:}

\begin{tcolorbox}[colback=white, colframe=red!60!black, fontupper=\ttfamily\small]
CC(C)(C)OC(=O)N1CCC(OC2=CC=C(C\#N)C=C2)CC1
\end{tcolorbox}

\end{tcolorbox}
\newpage
\begin{tcolorbox}[
    title=Case Example: Clinical Toxicity Prediction,
    colback=gray!5!white,
    colframe=black!80!black,
    fonttitle=\bfseries,
    breakable,
    enhanced
]

\textbf{Prompt:}

\begin{tcolorbox}[
    colback=white,
    colframe=gray!60!black,
    boxrule=0.4pt,
    left=2pt, right=2pt, top=2pt, bottom=2pt,
    fontupper=\ttfamily\small,
    breakable
]
\#\#\# Instruction:  
Does the molecule exhibit clinical toxicity?

\#\#\# Input:  
Molecule (SMILES):  
C=CC[NH+]1CC[C@]23CCCC[C@H]2[C@H]1Cc4c3cc(cc4)O  

Molecule (Structure):  
<mol> <atom\_273> = <atom\_273> <atom\_303> <atom\_291> 1 <atom\_397> <atom\_296> <atom\_382> 2 3 <atom\_389> <atom\_190> <atom\_215> <atom\_389> <atom\_382> 2 <atom\_291> 1 <atom\_296> <atom\_424> 1 = <atom\_424> 3 <atom\_186> = <atom\_365> ( <atom\_186> ) <atom\_107> = <atom\_243> 1 </mol>

\#\#\# Response:  
\end{tcolorbox}

\vspace{1mm}
\textbf{Model Output:}

\begin{tcolorbox}[colback=white, colframe=yellow!60!black, fontupper=\ttfamily\small]
no
\end{tcolorbox}

\vspace{1mm}
\textbf{Reference Answer:}

\begin{tcolorbox}[colback=white, colframe=red!60!black, fontupper=\ttfamily\small]
no
\end{tcolorbox}

\end{tcolorbox}
\newpage
\begin{tcolorbox}[
    title=Case Example: Retrosynthesis Prediction,
    colback=gray!5!white,
    colframe=black!80!black,
    fonttitle=\bfseries,
    breakable,
    enhanced
]

\textbf{Prompt:}

\begin{tcolorbox}[
    colback=white,
    colframe=gray!60!black,
    boxrule=0.4pt,
    left=2pt, right=2pt, top=2pt, bottom=2pt,
    fontupper=\ttfamily\small,
    breakable
]
\#\#\# Instruction:  

Please suggest potential reactants used in the synthesis of the provided product.

\#\#\# Input:  

Product (SMILES):  

C\#CC1=CC=C(C(=O)OCCCCCCC)C=C1

Product (Structure):  

<mol> <atom\_227> \# <atom\_227> <atom\_324> 1 = <atom\_77> <atom\_372> = <atom\_409> ( <atom\_241> ( = <atom\_340> ) <atom\_171> <atom\_171> <atom\_171> <atom\_208> <atom\_283> <atom\_202> <atom\_144> <atom\_308> ) <atom\_372> = <atom\_372> 1 </mol>

\#\#\# Response:  
\end{tcolorbox}

\vspace{1mm}
\textbf{Model Output:}

\begin{tcolorbox}[colback=white, colframe=yellow!60!black, fontupper=\ttfamily\small]
CCCCCCCOC(=O)C1=CC=C(C\#C[Si](C)(C)C)C=C1
\end{tcolorbox}

\vspace{1mm}
\textbf{Reference Answer:}

\begin{tcolorbox}[colback=white, colframe=red!60!black, fontupper=\ttfamily\small]
CCCCCCCOC(=O)C1=CC=C(C\#C[Si](C)(C)C)C=C1
\end{tcolorbox}

\end{tcolorbox}
\newpage
\begin{tcolorbox}[
    title=Case Example: Reagent Prediction,
    colback=gray!5!white,
    colframe=black!80!black,
    fonttitle=\bfseries,
    breakable,
    enhanced
]

\textbf{Prompt:}

\begin{tcolorbox}[
    colback=white,
    colframe=gray!60!black,
    boxrule=0.4pt,
    left=2pt, right=2pt, top=2pt, bottom=2pt,
    fontupper=\ttfamily\small,
    breakable
]
\#\#\# Instruction:  

From the provided chemical reaction, propose some possible reagents that could have been used.

\#\#\# Input:  

Reactants (SMILES):  

O=C(OCC1=CC=CC=C1)N2CCCC(CN3C=CC=CC3=O)C2

Reactants (Structure):  

<mol> <atom\_225> = <atom\_225> 1 <atom\_225> = <atom\_394> <atom\_123> = <atom\_77> <atom\_452> 1 <atom\_452> <atom\_452> 1 <atom\_77> <atom\_392> <atom\_392> <atom\_478> ( <atom\_478> ( = <atom\_392> ) <atom\_392> <atom\_392> <atom\_481> 2 = <atom\_481> <atom\_64> = <atom\_64> <atom\_168> = <atom\_388> 2 ) <atom\_392> 1 </mol>

Product (SMILES):  

O=C1C=CC=CN1CC2CCCNC2

Product (Structure):  

<mol> <atom\_388> = <atom\_168> 1 <atom\_64> = <atom\_64> <atom\_481> = <atom\_481> <atom\_392> 1 <atom\_392> <atom\_397> 1 <atom\_478> <atom\_478> <atom\_478> <atom\_478> <atom\_478> 1 </mol>

\#\#\# Response:  
\end{tcolorbox}

\vspace{1mm}
\textbf{Model Output:}

\begin{tcolorbox}[colback=white, colframe=yellow!60!black, fontupper=\ttfamily\small]
CCO.[Pd]
\end{tcolorbox}

\vspace{1mm}
\textbf{Reference Answer:}

\begin{tcolorbox}[colback=white, colframe=red!60!black, fontupper=\ttfamily\small]
CCO.[Pd]
\end{tcolorbox}
\end{tcolorbox}
\newpage
\begin{tcolorbox}[
    title=Case Example: Molecule Captioning,
    colback=gray!5!white,
    colframe=black!80!black,
    fonttitle=\bfseries,
    breakable,
    enhanced
]

\textbf{Prompt:}

\begin{tcolorbox}[
    colback=white,
    colframe=gray!60!black,
    boxrule=0.4pt,
    left=2pt, right=2pt, top=2pt, bottom=2pt,
    fontupper=\ttfamily\small,
    breakable
]
\#\#\# Instruction: 

Generate a natural-language caption that describes the molecule.

\#\#\# Input:  

Molecule (SMILES): 

C[C@H](CCCC(CO)CO)[C@H]1CC[C@@H]2[C@@]1([C@H](C[C@H][C@H]2[C@@H](C[C@H]4[C@@]3 
(CC[C@H](C4)O)C)O)O)C

Structure tokens:  

<mol> <atom\_261> <atom\_217> ( <atom\_217> <atom\_231> <atom\_420> <atom\_348> ( <atom\_348> <atom\_511> ) <atom\_348> <atom\_511> ) <atom\_450> 1 <atom\_450> <atom\_156> <atom\_156> 2 <atom\_18> 3 <atom\_483> ( <atom\_483> ) <atom\_156> <atom\_483> 4 <atom\_483> <atom\_483> ( <atom\_483> ) <atom\_18> <atom\_483> <atom\_483> 4 ( <atom\_74> ) <atom\_483> 3 <atom\_268> <atom\_221> ( <atom\_423> ) <atom\_423> 1 2 <atom\_221> </mol>

\#\#\# Response:  
\end{tcolorbox}

\vspace{1mm}
\textbf{Model Output:}

\begin{tcolorbox}[colback=white, colframe=yellow!60!black, fontupper=\ttfamily\small]
The molecule is a \textbf{3alpha-hydroxy steroid}, a \textbf{7alpha-hydroxy steroid}, a \textbf{12alpha-hydroxy steroid} and a \textbf{26-hydroxy steroid}. It has a role as a bile acid metabolite. It derives from a \textbf{hydride of a 5beta-cholestane}.
\end{tcolorbox}

\vspace{1mm}
\textbf{Reference Answer:}

\begin{tcolorbox}[colback=white, colframe=red!60!black, fontupper=\ttfamily\small]
The molecule is a \textbf{3alpha-hydroxy steroid}, a \textbf{7alpha-hydroxy steroid}, a \textbf{12alpha-hydroxy steroid}, a \textbf{26-hydroxy steroid}, a 27-hydroxy steroid and a cyprinol. It derives from a \textbf{hydride of a 5beta-cholestane}.
\end{tcolorbox}

\vspace{1mm}
\textbf{Evaluation Metrics:}  

ROUGE-L: 0.7931 \quad  BLEU-2: 0.6881 \quad BLEU-4: 0.5705 \quad METEOR: 0.8493

\end{tcolorbox}
\newpage

\subsection{Generalization in Retrosynthesis}
\label{sec:Generalization_retro}

To further assess AtomDisc’s capacity for genuine chemical generalization beyond memorized examples, we curated a set of retrosynthesis case studies (see Table~\ref{tab:retrosyn_cases}). Each case presents a target molecule (first column, labeled as “Product”) for which the model proposes plausible alternative disconnections or synthetic routes not present in the reference dataset. For each target, the second and third rows (“Pred. Rec. 1” and “Pred. Rec. 2”) show the reactants predicted by our model, while the following two rows (“GT. Rec. 1” and “GT. Rec. 2”) give the ground-truth reactants from the dataset.

These examples highlight the model’s ability to suggest diverse, mechanistically sound strategies—demonstrating flexibility in functional group disconnections, leaving group selection, and reaction order, rather than simply reproducing ground-truth solutions. Below, we discuss six representative cases in detail, each exemplifying a distinct mode of model generalization.

\paragraph{Case 1 for retrosynthesis.}  
This case illustrates a chemically plausible generalization: the model predicts a retrosynthetic disconnection at the amide bond, yielding two distinct reactants. While the reference route constructs the core via cyclization between an amine and an aldehyde, the model instead suggests a condensation between a brominated aromatic amine and a substituted aromatic aldehyde. This disconnection is mechanistically distinct and appears chemically valid, as both routes are standard in heterocyclic synthesis. The prediction thus demonstrates the model’s ability to propose viable alternatives, reflecting true generalization beyond surface-level memorization.

\paragraph{Case 2 for retrosynthesis.}  
This is a textbook example of valid generalization. The model replaces the brominated aromatic precursor in the reference with a chlorinated analogue. Since both bromine and chlorine are commonly used leaving groups in cross-coupling reactions, the substitution appears chemically reasonable and widely practiced, albeit with different catalysts or conditions. This prediction underscores AtomDisc’s capacity to generate alternative yet realistic synthetic routes.

\paragraph{Case 3 for retrosynthesis.}  
Here, the model proposes a markedly different set of reactants: instead of a simple phenyl isocyanate, it suggests a more elaborate urea derivative that already contains key heterocyclic and vinyl moieties. Despite deviating from the reference, the proposed transformation appears chemically feasible and achieves the same target. This highlights the model’s creativity in generating novel, valid strategies, even when diverging from canonical ground-truth pathways.


\begin{table*}[htbp]
  \centering
  \renewcommand{\arraystretch}{1.2}
  \begin{tabular}{
    >{\centering\arraybackslash}m{0.17\textwidth}
    >{\centering\arraybackslash}m{0.17\textwidth}
    >{\centering\arraybackslash}m{0.17\textwidth}
    >{\centering\arraybackslash}m{0.17\textwidth}
    >{\centering\arraybackslash}m{0.17\textwidth}
  }
    \toprule
    \textbf{Product}
      & \textbf{Pred.\ Rct.\ 1}
      & \textbf{Pred.\ Rct.\ 2}
      & \textbf{GT Rct.\ 1}
      & \textbf{GT Rct.\ 2} \\
    \midrule

    \includegraphics[width=\linewidth]{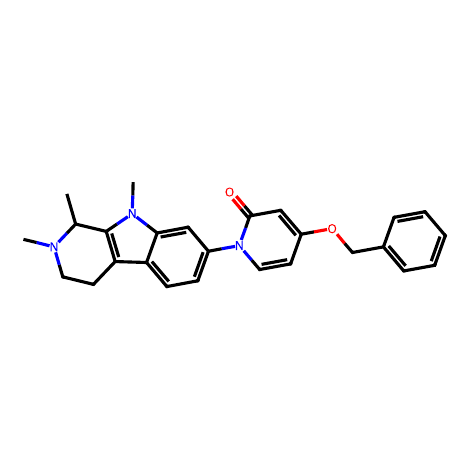}
    &
    \includegraphics[width=\linewidth]{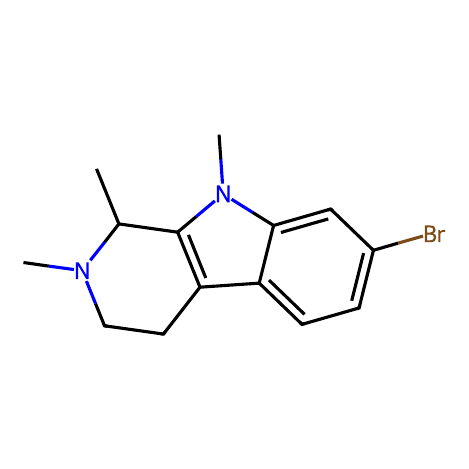}
    &
    \includegraphics[width=\linewidth]{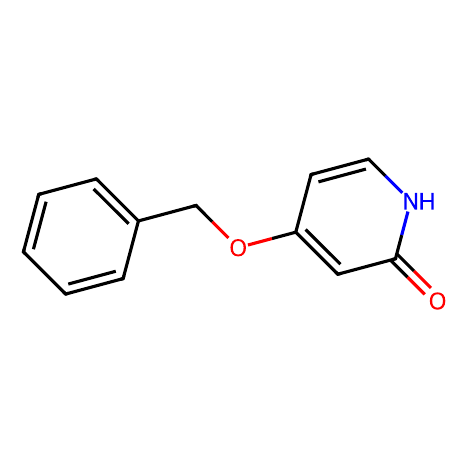}
    &
    \includegraphics[width=\linewidth]{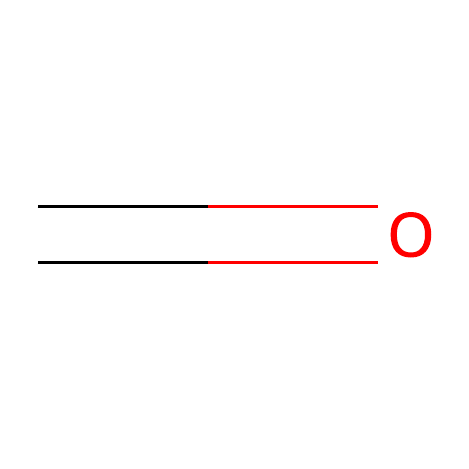}
    &
    \includegraphics[width=\linewidth]{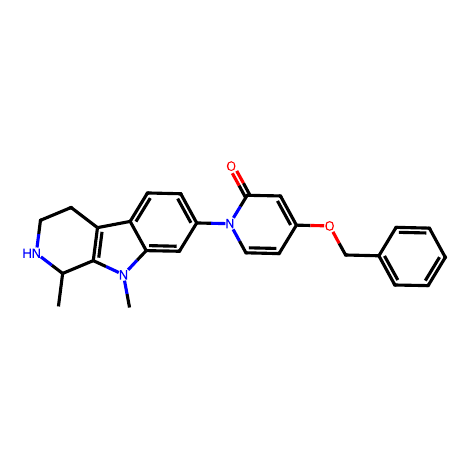}
    \\

    \includegraphics[width=\linewidth]{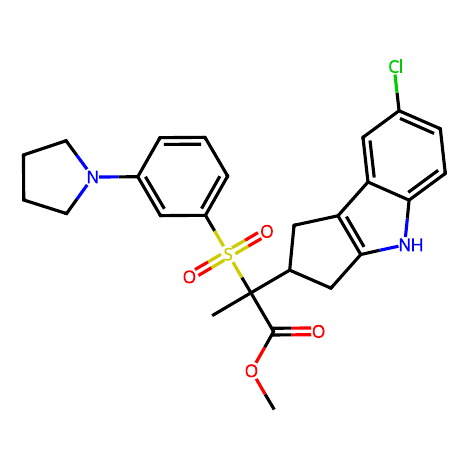}
    &
    \includegraphics[width=\linewidth]{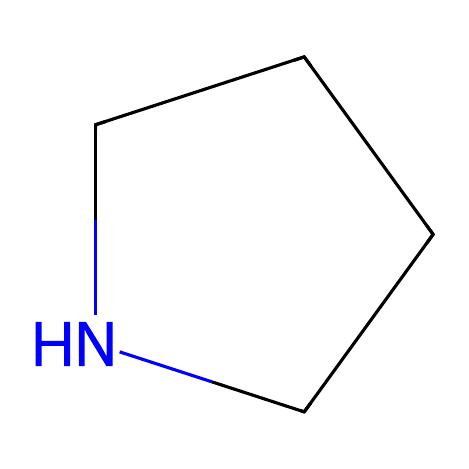}
    &
    \includegraphics[width=\linewidth]{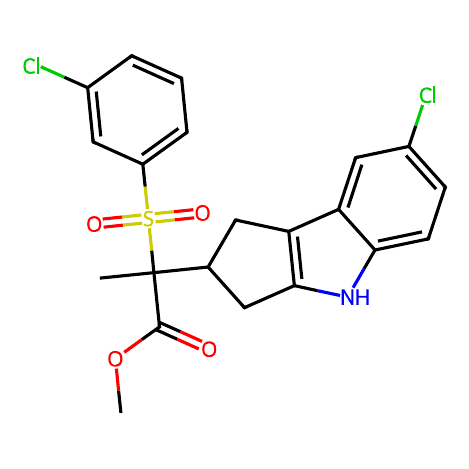}
    &
    \includegraphics[width=\linewidth]{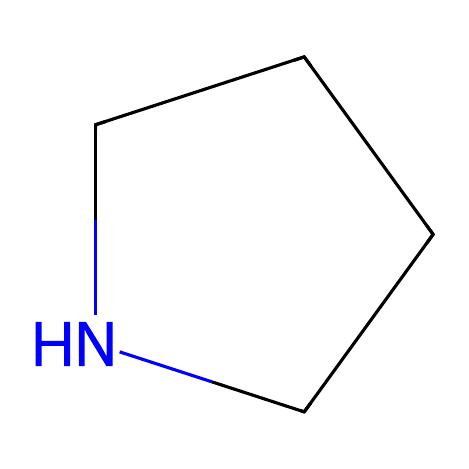}
    &
    \includegraphics[width=\linewidth]{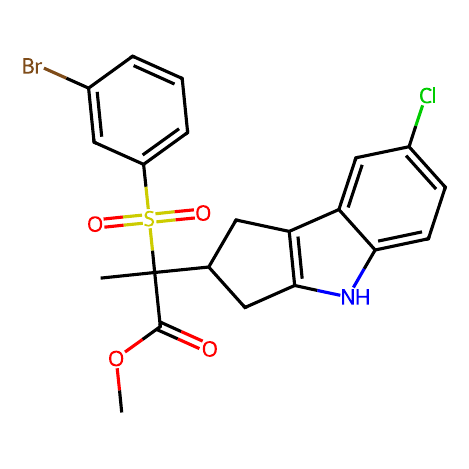}
    \\

    \includegraphics[width=\linewidth]{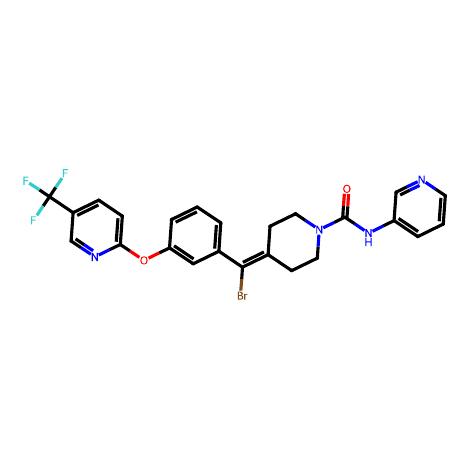}
    &
    \includegraphics[width=\linewidth]{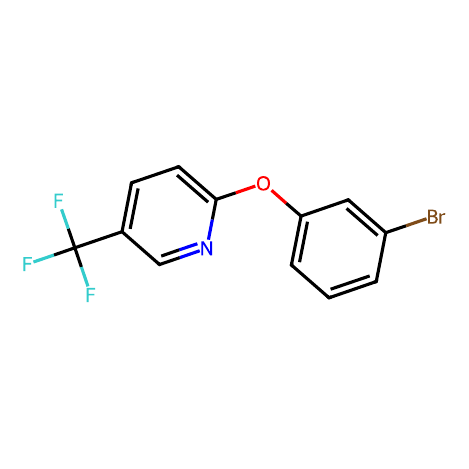}
    &
    \includegraphics[width=\linewidth]{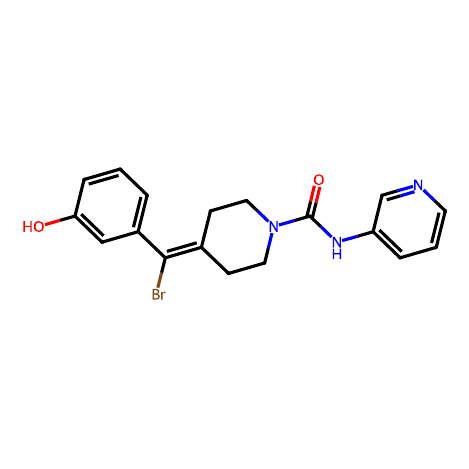}
    &
    \includegraphics[width=\linewidth]{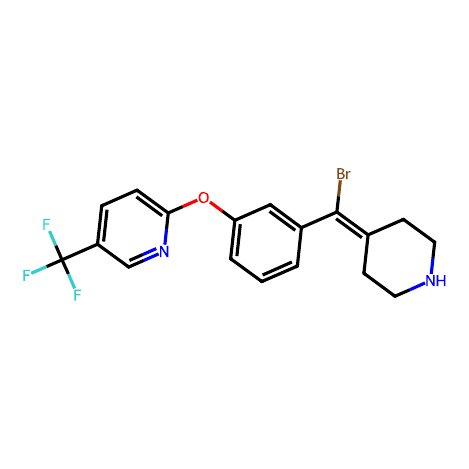}
    &
    \includegraphics[width=\linewidth]{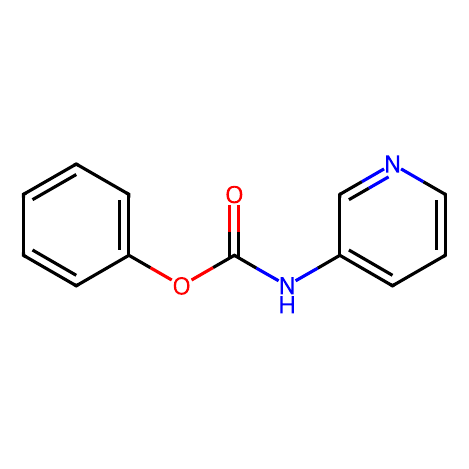}
    \\

    \includegraphics[width=\linewidth]{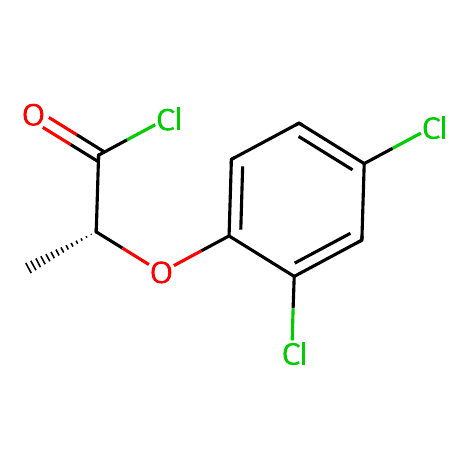}
    &
    \includegraphics[width=\linewidth]{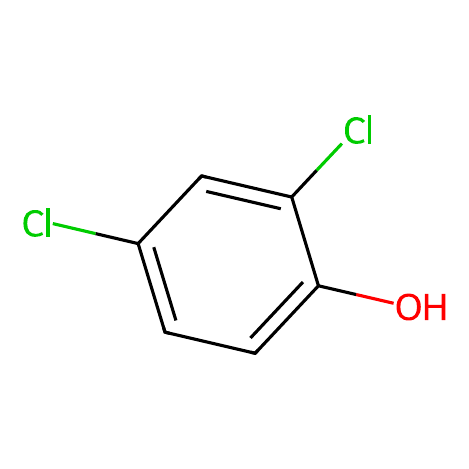}
    &
    \includegraphics[width=\linewidth]{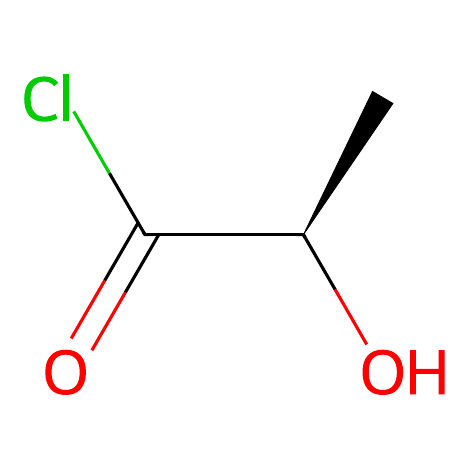}
    &
    \includegraphics[width=\linewidth]{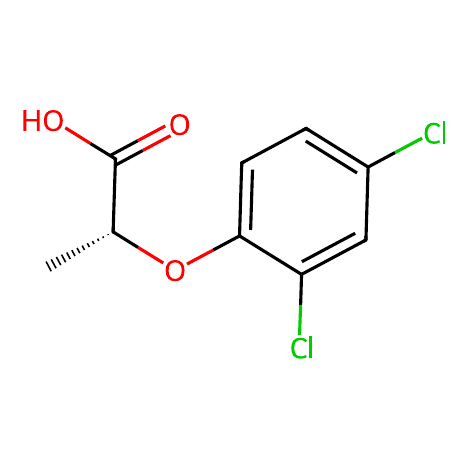}
    &
    \includegraphics[width=\linewidth]{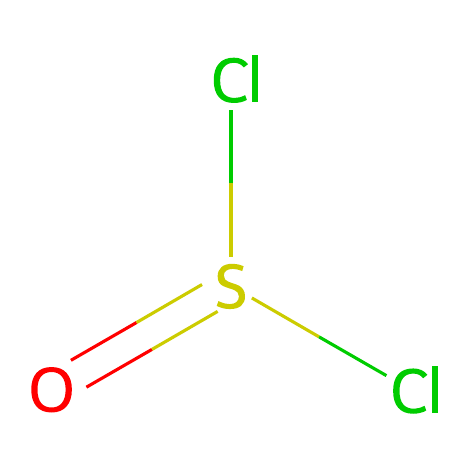}
    \\

    \includegraphics[width=\linewidth]{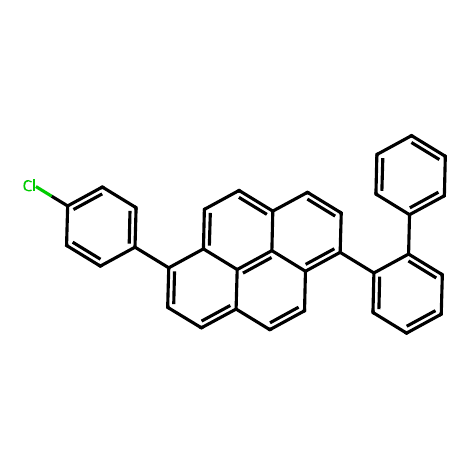}
    &
    \includegraphics[width=\linewidth]{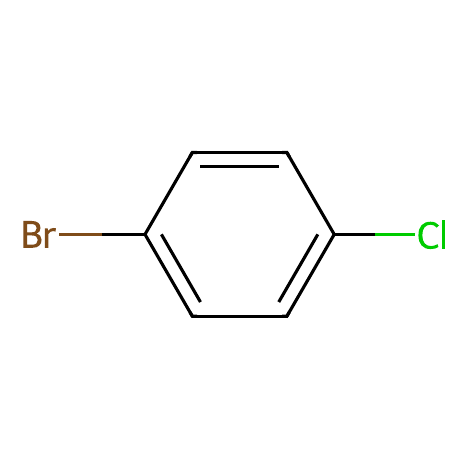}
    &
    \includegraphics[width=\linewidth]{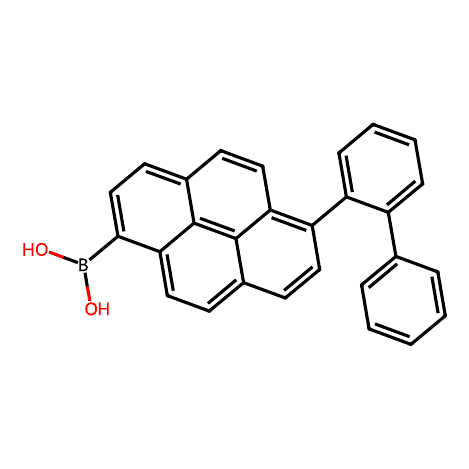}
    &
    \includegraphics[width=\linewidth]{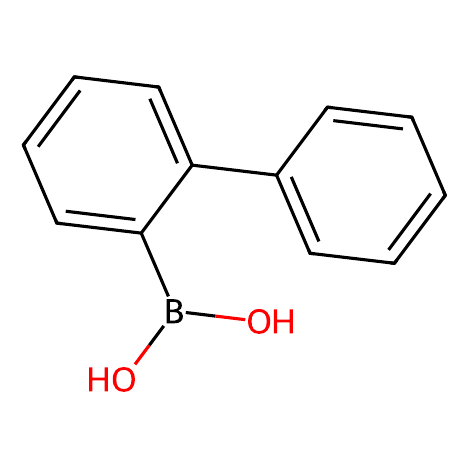}
    &
    \includegraphics[width=\linewidth]{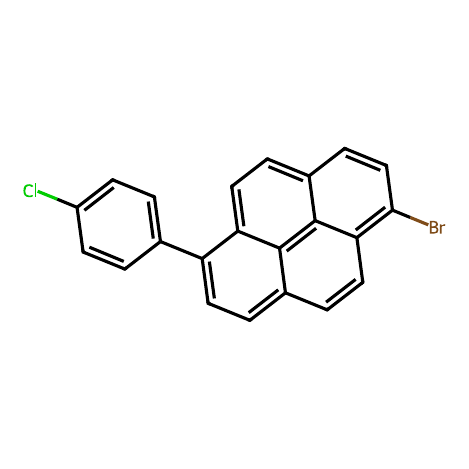}
    \\

    \includegraphics[width=\linewidth]{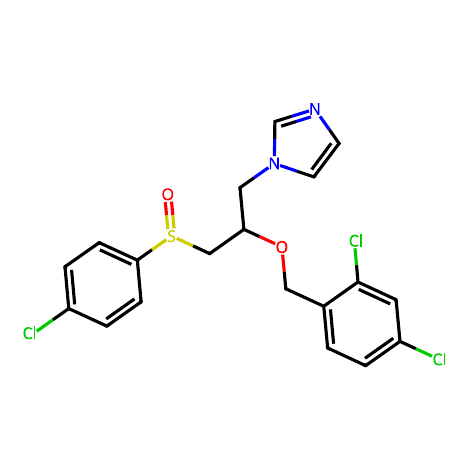}
    &
    \includegraphics[width=\linewidth]{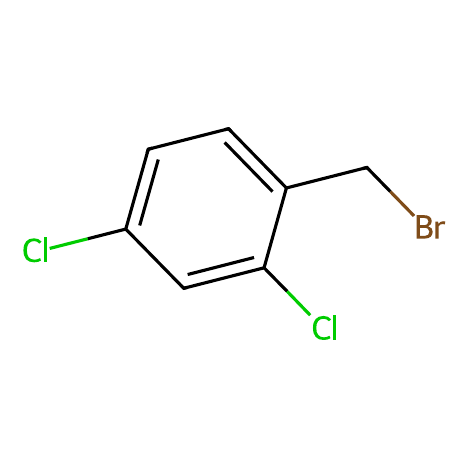}
    &
    \includegraphics[width=\linewidth]{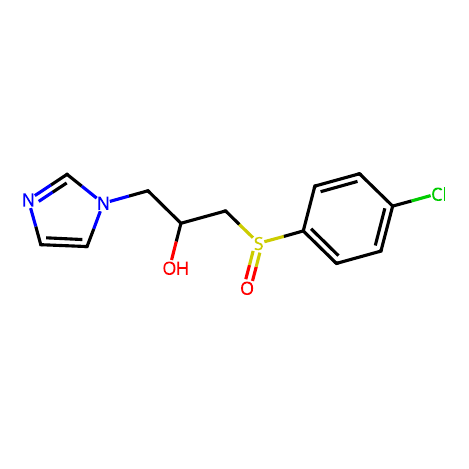}
    &
    \includegraphics[width=\linewidth]{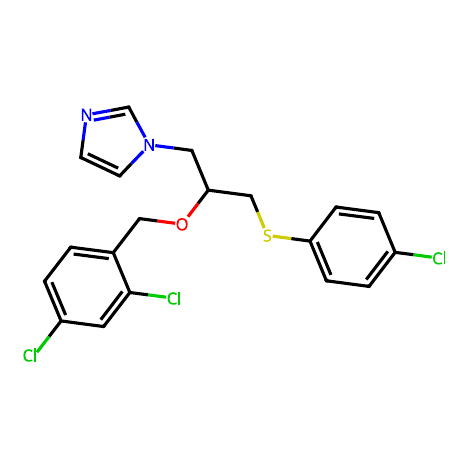}
    &
    \includegraphics[width=\linewidth]{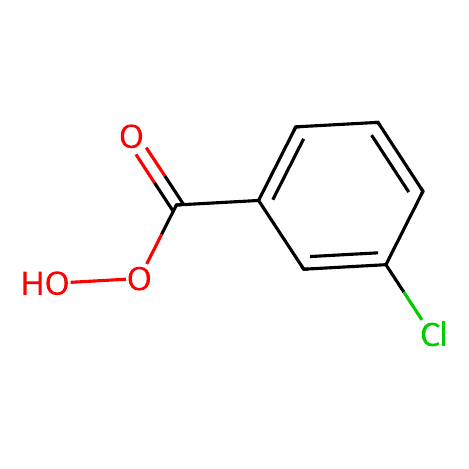}
    \\
    
    \bottomrule
  \end{tabular}
  \caption{Generalize Case for Retrosynthesis.}
  \label{tab:retrosyn_cases}
\end{table*}

\paragraph{Case 4 for retrosynthesis.}  
While the ground truth reflects a direct chlorination of a carboxylic acid using thionyl chloride, the model proposes an alternative strategy: using a pre-formed acid chloride and a chlorinated phenol. Such inversion of synthetic order has precedent in practice, especially when certain intermediates are more reactive or readily available. This prediction demonstrates AtomDisc’s capacity to generalize over synthetic logic, not just memorized steps.

\paragraph{Case 5 for retrosynthesis.}  
In this example, the model predicts a Suzuki–Miyaura coupling with roles of the boronic acid and halide reversed compared to the reference. Despite this inversion, the product is the same and the reaction would appear chemically sound. This case showcases AtomDisc’s understanding of functional group symmetry and its ability to generalize over interchangeable coupling partners.

\paragraph{Case 6 for retrosynthesis.}  
Here, the model shifts the oxidation step to an earlier stage in synthesis, contrasting with the reference’s late-stage sulfone formation. The aryl component is instead introduced via electrophilic substitution onto the preformed sulfone. This reversed strategy appears chemically viable, suggesting the model has learned to generalize reaction order beyond rigid, textbook-like templates.
\newpage
\subsection{Generalization in Reagent Prediction}
\label{sec:Generalization_reagent}

To further evaluate AtomDisc’s ability to generalize beyond memorized training examples, we present a series of reagent prediction case studies (see Table~\ref{tab:reagent_pred_cases}). In each case, the model proposes alternative reagents or reaction pathways not found in the reference data. These examples demonstrate AtomDisc’s flexibility in recognizing chemically valid alternatives, such as variations in reagent choice, functional group transformations, and synthetic order—reflecting mechanistic understanding rather than rote recall.

Below, we provide detailed discussion of six representative cases, each illustrating a different aspect of model generalization. For clarity, in Table~\ref{tab:reagent_pred_cases}, the “Product” column shows the target molecule to be synthesized, “Rec. 1” and “Rec. 2” denote the required reactants, “GT Reagent” lists the ground-truth reagent from the dataset, and “Pred. Reagent” shows the alternative reagent(s) proposed by our model.

\begin{table*}[htbp]
  \centering
  \renewcommand{\arraystretch}{1.2}
  \begin{tabular}{
    >{\centering\arraybackslash}m{0.18\textwidth}
    >{\centering\arraybackslash}m{0.18\textwidth}
    >{\centering\arraybackslash}m{0.18\textwidth}
    >{\centering\arraybackslash}m{0.18\textwidth}
    >{\centering\arraybackslash}m{0.18\textwidth}
  }
    \toprule
    \textbf{Product}
      & \textbf{Rec.\ 1}
      & \textbf{Rec.\ 2}
      & \textbf{GT Reagent}
      & \textbf{Pred.\ Reagent} \\
    \midrule

    \includegraphics[width=\linewidth]{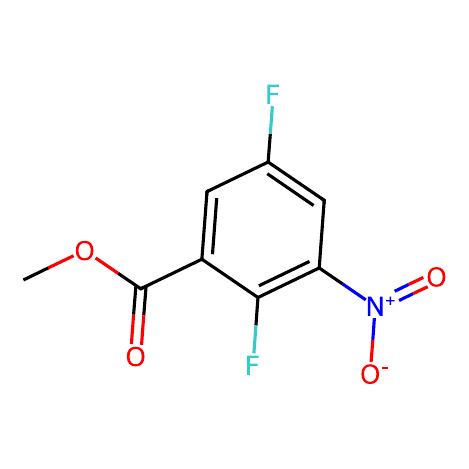}
    &
    \includegraphics[width=\linewidth]{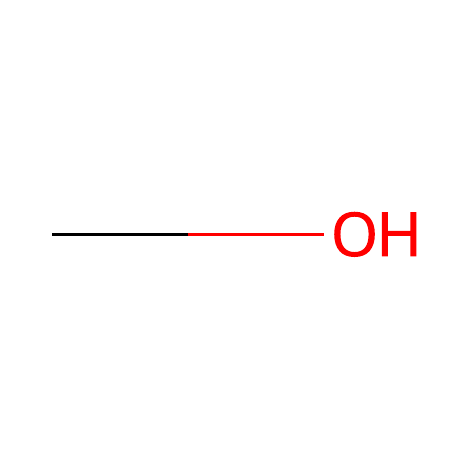}
    &
    \includegraphics[width=\linewidth]{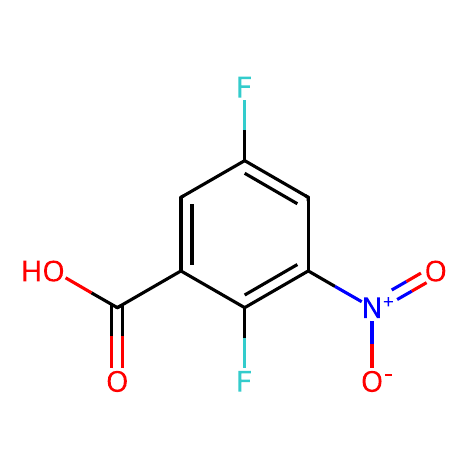}
    &
    \includegraphics[width=\linewidth]{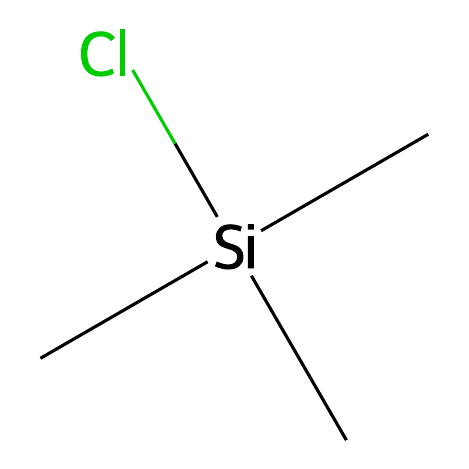}
    &
    \includegraphics[width=\linewidth]{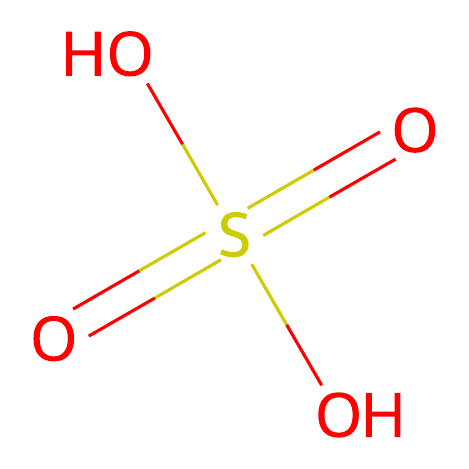}
    \\

    \includegraphics[width=\linewidth]{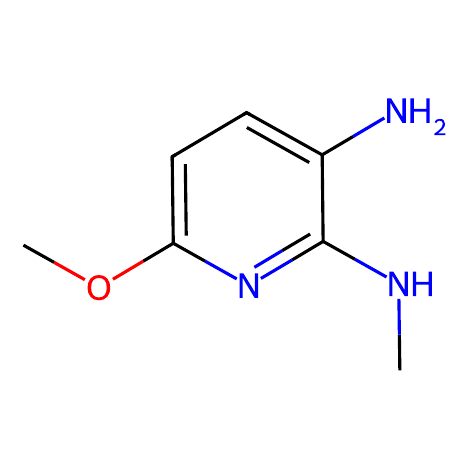}
    &
    \includegraphics[width=\linewidth]{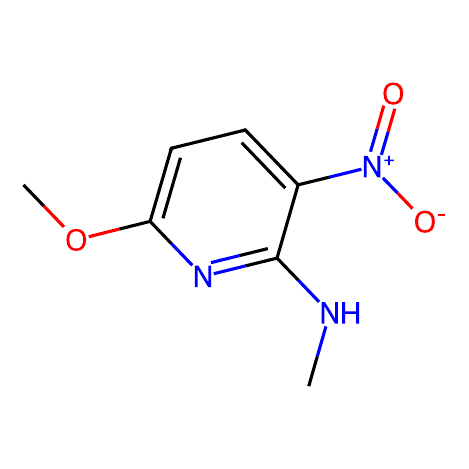}
    &
    
    &
    \includegraphics[width=\linewidth]{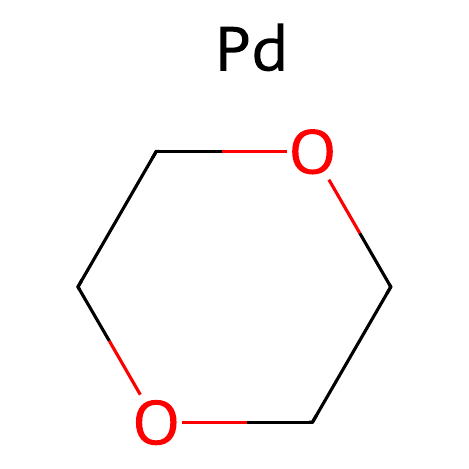}
    &
    \includegraphics[width=\linewidth]{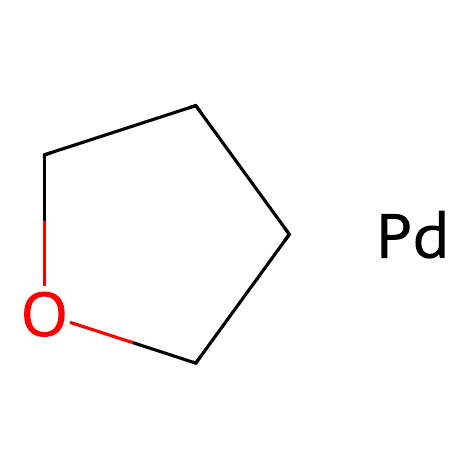}
    \\

    \includegraphics[width=\linewidth]{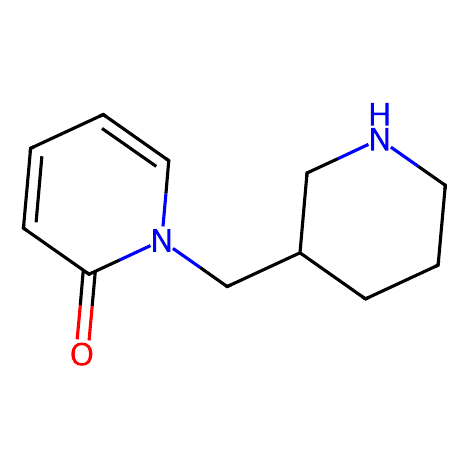}
    &
    \includegraphics[width=\linewidth]{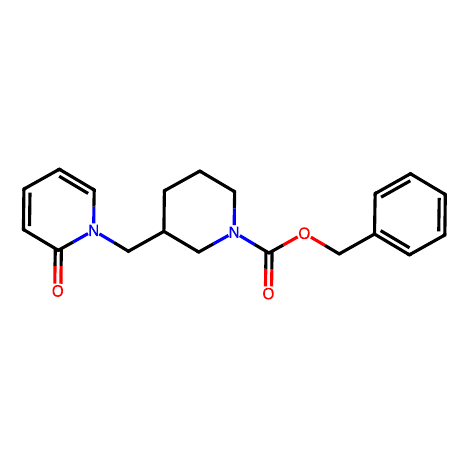}
    &

    &
    \includegraphics[width=\linewidth]{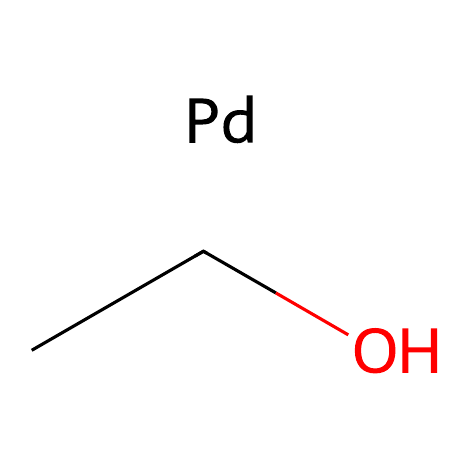}
    &
    \includegraphics[width=\linewidth]{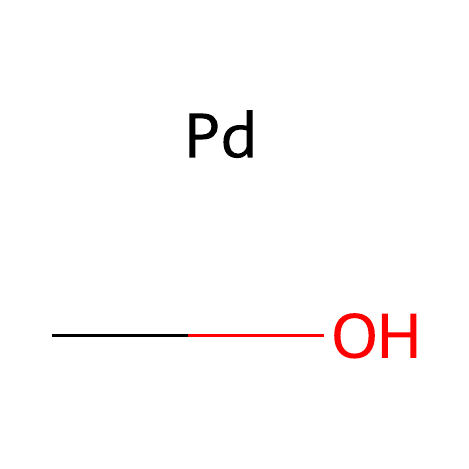}
    \\

    \includegraphics[width=\linewidth]{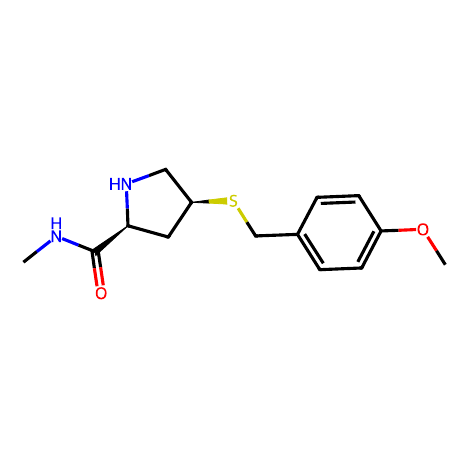}
    &
    \includegraphics[width=\linewidth]{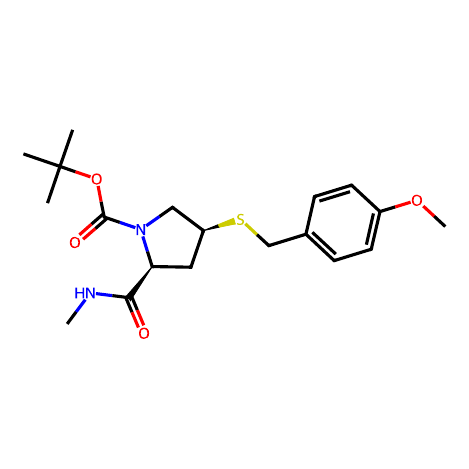}
    &

    &
    \includegraphics[width=\linewidth]{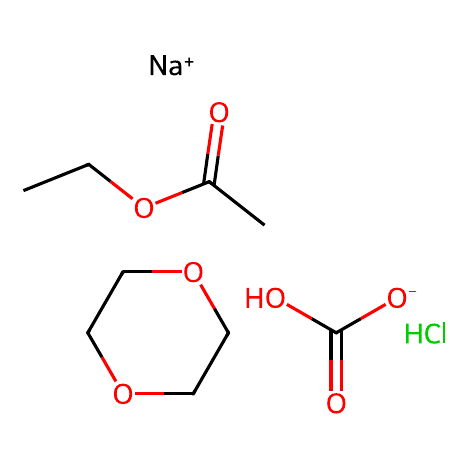}
    &
    \includegraphics[width=\linewidth]{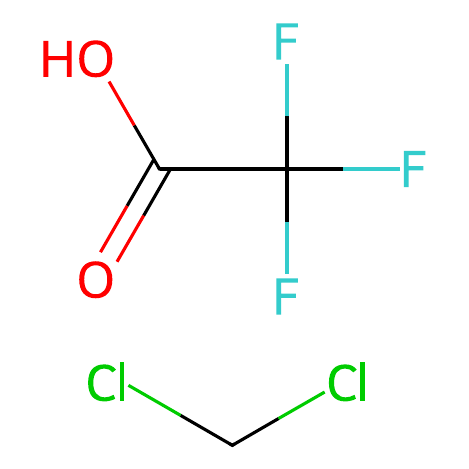}
    \\

    \includegraphics[width=\linewidth]{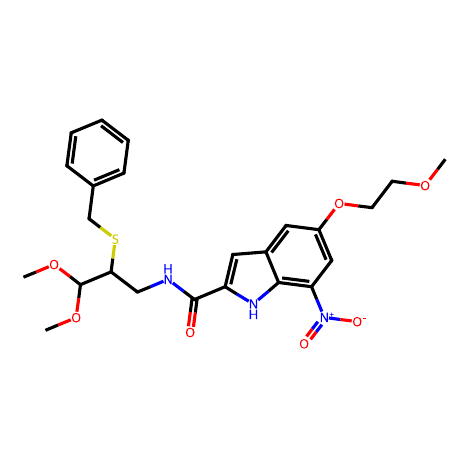}
    &
    \includegraphics[width=\linewidth]{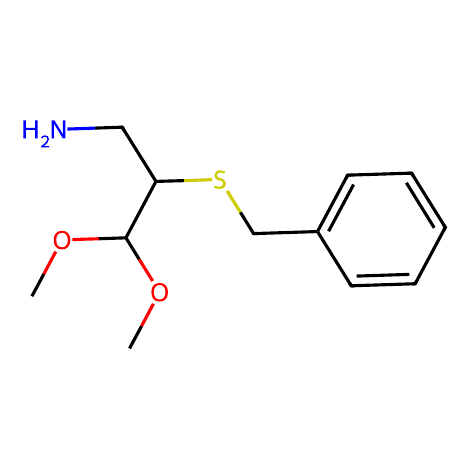}
    &
    \includegraphics[width=\linewidth]{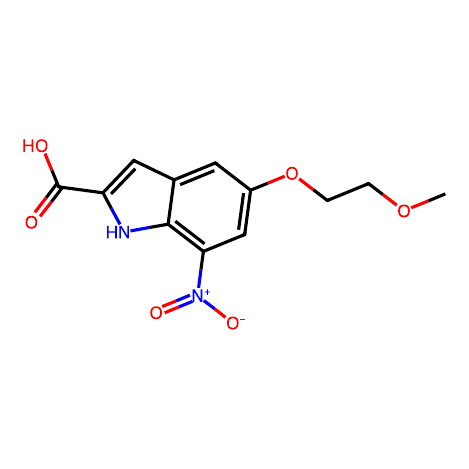}
    &
    \includegraphics[width=\linewidth]{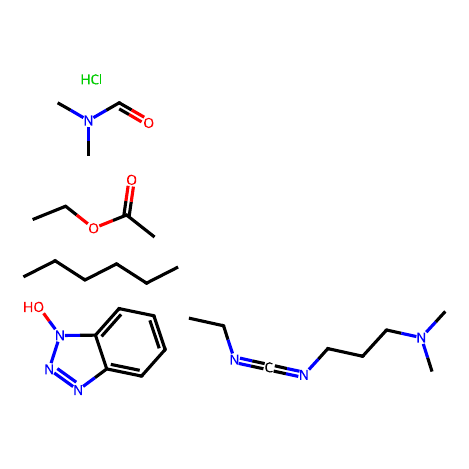}
    &
    \includegraphics[width=\linewidth]{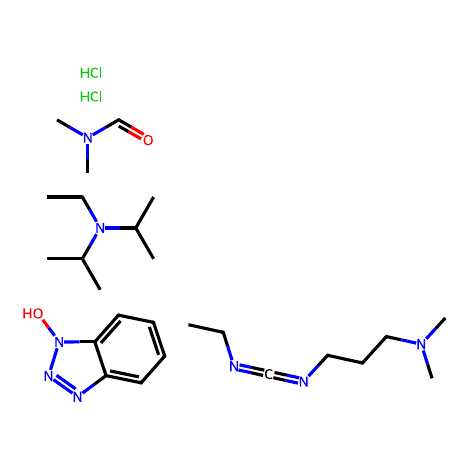}
    \\

    \includegraphics[width=\linewidth]{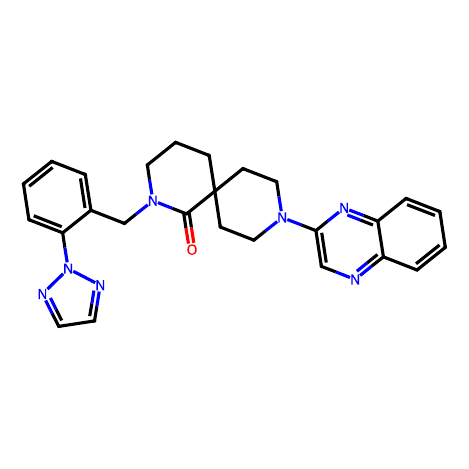}
    &
    \includegraphics[width=\linewidth]{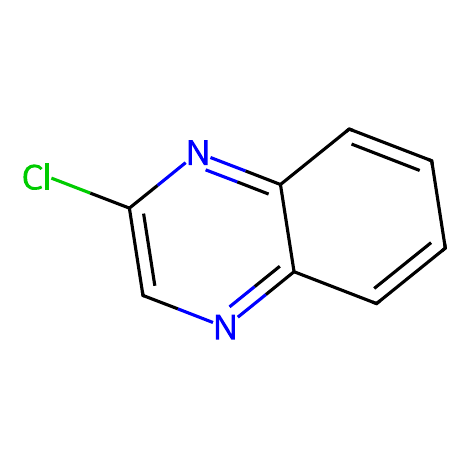}
    &
    \includegraphics[width=\linewidth]{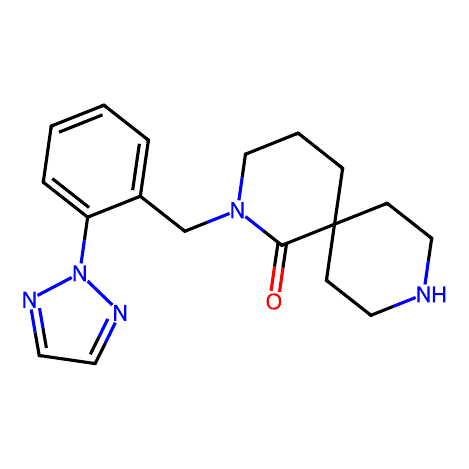}
    &
    \includegraphics[width=\linewidth]{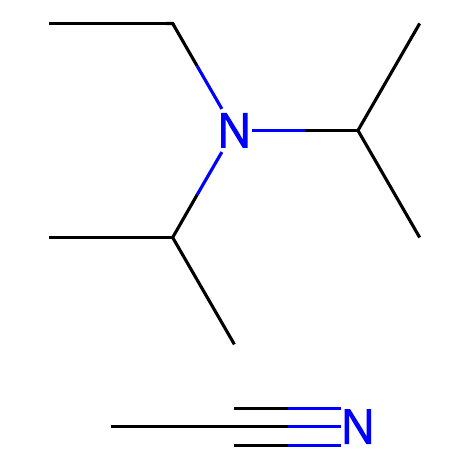}
    &
    \includegraphics[width=\linewidth]{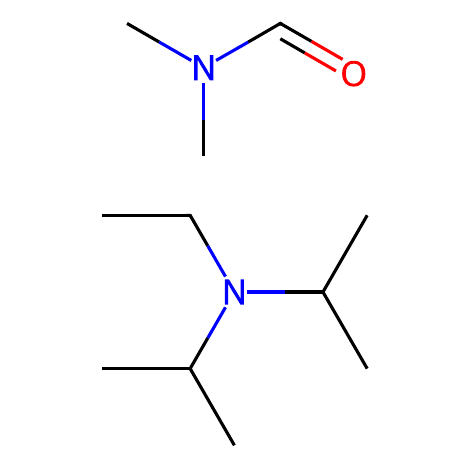}
    \\

    \bottomrule
  \end{tabular}
  \caption{Generalized cases on reagent prediction task.}
  \label{tab:reagent_pred_cases}
\end{table*}

\paragraph{Case 1 for reagent prediction.}
The model predicts sulfuric acid ($H_2SO_4$) as a catalyst for esterification — a classic choice that appears chemically reasonable. Though different from the reference (possibly a silylating agent), sulfuric acid is widely used in both academic and industrial contexts. This prediction suggests plausible generalization in identifying functionally related reagents.

\paragraph{Case 2 for reagent prediction.}
AtomDisc identifies palladium catalysis and a polar aprotic solvent (THF) for nitro reduction, which captures aspects of the reaction class. However, the absence of a reducing agent (e.g., $H_2$ or hydride) makes the prediction incomplete. This indicates that the model can point to mechanistically relevant features, though with occasional gaps in specificity.

\paragraph{Case 3 for reagent prediction.}
The model substitutes methanol for ethanol while retaining palladium catalysis. This minor modification is consistent with known solvent flexibility in such reactions and appears chemically plausible, suggesting some recognition of reagent equivalence.

\paragraph{Case 4 for reagent prediction.}
AtomDisc suggests trifluoroacetic acid (TFA)/DCM for Boc deprotection instead of the reference HCl/dioxane system. Both acidic systems are commonly used in this context, and the substitution appears reasonable. The prediction therefore reflects adaptation to functionally similar strategies rather than strict memorization.

\paragraph{Case 5 for reagent prediction.}
This example shows that the model identifies key components for amide coupling. Although minor structural differences exist (e.g., TEA analogs) and some redundancies appear (e.g., duplicated halides), the predicted reagents seem broadly consistent with known practice, indicating functional generalization rather than exact recall.

\paragraph{Case 6 for reagent prediction.}
The model proposes a non-nucleophilic base (DIPEA) and polar aprotic solvents (DMF/DMAc) for a C–N coupling, aligning with elements of established protocols such as Buchwald–Hartwig or Ullmann conditions. However, the omission of a catalytic species makes the prediction only partially complete, highlighting dataset-driven constraints in catalyst representation.

\newpage
\subsection{Interpretable Tokenization Reveals Structure–Property Associations}
The interpretable, atom-level tokenization provided by AtomDisc offers a unique window into the underlying chemical logic learned by molecular language models. Rather than treating atomic environments as black boxes, our framework enables systematic attribution of both model predictions and emergent token boundaries to specific, quantifiable chemical properties. In this section, we move beyond aggregate benchmarks to perform a series of case studies, demonstrating how discrete atom tokens can uncover subtle and non-trivial structure–property associations directly from data.

To facilitate this analysis, we design a unified experimental setup and define a set of rigorous atomic metrics, enabling quantitative comparison of property distributions across different tokens and functional groups. The following subsection details our methodology for atom-level property calculation and dataset curation, providing a foundation for the scientific discoveries presented in subsequent analyses.
\label{sec:scientific_discovery}
\subsubsection{Setup and Atomic Property Metrics}
\label{sec:case_scientific_setup}

To systematically analyze the chemical meaning and discovery potential of AtomDisc’s interpretable tokens, we utilized three key atomic properties—Mulliken atomic charge, local polar surface area (PSA), and local $\pi$-electron occupancy—as described in \blue{Supplementary Information B.4}. These properties capture complementary aspects of atomic electronic structure and local environment, providing a robust basis for comparing property distributions across different token categories and functional groups.

For each target token or functional group (e.g., mixture tokens or specific OH tokens), we sampled 100–500 representative atoms from the dataset, computed their properties as described above, and visualized the resulting distributions. Statistical analyses—including U-tests, Kolmogorov–Smirnov statistics, Wasserstein distances, overlap of KDE,and Jensen–Shannon divergence—were applied to compare property distributions across token categories and functional groups.

\subsubsection{Chemical Interpretability of Discrete Atom Tokens}

Although our codebook was learned in a self-supervised way without explicit chemical labels, we find that the resulting discrete tokens (i.e., \texttt{<atom\_$k$>}) are highly correlated with chemically meaningful patterns. Specifically, the tokenizer naturally assigns different tokens to atoms of the same element type if they are in distinct chemical environments (e.g., a carbon in a carboxylic acid vs. a carbon in a methyl group), enabling fine-grained differentiation beyond traditional atom types.

To quantitatively assess the chemical interpretability of our atom-level tokenizer, we performed a large-scale analysis of the conditional probability $P(\mathrm{FG}|\text{token})$—that is, the probability that a given discrete token corresponds to a specific functional group (FG) in real molecules.

Specifically, we sampled \textbf{220,000} molecules from the PubChem database, ensuring a diverse coverage of chemical scaffolds and functional groups. For each molecule, we performed the following steps:
\begin{enumerate}
\item \textbf{Atom Representation Extraction:} Each molecule was converted to a molecular graph and fed through a {MoleculeSTM} GNN encoder (pretrained on GEOM and PubChem), resulting in a set of continuous atom embeddings ${\mathbf{h}_i}$.
\item \textbf{Vector Quantization:} Each atom embedding $\mathbf{h}_i$ was discretized to its nearest codebook entry via vector quantization (VQ), yielding a discrete structural token $\mathrm{token}_i$ (i.e., a code index).
\item \textbf{Functional Group Annotation:} To assign ground-truth functional group labels, we extracted functional groups from each molecule using a curated set of SMARTS patterns. To ensure {exclusive and unambiguous} assignment, we matched functional groups in a prioritized order (from largest to smallest), so that each atom is assigned to at most one functional group—the first pattern it matches.
\item \textbf{Token-FG Mapping:} We collected statistics only on {core atoms of functional groups} (e.g., the oxygen in a hydroxyl group, the nitrogen in an amide, etc.). For each token index, we counted how often it appeared as the structural token for each FG core atom, aggregating across the entire dataset.
\end{enumerate}
Formally, for each token $t$ and functional group $\mathrm{FG}$, we computed:
$$P(\mathrm{FG}|t) = \frac{\text{Count}(t,\mathrm{FG})}{\sum_{\mathrm{FG}’} \text{Count}(t, \mathrm{FG}’)}$$
where $\text{Count}(t,\mathrm{FG})$ is the number of times token $t$ is assigned to an atom annotated as belonging to functional group $\mathrm{FG}$ across the corpus.

\paragraph{Case Study: Hydroxyl Groups (\texttt{OH})}

\begin{figure}[htbp]
    \centering
    \includegraphics[width=\textwidth]{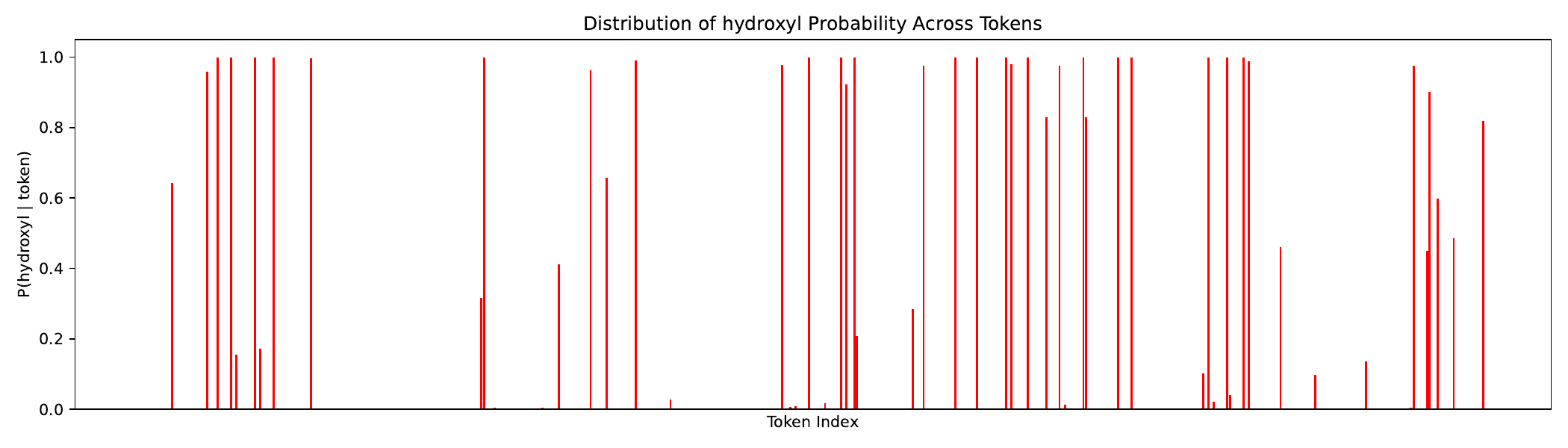}
    \caption{Distribution of $P(\mathrm{hydroxyl}|\text{token})$ across all 512 atom tokens. Tokens with probability $>0.8$ can be interpreted as encoding "pure" hydroxyl atoms, while others represent more diverse or mixed chemical contexts.}
    \label{fig:token_fg_prob_oh}
\end{figure}

We take the hydroxyl group as a representative example. In Figure~\ref{fig:token_fg_prob_oh}, we visualize $P(\mathrm{hydroxyl}|\text{token})$ for all atom tokens. Several tokens exhibit a strong association with the hydroxyl group (probability $>0.8$), indicating that these codes are consistently assigned to oxygen atoms in clear hydroxyl environments. In contrast, other tokens have much lower, yet nonzero, probabilities. This suggests that these tokens are shared across more chemically ambiguous or mixed contexts, where the oxygen atom of a hydroxyl group may resemble other functional groups in its local environment—highlighting the tokenizer’s ability to reflect context-dependent overlap and chemical nuance.

\newpage
\begin{table}[htbp]
\centering
\caption{
\textbf{{Statistical comparison of predicted LogS values for original versus swapped tokens.}}
For each comparison, two groups are analyzed: (\textbf{{Original}}) predictions from original representation sequence, and (\textbf{{Swapped}}) predictions from swapped sequence. Reported metrics include mean $\pm$ standard deviation, distribution comparison metrics (\textbf{{Wasserstein distance}}, \textbf{{Jensen–Shannon divergence}}, \textbf{{Overlap area}}), and results of paired significance tests (\textbf{{Wilcoxon}} and \textbf{{Paired t-test}}).
}
\label{tab:swap_token_prediction_comparison}
\resizebox{\textwidth}{!}{%
\begin{tabular}{@{}lcccccccccc@{}}
\toprule
& \multicolumn{2}{c}{\textbf{Distribution}} & \multicolumn{3}{c}{\textbf{Distribution Comparison}} & \multicolumn{4}{c}{\textbf{Paired Tests}}  \\
\cmidrule(lr){2-3} \cmidrule(lr){4-6} \cmidrule(lr){7-10} \cmidrule(lr){11-11}
\textbf{Comparison} & \textbf{Original} & \textbf{Swapped} & \textbf{Wass.} & \textbf{JS Div.} & \textbf{Overlap} & \textbf{Wilcoxon-stat} & \textbf{W p-val} & \textbf{t-stat} & \textbf{t p-val}\\
\midrule

\textbf{Token 319 vs Swapped to Token 338}
& 3.03 $\pm$ 2.47 & 2.64 $\pm$ 2.40 & 0.40 & 0.03 & 0.78 & 6.5e+04 & $<$ 1e-10 & 21.08 & $<$ 1e-10 \\

\textbf{Token 338 vs Swapped to Token 319}
& 3.34 $\pm$ 1.74 & 3.41 $\pm$ 1.80 & 0.09 & 0.02 & 0.84 & 3.2e+04 & $<$ 1e-10 & -7.25 & $<$ 1e-10 \\
\bottomrule
\end{tabular}%
}
\end{table}
\newpage
\paragraph{Case Study: Mixture Token.}
\label{sec:mixture_token}
To systematically explore the interpretability of our discrete atom tokenizer, we focus on the so-called \emph{mixture tokens}---i.e., those atom-level structure tokens that exhibit significant association with more than one functional group (FG). Specifically, for each token $t$, we compute the conditional probability $P(\mathrm{FG} \mid t)$, defined as the proportion of times token $t$ is assigned to an atom annotated with a given FG label, using SMARTS matching on a large set of molecules from PubChem.

\begin{figure}[htbp]
    \centering
    \includegraphics[width=\linewidth]{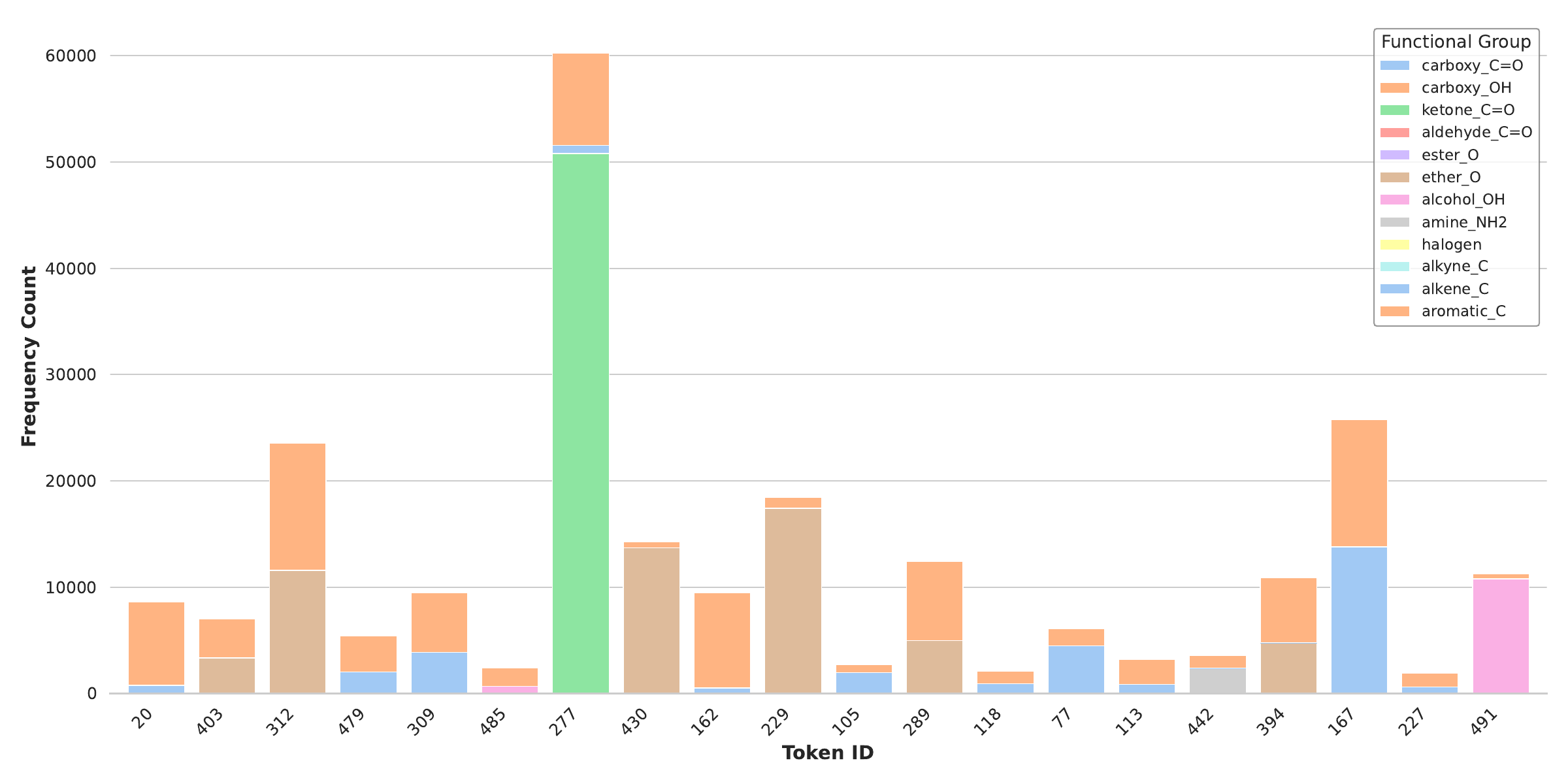}
    \caption{Functional group composition of selected \emph{mixture tokens}, visualized as stacked bar charts for each token. Each bar represents a discrete structure token (identified as a "mixture token" based on its high association with multiple functional groups), with colored segments showing the conditional probability $P(\mathrm{FG} \mid \mathrm{token})$ for each functional group (FG). The diversity of colors within a single bar highlights that these tokens capture atomic environments overlapping between traditional functional group categories, rather than being exclusive to a single moiety. This reveals that our atom tokenizer can identify chemically ambiguous or transitional regions of molecular space, which are difficult to capture using classic deterministic rules.}
    \label{fig:mixture_token}
\end{figure}

A token is considered a mixture token if it satisfies the criterion that $P(\mathrm{FG} \mid t) > 0.1$ for at least two different FGs. This process is performed for all 220,000 molecules, using the following pipeline:
\begin{enumerate}
    \item Each molecule is parsed with RDKit and encoded by a GNN-based structure encoder (MoleculeSTM), yielding atom-level representations.
    \item The atom representations are discretized via a vector-quantized codebook, assigning each atom to a discrete structure token.
    \item Functional groups are extracted for each atom using SMARTS-based pattern matching, with exclusive assignment achieved via a priority-ordered matching scheme to ensure that each atom is assigned to at most one FG.
    \item The $P(\mathrm{FG} \mid t)$ matrix is computed by counting, for each token, the frequency with which it is mapped to atoms of each FG type, normalized by the total number of assignments of that token.
\end{enumerate}
As shown in Figure~\ref{fig:mixture_token}, each bar corresponds to a mixture token, with colored segments representing the conditional probability of each functional group. Unlike "pure" tokens that are highly specific to a single FG, mixture tokens display a heterogeneous FG composition, suggesting that these tokens capture chemically ambiguous or transitional atomic environments---such as those appearing at the boundaries of functional groups or in conjugated systems.

The emergence of such mixture tokens provides a data-driven perspective on the fuzzy boundaries and continuum between canonical functional group categories. Rather than imposing sharp, rule-based partitions, our learned tokenizer discovers latent structure that can reflect the subtleties of real chemical environments, potentially offering new insights for downstream chemical understanding, molecular design, and property prediction tasks.

\subsubsection{Data-Driven Discovery of Token–Functional Group Relationships}
From the above statistical analysis, we observe that most atom tokens exhibit strong specificity—each token is predominantly associated with a single functional group, reflecting high selectivity in how the tokenizer organizes chemical information. However, there also exist a subset of “mixture” tokens, which are mapped to multiple functional groups with significant probability, indicating the presence of shared or ambiguous atomic environments across different functional motifs. On the other hand, even within a single functional group such as hydroxyl (OH), we find that the oxygen atom can be represented by several different tokens depending on its local structural context. This suggests that the tokenizer is sensitive to subtle differences in environment—such as whether the OH is attached to a ring, a chain, or is adjacent to other polar or electronic groups.

From a technical perspective, these patterns arise because the tokenizer clusters atom-level representations in embedding space according to the similarities of their local environments, rather than relying solely on element or functional group labels. As a result, atoms with similar neighborhood topology, hybridization, and electronic context are assigned the same token, while differences—however subtle—may lead to distinct token assignments. Based on these observations, we hypothesize the following:
\begin{itemize}
    \item For hydroxyl groups, different token assignments (e.g., ring-attached vs. chain-attached OH) may capture variations in electronic properties, hydrogen-bonding potential, and solvent accessibility, which can influence molecular properties such as solubility.
    \item For mixture tokens, the presence of multiple functional group assignments suggests that these tokens encode chemically ambiguous or transitional environments—such as conjugated systems, boundary atoms between different functional motifs, or atoms influenced by multiple neighboring groups.
\end{itemize}

To test these hypotheses, we next perform detailed property analysis and statistical comparison across both “pure” and “mixture” tokens, examining the distribution of atomic properties such as Mulliken charge, $\pi$-electron occupancy, and local PSA for each token group.

\paragraph{Atomic Property Analysis of Hydroxyl Tokens}

\begin{figure}[htbp]
    \centering
    \includegraphics[width=\linewidth]{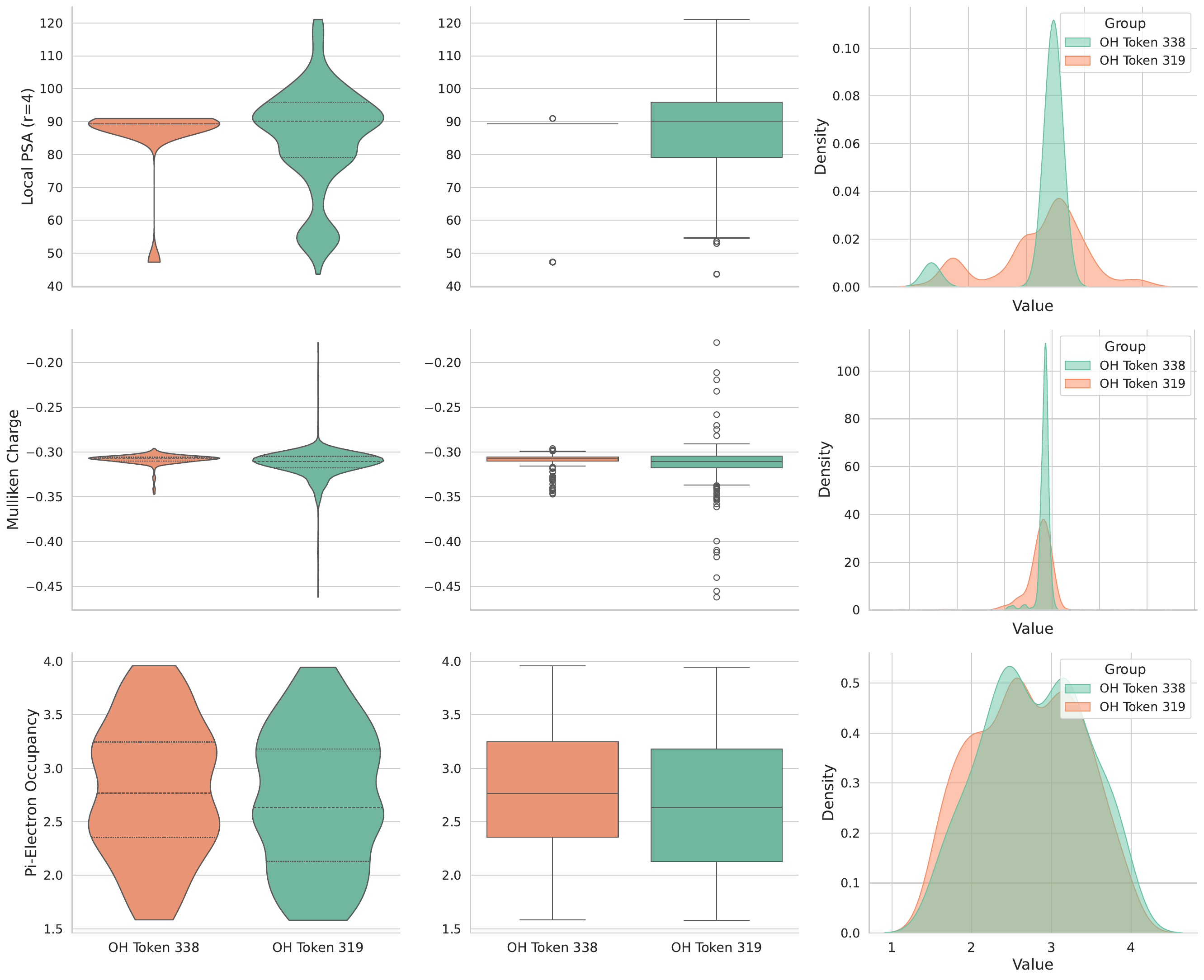}
    \caption{Comparison of key atomic properties—Mulliken charge, local $\pi$-electron occupancy, and local polar surface area (PSA)—for hydroxyl groups assigned to token 338 versus token 319. These distributions illustrate the environment-dependent differentiation learned by the tokenizer, with token 338 (chain OH) and token 319 (ring-attached OH) showing distinct property profiles.}
    \label{fig:oh_318_339}
\end{figure}

To elucidate why chemically equivalent hydroxyl groups may be assigned to different structure tokens by the tokenizer, we systematically compared the distributions of key atomic properties—Mulliken charge, local polar surface area (PSA), and local $\pi$-electron occupancy—for OH groups assigned to token 338 versus token 319. As shown in Figure~\ref{fig:oh_318_339}, these distributions exhibit notable, environment-dependent differences.

Figure~\ref{fig:oh_318_339} presents the main distributions of key atomic properties for hydroxyl groups assigned to tokens 338 and 319, illustrating the context-dependent differentiation learned by our tokenizer. To further supplement these findings, Figure~\ref{fig:appendix_OH} provides additional representative examples and more comprehensive distributional analyses. This includes more molecular samples, extended LogS prediction distributions under token swapping, as well as alternative visualization formats such as violin plots and boxplots in addition to KDE. Together, these figures offer a complete picture of the environment-dependent behavior of the token assignments and further validate the interpretability of AtomDisc on chemically equivalent functional groups.

\begin{figure}
    \centering
    \includegraphics[width=\linewidth]{fig/case_1_OH.pdf}
    \caption{AtomDisc interpretability for equivalent hydroxyl group tokens. (a) Distribution of LogS predictions when exchanging two distinct OH tokens (319 and 338) in test molecules. (b) Representative molecules with 319 and 338 highlighted. (c) Visualization of top-3 attended atoms for LogS prediction in selected molecules (blue circles).}
    \label{fig:appendix_OH}
\end{figure}

To quantitatively assess whether these differences are statistically significant, we performed comprehensive distributional comparisons using the Wasserstein distance, Jensen–Shannon divergence, and distribution overlap area. In addition, we conducted non-parametric significance tests, including the Mann–Whitney U test and Kolmogorov–Smirnov (KS) test, to evaluate whether the atomic properties assigned to each token originate from the same population. The results are shown in Table~\ref{tab:oh_token_property_comparison_optimized}.

\begin{table}[htbp]
\centering
\caption{
\textbf{Statistical comparison of key atomic properties for OH tokens 338 versus 319.}
For each property, two groups are compared: (\textbf{OH 338}) data assigned to token 338, and (\textbf{OH 319}) data assigned to token 319. Reported metrics include mean $\pm$ standard deviation, distribution comparison metrics (\textbf{Wasserstein distance}, \textbf{Jensen–Shannon divergence}, \textbf{Overlap area}), and results of non-parametric significance tests (\textbf{Mann–Whitney U} and \textbf{Kolmogorov–Smirnov}).
}
\label{tab:oh_token_property_comparison_optimized}
\resizebox{\textwidth}{!}{%
\begin{tabular}{@{}lccccccccc@{}}
\toprule
& \multicolumn{2}{c}{\textbf{Distribution}} & \multicolumn{3}{c}{\textbf{Distribution Comparison}} & \multicolumn{4}{c}{\textbf{Significance Tests}} \\
\cmidrule(lr){2-3} \cmidrule(lr){4-6} \cmidrule(lr){7-10}
\textbf{Property} & \textbf{OH 338} & \textbf{OH 319} & \textbf{Wass.} & \textbf{JS Div.} & \textbf{Overlap} & \textbf{U-stat} & \textbf{U p-val} & \textbf{KS-stat} & \textbf{KS p-val} \\
\midrule

\textbf{Local PSA (\AA$^2$, r=4)}
& 85.87 $\pm$ 11.62 & 85.16 $\pm$ 15.73 & 8.99 & 0.99 & 0.48 & 1.55e+5 & $<$1e-10 & 0.58 & $<$1e-10 \\

\textbf{Mulliken Charge (a.u.)}
& -0.31 $\pm$ 0.01 & -0.31 $\pm$ 0.02 & 0.01 & 0.44 & 0.56 & 2.45e+5 & $<$1e-10 & 0.33 & $<$1e-10 \\

\textbf{$\pi$-Electron Occupancy}
& 2.79 $\pm$ 0.63 & 2.71 $\pm$ 0.65 & 0.09 & 0.49 & 0.94 & 2.32e+4 & 0.22 & 0.11 & 0.15 \\
\bottomrule
\end{tabular}%
}
\end{table}

The results demonstrate that, for both local PSA and Mulliken charge, the differences between OH338 and OH319 are highly significant (all $p$-values $< 10^{-10}$; see Table~\ref{tab:oh_token_property_comparison_optimized}). Specifically, token 338 is predominantly associated with hydroxyl groups located on flexible, chain-like molecular fragments, which tend to have higher PSA and slightly lower Mulliken charge—indicating greater solvent accessibility and a more pronounced ability to form hydrogen bonds. In contrast, token 319 is more frequently assigned to hydroxyl groups attached to cyclic, ether-containing rings; these environments are often more sterically hindered and less exposed to solvent, resulting in lower PSA and slightly higher Mulliken charge.

For local $\pi$-electron occupancy, however, the distributions between the two tokens show only modest differences, and the statistical tests do not indicate a significant separation ($p$-value $= 0.22$ for Mann–Whitney U). This may be attributed to the fact that $\pi$-electron occupancy for hydroxyl oxygen atoms is generally less sensitive to the larger molecular environment—since the lone pairs and bonding configuration of the oxygen are largely retained regardless of whether the OH group is on a ring or a chain. In contrast, both solvent accessibility (PSA) and electronic polarization (Mulliken charge) are more directly affected by local steric and electronic context, leading to clearer token-level differentiation for these properties.

The distributional metrics (e.g., Wasserstein distance for PSA = 8.99, JS divergence = 0.99, and overlap area = 0.48) further confirm the substantial difference in local environment between these two tokens, while the relatively small overlap in Mulliken charge distribution reflects the tokenizer’s capacity to resolve subtle electronic differences even within the same functional group.

Taken together, these findings indicate that the tokenizer does not simply assign arbitrary codes, but learns to differentiate and specialize tokens according to nuanced aspects of local atomic environment—enabling the LLM to distinguish functionally relevant subtypes of hydroxyl groups in downstream tasks.

\paragraph{Property Distributions of Mixture Tokens Across Functional Groups}

To further elucidate the chemical meaning captured by mixture tokens, we systematically compared key atomic property distributions between “pure” tokens—those dedicated exclusively to aromatic or alkene carbons—and mixture token 20, which encodes both aromatic and alkene carbons in ambiguous environments (see Table~\ref{tab:property_comparison_detailed}). For each property, we report mean and standard deviation within each group, alongside quantitative metrics of distributional separation and overlap.

\begin{figure}
    \centering
    \includegraphics[width=\linewidth]{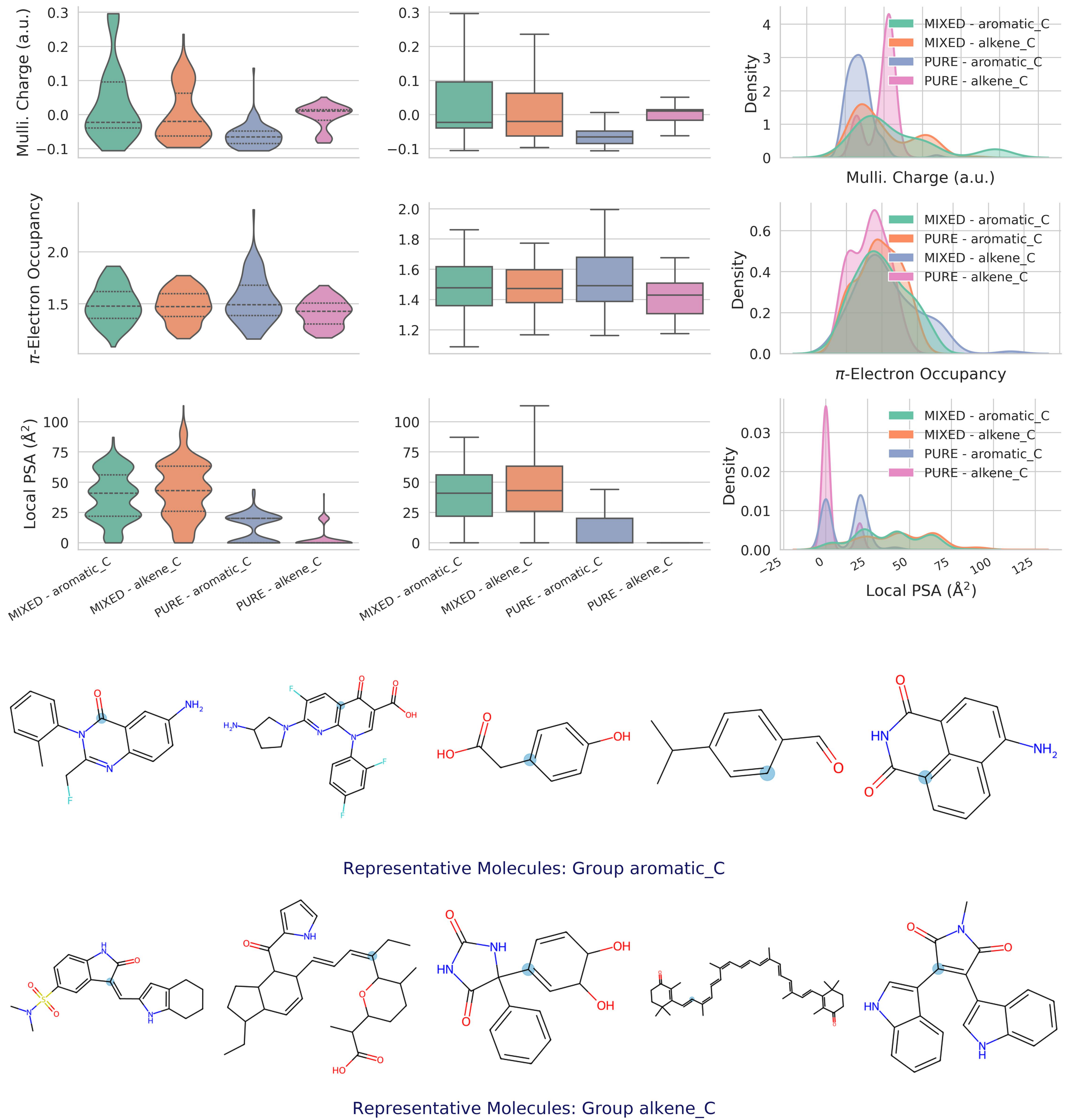}
    \caption{
    \textbf{Statistical analysis of mixture token 20 in relation to carbon atom properties.} 
    (Top) Violin, box, and kernel density plots show the distributions of Mulliken charge, $\pi$-electron occupancy, and local polar surface area (PSA) for four groups of carbon atoms: 
    \textbf{MIXED\_aromatic\_C} and \textbf{MIXED\_alkene\_C} represent aromatic and alkene carbons that are both assigned to mixture token 20. 
    \textbf{PURE\_aromatic\_C} and \textbf{PURE\_alkene\_C} denote carbons assigned to dedicated tokens that exclusively represent either aromatic or alkene carbons. 
    (Bottom) Example molecules from mixture token by different functional groups, with blue highlights indicating the atoms in token 20. These results illustrate the tokenizer’s ability to reveal chemical ambiguity and uncover non-trivial structure–property relationships.
    }
    \label{fig:case_2_appendix}
\end{figure}

The property distributions for mixture token 20 are illustrated in Figure~\ref{fig:case_2_appendix}. The upper panels display violin, box, and kernel density plots for Mulliken charge, $\pi$-electron occupancy, and local PSA across four atom groups: MIXED\_aromatic\_C and MIXED\_alkene\_C (carbons assigned to mixture token 20), and PURE\_aromatic\_C and PURE\_alkene\_C (carbons with dedicated pure tokens). Notably, mixture token groups exhibit significantly broader and more overlapping property distributions compared to their pure-token counterparts, confirming that token 20 is used in chemically ambiguous or transitional contexts rather than for classical, well-defined functional groups. The lower panels in Figure~\ref{fig:case_2_appendix} provide example molecules, with atoms assigned to token 20 highlighted in blue. These examples further demonstrate that mixture tokens frequently occur in structurally or electronically ambiguous environments—such as carbons adjacent to both aromatic and alkene character—underscoring the tokenizer’s ability to systematically capture nuanced, context-dependent atomic features.

\begin{table}[htbp]
\centering
\caption{
\textbf{Statistical comparison of key atomic properties for aromatic and alkene carbons assigned to pure versus mixture tokens.}
For each property, two groups are compared: 
(\textbf{PURE}) aromatic and alkene carbons assigned to dedicated tokens specific to each group, and 
(\textbf{MIXED}) aromatic and alkene carbons both assigned to mixture token 20. 
Reported metrics include mean $\pm$ standard deviation for each group, 
\textbf{Wasserstein distance} (Wass., a measure of distributional shift), 
\textbf{Jensen–Shannon divergence} (JS Div., quantifying distributional overlap), 
\textbf{Distribution overlap} (fractional overlap area), 
and results of parametric (\textbf{Welch's T-test}: T-stat, p-value) and non-parametric (\textbf{Kolmogorov–Smirnov test}: KS-stat, p-value) significance tests. 
Mulliken charge and $\pi$-electron occupancy are computed from quantum chemical calculations, while local PSA (polar surface area) is computed using RDKit within a 4-bond environment of each atom. 
Compared to pure tokens, mixture tokens display broader, less separable distributions and greater overlap, 
indicating that AtomDisc's tokenizer captures chemically ambiguous environments bridging conventional functional groups.
}
\label{tab:property_comparison_detailed}
\resizebox{\textwidth}{!}{%
\begin{tabular}{@{}llccccccccc@{}}
\toprule
& & \multicolumn{2}{c}{\textbf{Distribution}} & \multicolumn{3}{c}{\textbf{Distribution Comparison}} &  \multicolumn{4}{c}{\textbf{Significance Tests}} \\
\cmidrule(lr){3-4} \cmidrule(lr){5-7} \cmidrule(lr){8-11}  
\textbf{Property} & \textbf{Group} & \textbf{Aromatic C} & \textbf{Alkene C} & \textbf{Wass.} & \textbf{JS Div.} & \textbf{Overlap} & \textbf{U-stat} & \textbf{p-val} & \textbf{KS-stat} & \textbf{p-val} \\
\midrule

\multirow{2}{*}{\textbf{Mulliken Charge (a.u.)}}
& PURE  & -0.06 $\pm$ 0.03 & -0.01 $\pm$ 0.04 & 0.06 & 0.74 & 0.27 & 1.20e+3 & $<$1e-10 & 0.70 & $<$1e-10 \\
& MIXED &  0.03 $\pm$ 0.11 &  0.00 $\pm$ 0.08 & 0.03 & 0.35 & 0.43 &   5.86e+3 & 2.53e-2 & 0.28 & 5.22e-4 \\
\midrule

\multirow{2}{*}{\textbf{$\pi$-Electron Occupancy}}
& PURE  & 1.48 $\pm$ 0.15 & 1.41 $\pm$ 0.13 & 0.06 & 0.25 & 0.63 &   6.23e+3 & 2.60e-3 & 0.22 & 1.56e-2 \\
& MIXED & 1.49 $\pm$ 0.18 & 1.53 $\pm$ 0.22 & 0.04 & 0.21 & 0.70 &  4.56e+3 & 3.38e-1 & 0.12 & 4.15e-1 \\
\midrule

\multirow{2}{*}{\textbf{Local PSA (\AA$^2$)}}
& PURE  & 12.48 $\pm$ 11.44 &  3.39 $\pm$ 7.88 & 9.09 & 0.39 & 0.59 &  1.79e+5 & $<$1e-10 & 0.41 & $<$1e-10 \\
& MIXED & 37.40 $\pm$ 19.80 & 42.37 $\pm$ 22.57 & 5.54 & 0.22 & 0.43 &  1.07e+5 & 7.60e-5 & 0.18 & 1.71e-7 \\
\bottomrule
\end{tabular}%
}
\end{table}

First, for all three properties—Mulliken charge, $\pi$-electron occupancy, and local PSA—the separation between aromatic and alkene carbons is substantially larger for pure tokens than for the mixture token. For example, in pure tokens, the mean Mulliken charge of aromatic C is -0.06 versus -0.01 for alkene C, with a large Wasserstein distance (0.06), high Jensen-Shannon divergence (0.74), and minimal distributional overlap (0.27). The corresponding statistical tests (Mann–Whitney U, KS-test) all confirm highly significant differences ($p$-values $< 10^{-19}$). In contrast, when considering carbons assigned to mixture token 20, both the means and standard deviations of Mulliken charge converge (0.03 vs. 0.00), and the Wasserstein distance (0.03), Jensen-Shannon divergence (0.35), and overlap (0.43) all indicate substantially reduced separation and increased overlap. The significance of the difference is also markedly reduced.

A similar pattern is observed for $\pi$-electron occupancy and PSA. The pure tokens exhibit distinct and well-separated distributions (with significant $p$-values), while the mixture token groupings are both broader and more overlapping, with much larger areas of distributional overlap (0.70 for $\pi$-electron, 0.43 for PSA), and reduced significance in the statistical tests. Interestingly, for local PSA, while mixture tokens also show a large distributional overlap ($0.43$), this value is actually \textit{lower} than that of pure tokens ($0.59$). This at first seems counterintuitive, since mixture tokens generally correspond to broader and less well-separated distributions. However, a closer look at the data reveals that both aromatic and alkene carbons assigned to mixture token 20 have much higher PSA values on average than their pure-token counterparts. As a result, the PSA distributions of both groups are shifted to higher values, with more atoms in the high-PSA “tail” region. This upward shift reduces the region where the two distributions overlap, since both are now more “spread out” towards larger PSA, resulting in a smaller fractional overlap despite greater overall variance. This effect is especially pronounced for a small subset of atoms with extremely large PSA (outliers), which pull the means higher and compress the overlapping region.

These observations suggest that mixture tokens are assigned to atomic environments where the classic boundaries between aromatic and alkene character become blurred—often due to local context such as conjugation or the presence of strongly polar substituents (e.g., OH, NH). Visualization of representative cases confirms that mixture token 20 is frequently found in ring systems with nearby polar groups, where electronic structure is neither fully aromatic nor purely alkene-like, but rather intermediate. As a result, mixture tokens capture this chemical ambiguity, exhibiting broader, intermediate, and more overlapping property distributions compared to the sharply partitioned pure tokens.

In summary, our tokenizer not only reflects clear structure–property correlations where they exist (as in pure tokens), but also systematically identifies regions of chemical ambiguity and continuum. This emergent “boundary-aware” representation demonstrates the capacity of data-driven tokenization to move beyond rigid, human-defined categories and encode nuanced chemical reality.

\begin{figure}[htbp]
    \centering
    \includegraphics[width=\linewidth]{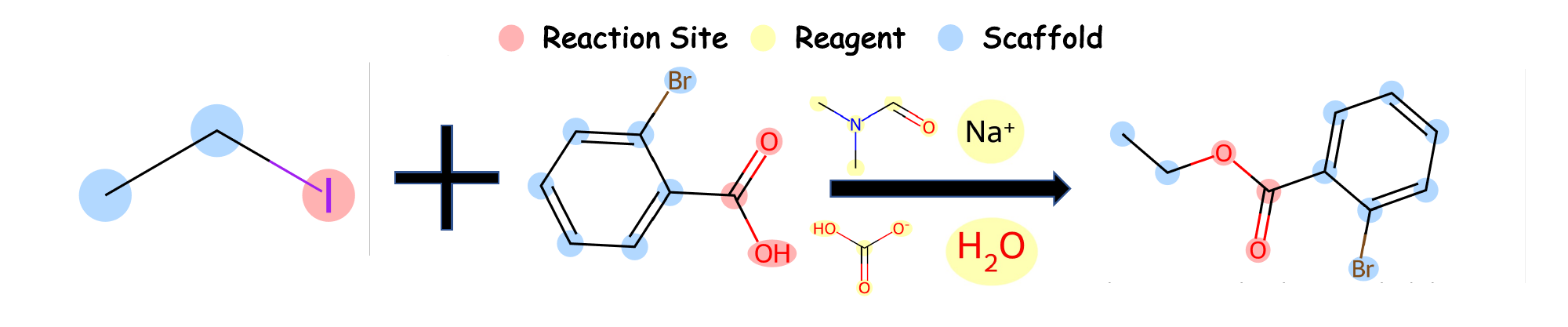}
    \caption{\textbf{Token annotation of an esterification reaction.} Reactants, reagents, and product are highlighted at the token level: reaction sites (red), reagents (yellow), and molecular scaffold (blue). This visualization demonstrates the explicit mapping of chemical roles in AtomDisc’s interpretable framework.}
    \label{fig:attn_example_forward}
\end{figure}

\paragraph{Attention Trends of Reaction, Reagent, and Scaffold Tokens during Decoding}

To further interpret the behavior of our model during product generation, we analyze the step-wise attention distribution over different classes of structure tokens in a representative case. We categorize atom tokens into three groups based on their chemical roles: \textbf{reaction site tokens} (corresponding to atoms involved in bond changes between reactants and products), \textbf{reagent tokens} (present only in reagent molecules), and \textbf{scaffold tokens} (other non-reacting atoms in the main reactant). The classification is automatically determined by comparing the atom mappings between reactants and products. For each decoding step, we record the attention weights assigned to the top-$3$ structure tokens in each class and normalize the sum within each step.

\begin{figure}[htbp]
    \centering
    \includegraphics[width=\linewidth]{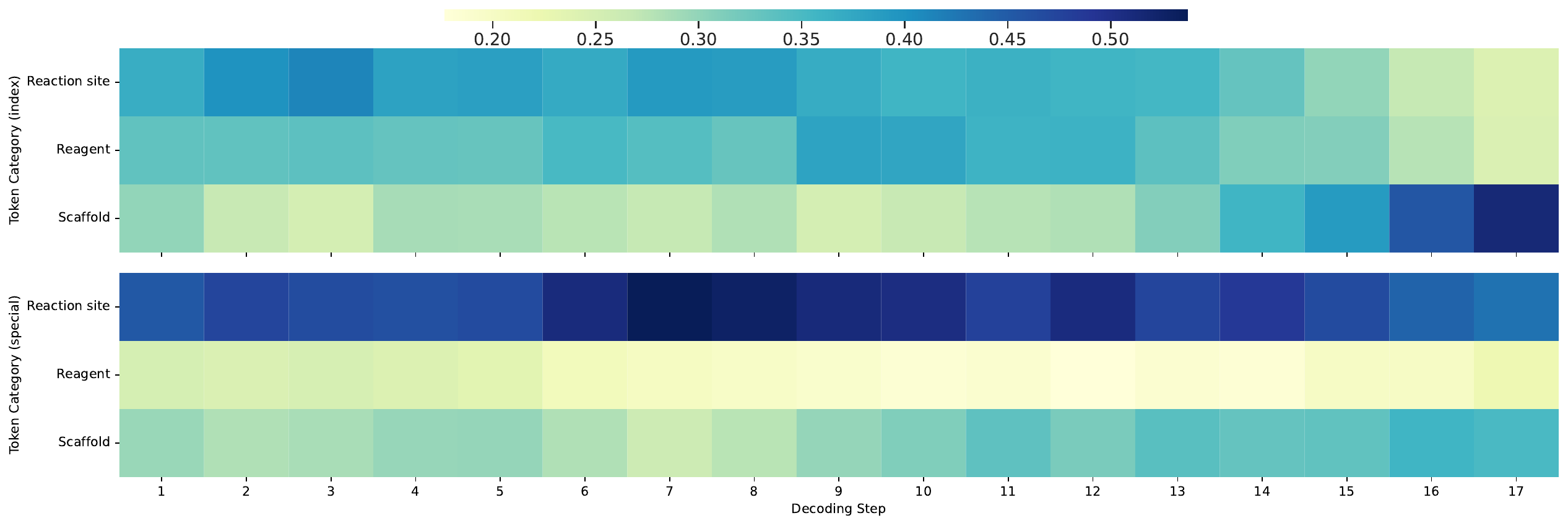}
    \caption{Comparison of step-wise relative attention distributions over reaction site, reagent, and scaffold tokens during product decoding in AtomDisc. For a given reaction case, we separately visualize the model’s attention to SMILES tokens (top) and to special atom tokens (bottom), both of which are included as part of the model’s input sequence. Both token types show similar global attention dynamics: attention is initially focused on reaction site tokens, and then progressively shifts to reagent and scaffold tokens as decoding proceeds. Notably, attention on reaction site information decays more rapidly when measured over SMILES tokens, while the special atom token representation enables more persistent focus on reaction centers at later generation steps. This difference may arise from tokenization granularity and the enhanced structural context provided by special tokens.}
    \label{fig:attn-type-heatmap-compare}
\end{figure}

As illustrated in Figure~\ref{fig:attn_example_forward}, we provide a concrete visualization of this token-level annotation in the context of an esterification reaction. Atoms are color-coded according to their assigned roles: \textcolor{red}{reaction site tokens} (red) highlight the atoms directly involved in bond formation or cleavage during the reaction; \textcolor{yellow}{reagent tokens} (yellow) denote atoms that appear exclusively in the reagent molecules; and \textcolor{blue}{scaffold tokens} (blue) correspond to the remaining, non-reactive backbone of the main reactant. This explicit mapping of chemical roles demonstrates how AtomDisc enables fine-grained, interpretable tracking of atom-level contributions throughout the reaction process. To assess the robustness of our findings, we compare the step-wise attention distributions using both \textbf{SMILES string tokens} and \textbf{special atom tokens} as representations of molecular structure (see Figure~\ref{fig:attn-type-heatmap-compare}).

Both representations show consistent attention dynamics: initial focus on reaction centers, followed by reagent and scaffold regions as decoding progresses. The subtle differences (e.g., faster decay of reaction site attention for SMILES) highlight the impact of tokenization, but the overall consistency demonstrates the robustness of the model’s chemically meaningful focus. 

The more rapid decay of reaction site attention in the SMILES-based analysis can be attributed to the nature of SMILES tokenization. In SMILES, a chemically important atom or group (such as a reaction center) may be split across multiple tokens, and the sequence context is more variable and redundant compared to the structured special atom tokens, which are in one-to-one correspondence with molecular graph atoms. Consequently, attention assigned to chemically meaningful atoms is diluted among several SMILES tokens, leading to a faster drop in the measured relative attention for the reaction site class. In contrast, the special atom token representation preserves atom-level resolution and allows the model to maintain focused attention on key chemical centers throughout the decoding process. This observation highlights the importance of molecular representation granularity in interpreting model behavior, and suggests that atom-level special tokens facilitate more chemically meaningful and interpretable attention dynamics.

\newpage
\subsection{More Cases}
\label{appendix:more_cases}

To further demonstrate the chemical interpretability and discovery potential of AtomDisc, we provide additional case studies in this section. These examples highlight the diverse types of atomic environments captured by our tokenizer and illustrate how AtomDisc uncovers nuanced, context-dependent structural patterns that go beyond conventional functional group definitions.

\subsubsection{Mixture Tokens}

\begin{figure}
    \centering
    \includegraphics[width=\linewidth]{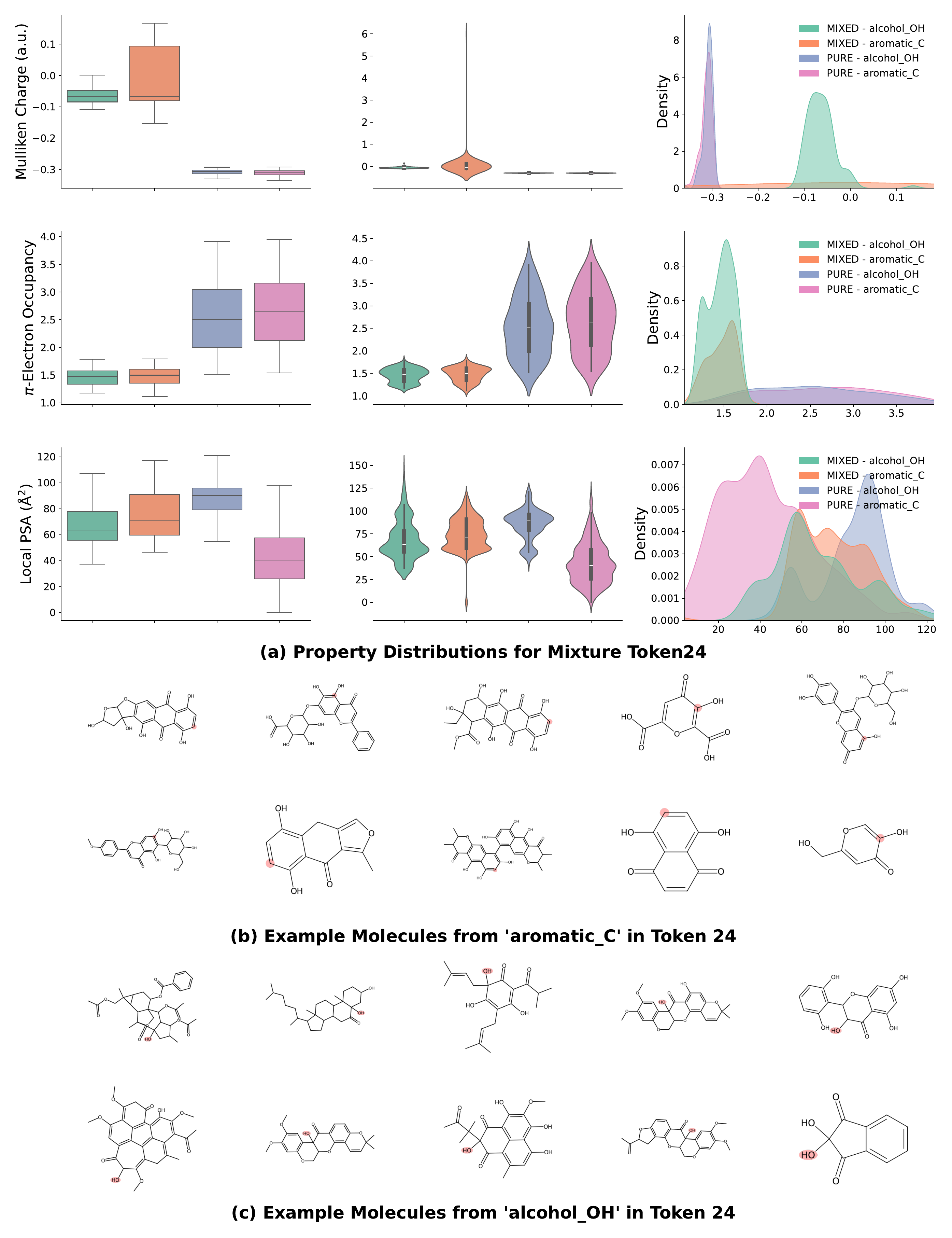}
    \caption{
        Distribution and molecular examples for mixture token 24, 
        showing properties including PSA, Mulliken charge, and $\pi$-electron occupancy.
    }
    \label{fig:appendix_mixture_24}
\end{figure}

\begin{figure}
    \centering
    \includegraphics[width=\linewidth]{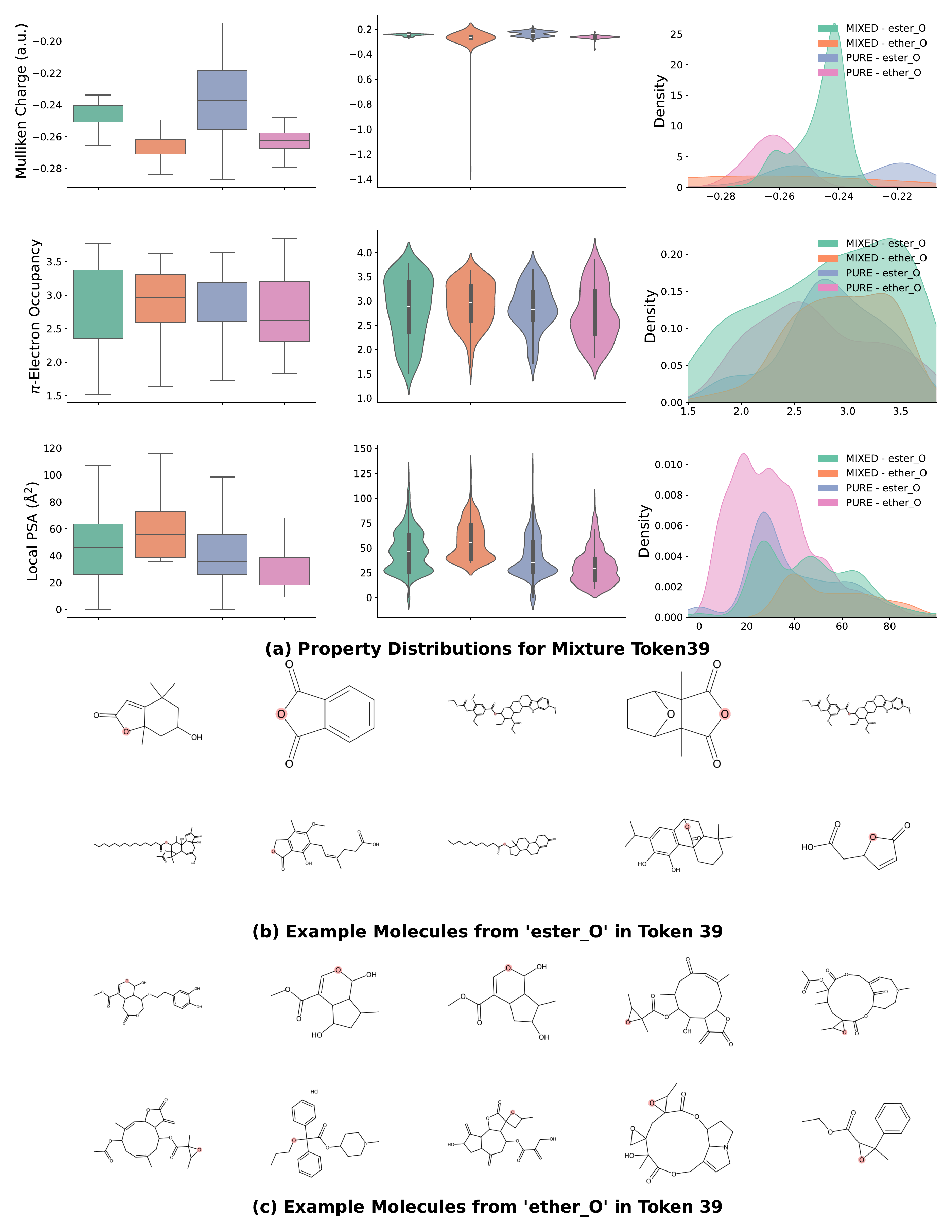}
    \caption{
        Distribution and molecular examples for mixture token 39, 
        showing properties including PSA, Mulliken charge, and $\pi$-electron occupancy.
    }
    \label{fig:appendix_mixture_39}
\end{figure}

We extend our analysis to additional mixture tokens, specifically tokens 24 and 39, and visualize their atomic assignments and environments. Token 24 is observed to encode both aromatic carbon atoms and hydroxyl oxygens in mixed local contexts, while token 39 captures atoms found in both ester and ether functional groups. By systematically analyzing these tokens (as shown in Table~\ref{tab:property_comparison_token24} and Table~\ref{tab:property_comparison_token39}), we further demonstrate that AtomDisc is able to identify and represent hybrid chemical environments—assigning shared codes to atoms that exhibit properties of multiple functional groups, and thus reflecting the tokenizer’s ability to model chemical ambiguity and subtle context-dependence at the atomic scale. The visualizations in Figure~\ref{fig:appendix_mixture_24} and Figure~\ref{fig:appendix_mixture_39} further illustrate these diverse atomic environments and highlight the interpretability of the learned token assignments.

\newpage

\begin{table}[htbp]
\centering
\caption{
\textbf{Statistical comparison of key atomic properties for aromatic and alcohol\_OH assigned to pure versus mixture token 24.}
For each property, two groups are compared: (\textbf{PURE}) aromatic\_C and alcohol\_OH assigned to dedicated tokens specific to each group, and (\textbf{MIXED}) aromatic\_C and alcohol\_OH both assigned to mixture token 24.
Reported metrics include mean $\pm$ standard deviation for each group, 
\textbf{Wasserstein distance} (Wass., a measure of distributional shift), 
\textbf{Jensen–Shannon divergence} (JS Div., quantifying distributional overlap), 
\textbf{Distribution overlap} (fractional overlap area), 
and results of parametric (\textbf{Welch's T-test}: T-stat, p-value) and non-parametric (\textbf{Kolmogorov–Smirnov test}: KS-stat, p-value) significance tests.
}
\label{tab:property_comparison_token24}
\resizebox{\textwidth}{!}{%
\begin{tabular}{@{}llccccccccc@{}}
\toprule
& & \multicolumn{2}{c}{\textbf{Distribution}} & \multicolumn{3}{c}{\textbf{Distribution Comparison}} &  \multicolumn{4}{c}{\textbf{Significance Tests}} \\
\cmidrule(lr){3-4} \cmidrule(lr){5-7} \cmidrule(lr){8-11}  
\textbf{Property} & \textbf{Group} & \textbf{aromatic\_C} & \textbf{alcohol\_OH} & \textbf{Wass.} & \textbf{JS Div.} & \textbf{Overlap} & \textbf{U-stat} & \textbf{p-val} & \textbf{KS-stat} & \textbf{p-val} \\
\midrule

\multirow{2}{*}{\textbf{Mulliken Charge (a.u.)}}
& PURE  & -0.06 $\pm$ 0.03 & -0.31 $\pm$ 0.01 & 0.25 & 1.00 & $<$1e-6 & 2.00e+4 & $<$1e-6 & 94.65 & $<$1e-6 \\
& MIXED &  0.05 $\pm$ 0.61 & -0.31 $\pm$ 0.01 & 0.36 & 1.00 & $<$1e-6 & 1.00e+4 & $<$1e-6 & 5.95 & $<$1e-6 \\
\midrule

\multirow{2}{*}{\textbf{$\pi$-Electron Occupancy}}
& PURE  & 1.46 $\pm$ 0.15 & 2.66 $\pm$ 0.66 & 1.20 & 0.89 & 0.10 & 321 & $<$1e-6 & -17.89 & $<$1e-6 \\
& MIXED & 1.48 $\pm$ 0.16 & 2.54 $\pm$ 0.65 & 1.06 & 0.85 & 0.13 & 337 & $<$1e-6 & -15.88 & $<$1e-6 \\
\midrule

\multirow{2}{*}{\textbf{Local PSA (\AA$^2$)}}
& PURE  & 43.21 $\pm$ 22.39 & 84.98 $\pm$ 16.66 & 41.77 & 0.75 & 0.19 & 4.01e+4 & $<$1e-6 & -40.65 & $<$1e-6 \\
& MIXED & 75.03 $\pm$ 17.26 & 68.43 $\pm$ 21.60 & 9.41 & 0.58 & 0.39 & 1.59e+5 & $<$1e-6 & 5.34 & $<$1e-6 \\

\bottomrule
\end{tabular}%
}
\end{table}

\begin{table}[htbp]
\centering
\caption{
\textbf{Statistical comparison of key atomic properties for ester\_O and ether\_O assigned to pure versus mixture token 39.}
For each property, two groups are compared: (\textbf{PURE}) ester\_O and ether\_O assigned to dedicated tokens, and (\textbf{MIXED}) ester\_O and ether\_O both assigned to mixture token 39.
Reported metrics include mean $\pm$ standard deviation for each group, 
\textbf{Wasserstein distance} (Wass., a measure of distributional shift), 
\textbf{Jensen–Shannon divergence} (JS Div., quantifying distributional overlap), 
\textbf{Distribution overlap} (fractional overlap area), 
and results of parametric (\textbf{Welch's T-test}: T-stat, p-value) and non-parametric (\textbf{Kolmogorov–Smirnov test}: KS-stat, p-value) significance tests.
}
\label{tab:property_comparison_token39}
\resizebox{\textwidth}{!}{%
\begin{tabular}{@{}llccccccccc@{}}
\toprule
& & \multicolumn{2}{c}{\textbf{Distribution}} & \multicolumn{3}{c}{\textbf{Distribution Comparison}} &  \multicolumn{4}{c}{\textbf{Significance Tests}} \\
\cmidrule(lr){3-4} \cmidrule(lr){5-7} \cmidrule(lr){8-11}
\textbf{Property} & \textbf{Group} & \textbf{ester\_O} & \textbf{ether\_O} & \textbf{Wass.} & \textbf{JS Div.} & \textbf{Overlap} & \textbf{U-stat} & \textbf{p-val} & \textbf{KS-stat} & \textbf{p-val} \\
\midrule

\multirow{2}{*}{\textbf{Mulliken Charge (a.u.)}}
& PURE  & -0.26 $\pm$ 0.01 & -0.25 $\pm$ 0.01 & 0.02 & 0.75 & 0.27 & 1.70e+03 & $<$1e-6 & 11.79 & $<$1e-6 \\
& MIXED & -0.28 $\pm$ 0.11 & -0.24 $\pm$ 0.02 & 0.04 & 0.69 & 0.32 & 9.22e+02 & $<$1e-6 & 3.73 & 3.11e-4 \\
\midrule

\multirow{2}{*}{\textbf{$\pi$-Electron Occupancy}}
& PURE  & 2.86 $\pm$ 0.47 & 2.83 $\pm$ 0.64 & 0.18 & 0.25 & 0.58 & 9.26e+03 & 0.83 & 0.42 & 0.68 \\
& MIXED & 2.93 $\pm$ 0.45 & 2.72 $\pm$ 0.56 & 0.24 & 0.35 & 0.52 & 6.19e+03 & 0.01 & 2.87 & 4.61e-3 \\
\midrule

\multirow{2}{*}{\textbf{Local PSA (\AA$^2$)}}
& PURE  & 40.63 $\pm$ 20.86 & 31.52 $\pm$ 17.79 & 10.09 & 0.73 & 0.23 & 3.10e+05 & $<$1e-6 & 8.36 & $<$1e-6 \\
& MIXED & 46.90 $\pm$ 20.60 & 57.26 $\pm$ 19.21 & 10.40 & 0.60 & 0.35 & 4.30e+04 & $<$1e-6 & -6.75 & $<$1e-6 \\

\bottomrule
\end{tabular}%
}
\end{table}

\newpage
\subsubsection{Functional Groups Assigned to Different Tokens}

\begin{figure}[htbp]
    \centering
    \includegraphics[width=\linewidth]{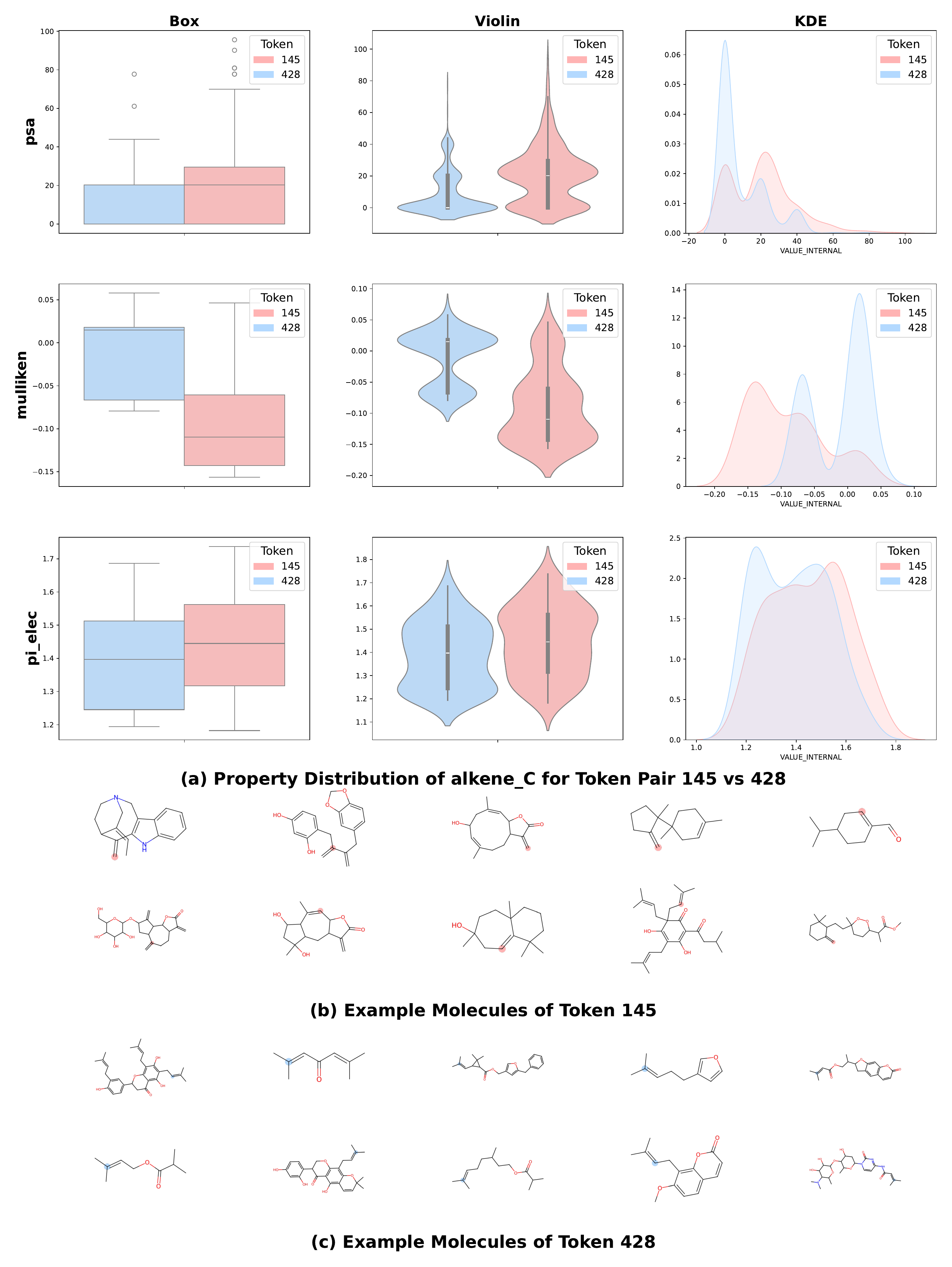}
    \caption{Distribution and molecular examples for alkene assigned to token 145 and 428, 
        showing properties including PSA, Mulliken charge, and $\pi$-electron occupancy.}
    \label{fig:appendix_alkene_145_428}
\end{figure}

\begin{figure}[htbp]
    \centering
    \includegraphics[width=\linewidth]{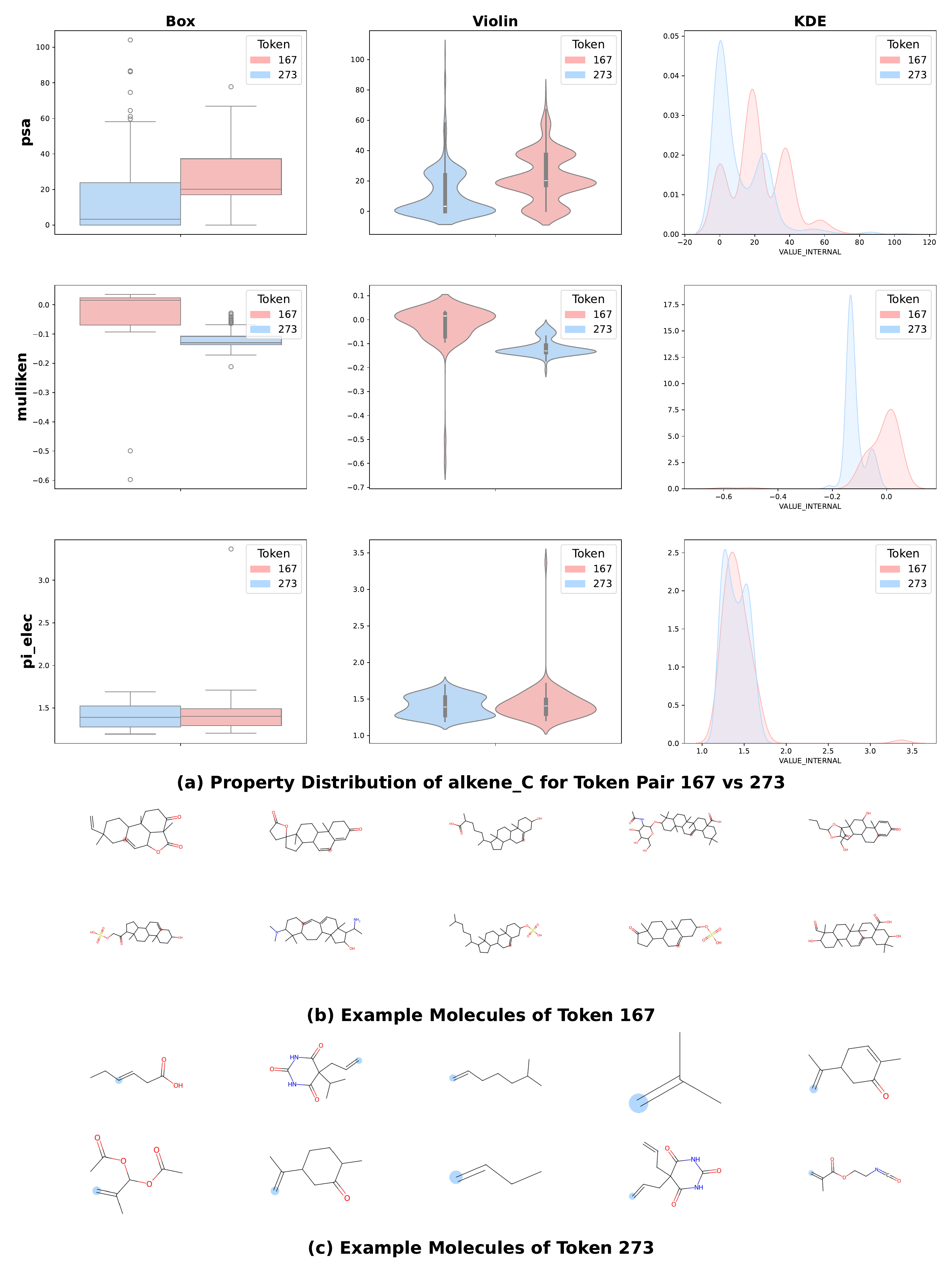}
    \caption{Distribution and molecular examples for alkene assigned to token 167 and 273, 
        showing properties including PSA, Mulliken charge, and $\pi$-electron occupancy.}
    \label{fig:appendix_alkene_167_273}
\end{figure}

\begin{figure}[htbp]
    \centering
    \includegraphics[width=\linewidth]{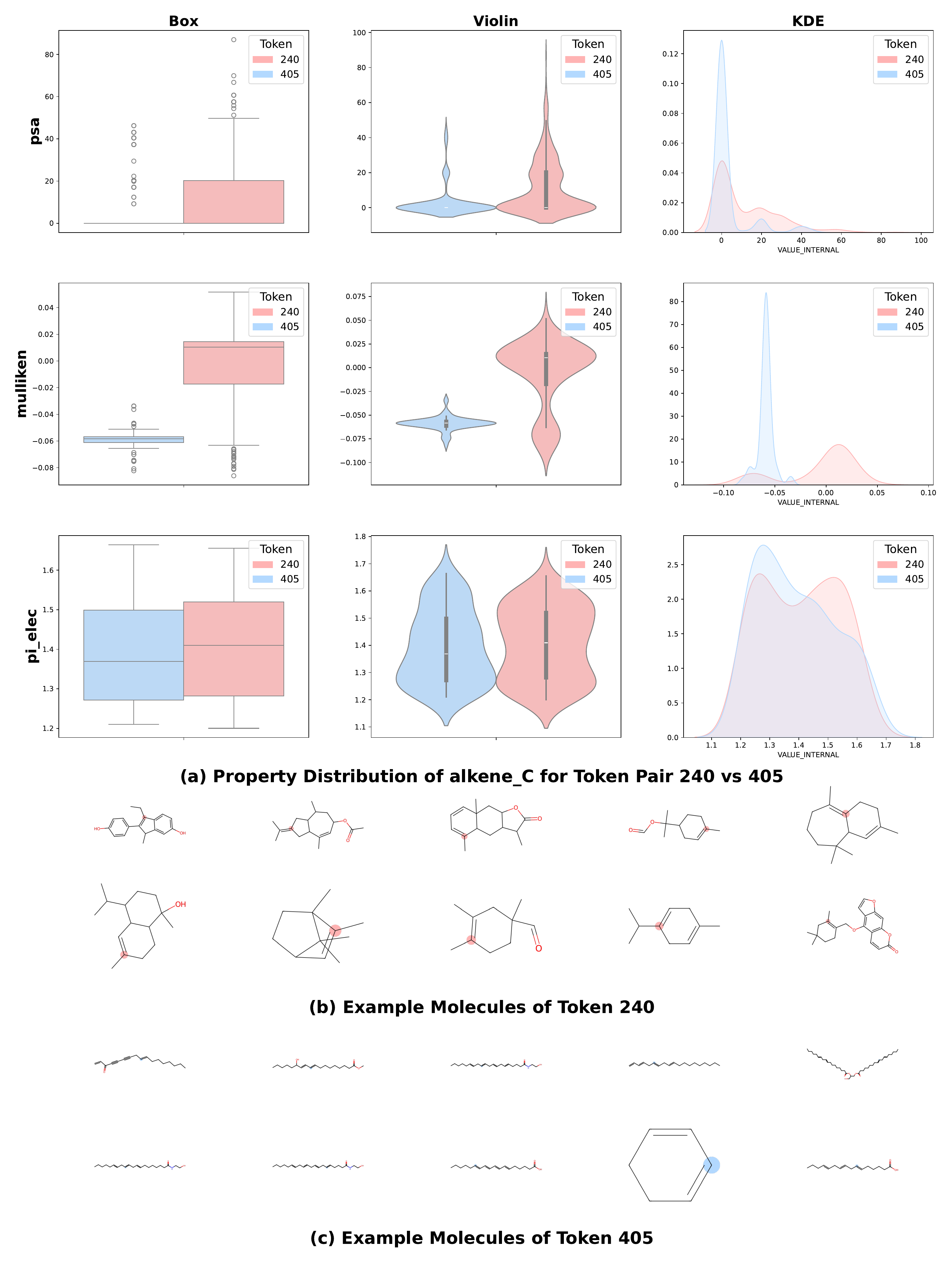}
    \caption{Distribution and molecular examples for alkene assigned to token 240 and 405, 
        showing properties including PSA, Mulliken charge, and $\pi$-electron occupancy.}
    \label{fig:appendix_alkene_240_405}
\end{figure}

\begin{figure}[htbp]
    \centering
    \includegraphics[width=\linewidth]{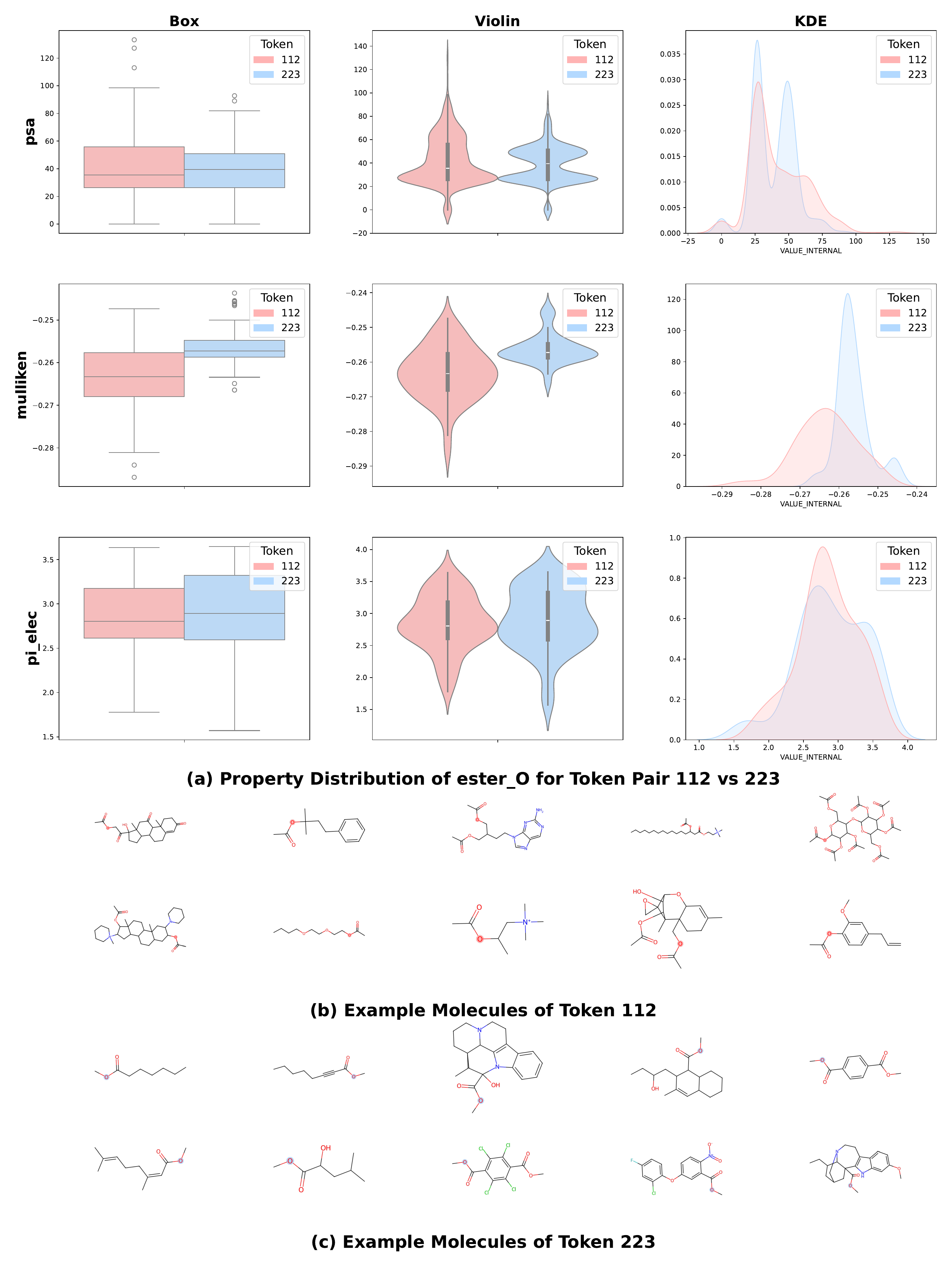}
    \caption{Distribution and molecular examples for ester assigned to token 112 and 223, 
        showing properties including PSA, Mulliken charge, and $\pi$-electron occupancy.}
    \label{fig:appendix_ester_112_223}
\end{figure}

\begin{figure}[htbp]
    \centering
    \includegraphics[width=\linewidth]{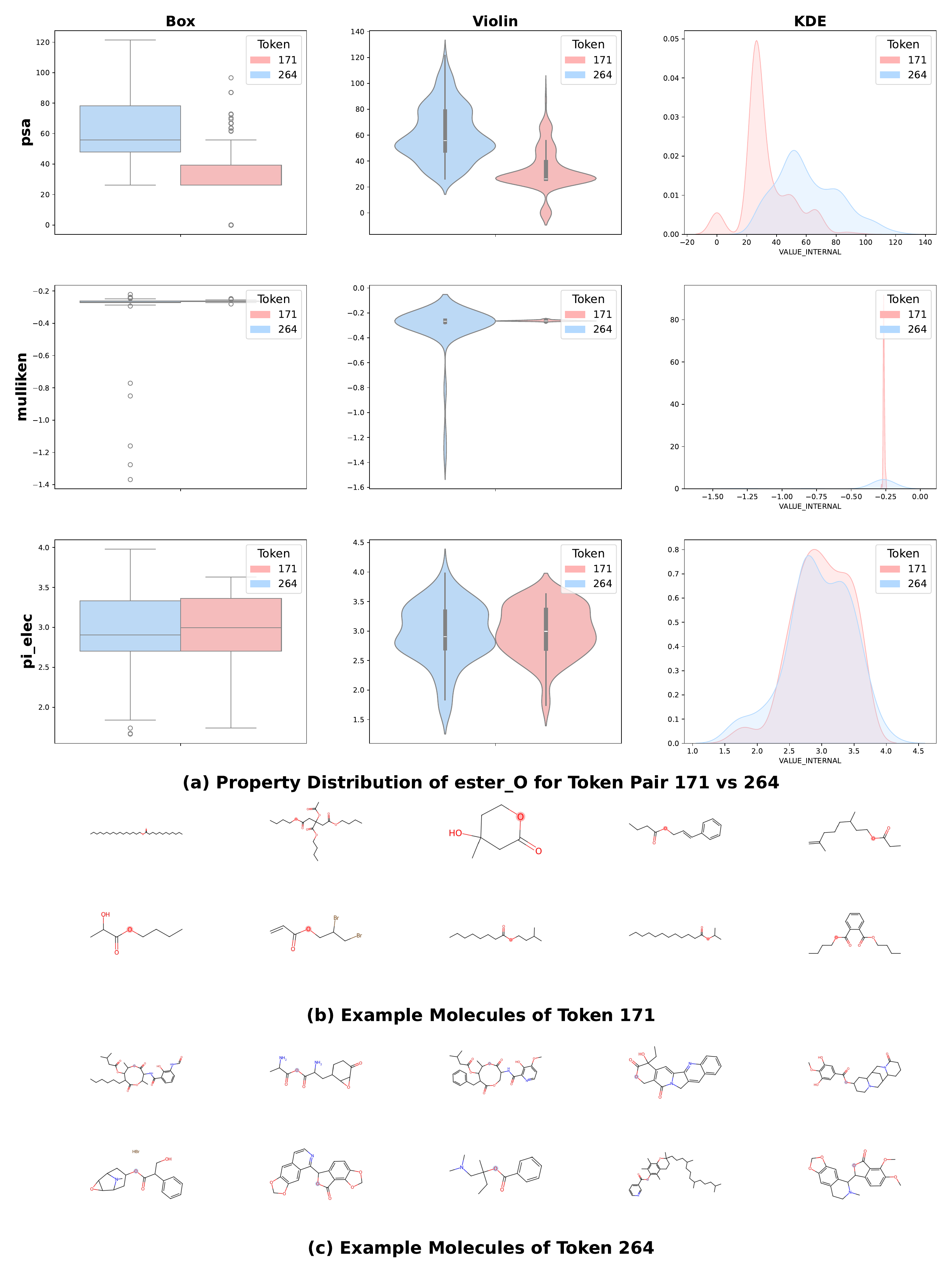}
    \caption{Distribution and molecular examples for ester assigned to token 171 and 264, 
        showing properties including PSA, Mulliken charge, and $\pi$-electron occupancy.}
    \label{fig:appendix_ester_171_264}
\end{figure}

\begin{figure}[htbp]
    \centering
    \includegraphics[width=\linewidth]{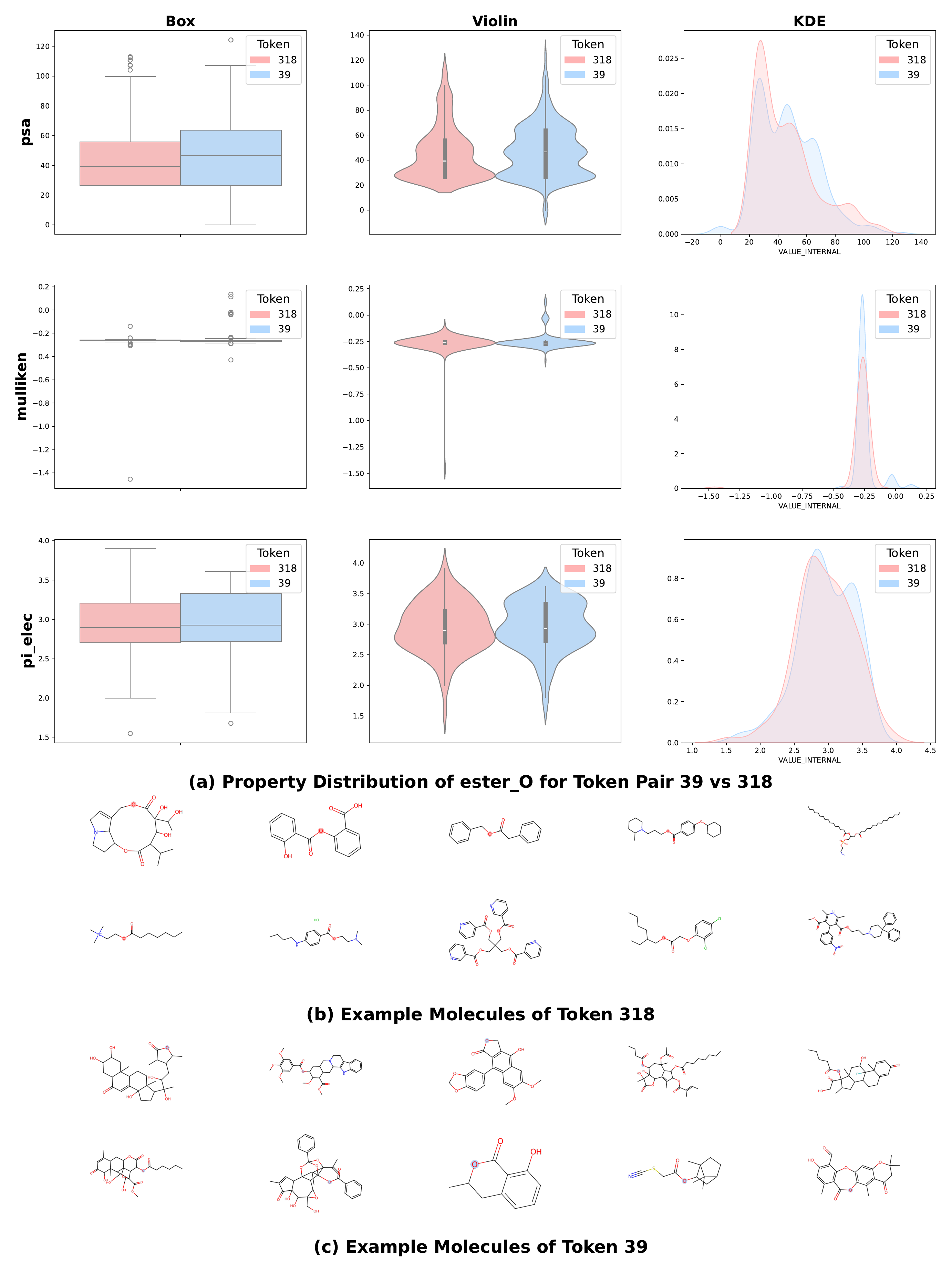}
    \caption{Distribution and molecular examples for ester assigned to token 39 and 318, 
        showing properties including PSA, Mulliken charge, and $\pi$-electron occupancy.}
    \label{fig:appendix_ester_O_39_318}
\end{figure}

\begin{figure}[htbp]
    \centering
    \includegraphics[width=\linewidth]{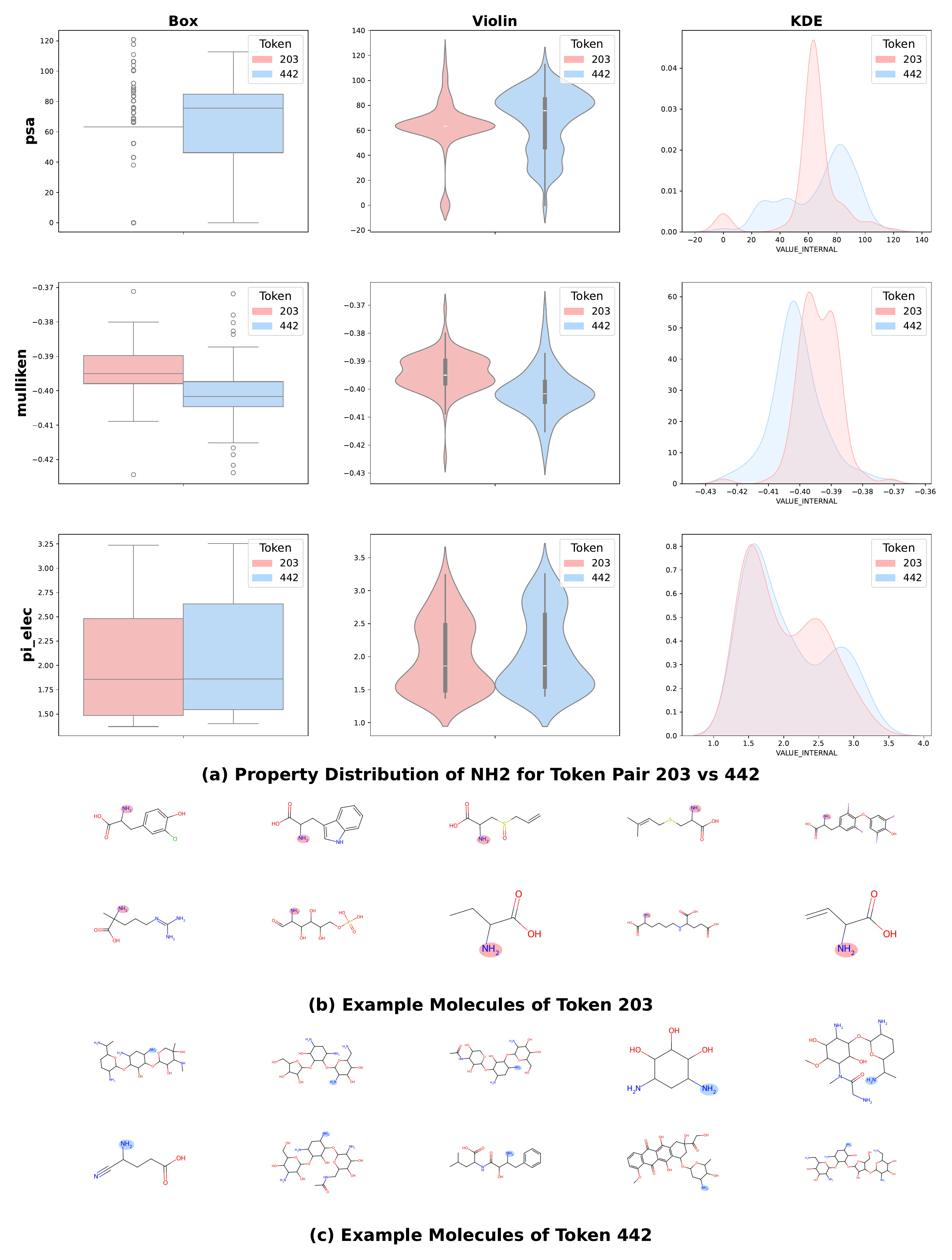}
    \caption{Distribution and molecular examples for amide assigned to token 203 and 442, 
        showing properties including PSA, Mulliken charge, and $\pi$-electron occupancy.}
    \label{fig:appendix_NH2_203_442}
\end{figure}

\begin{figure}[htbp]
    \centering
    \includegraphics[width=\linewidth]{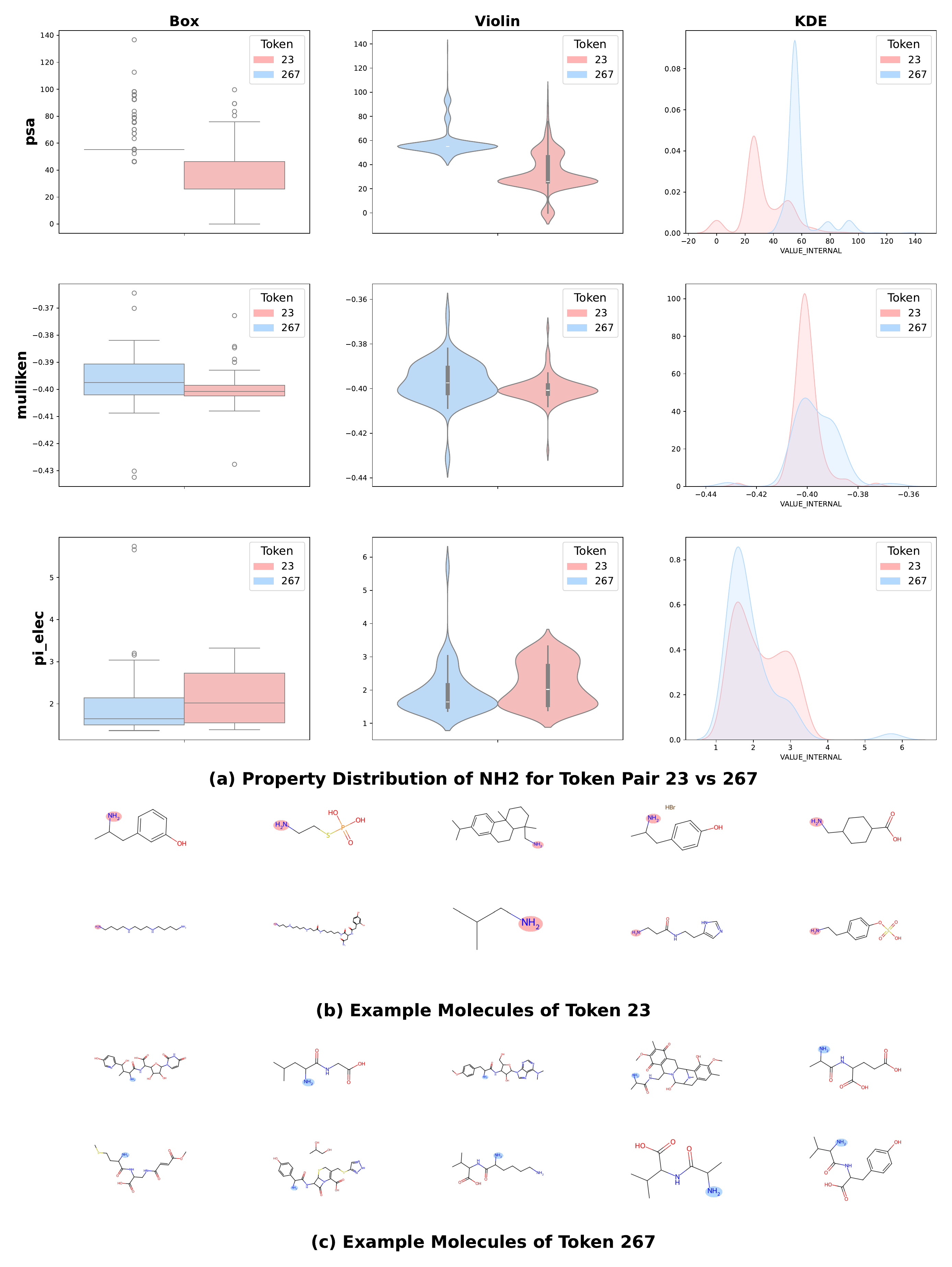}
    \caption{Distribution and molecular examples for amide assigned to token 23 and 267, 
        showing properties including PSA, Mulliken charge, and $\pi$-electron occupancy.}
    \label{fig:appendix_NH2_23_267}
\end{figure}

\begin{figure}[htbp]
    \centering
    \includegraphics[width=\linewidth]{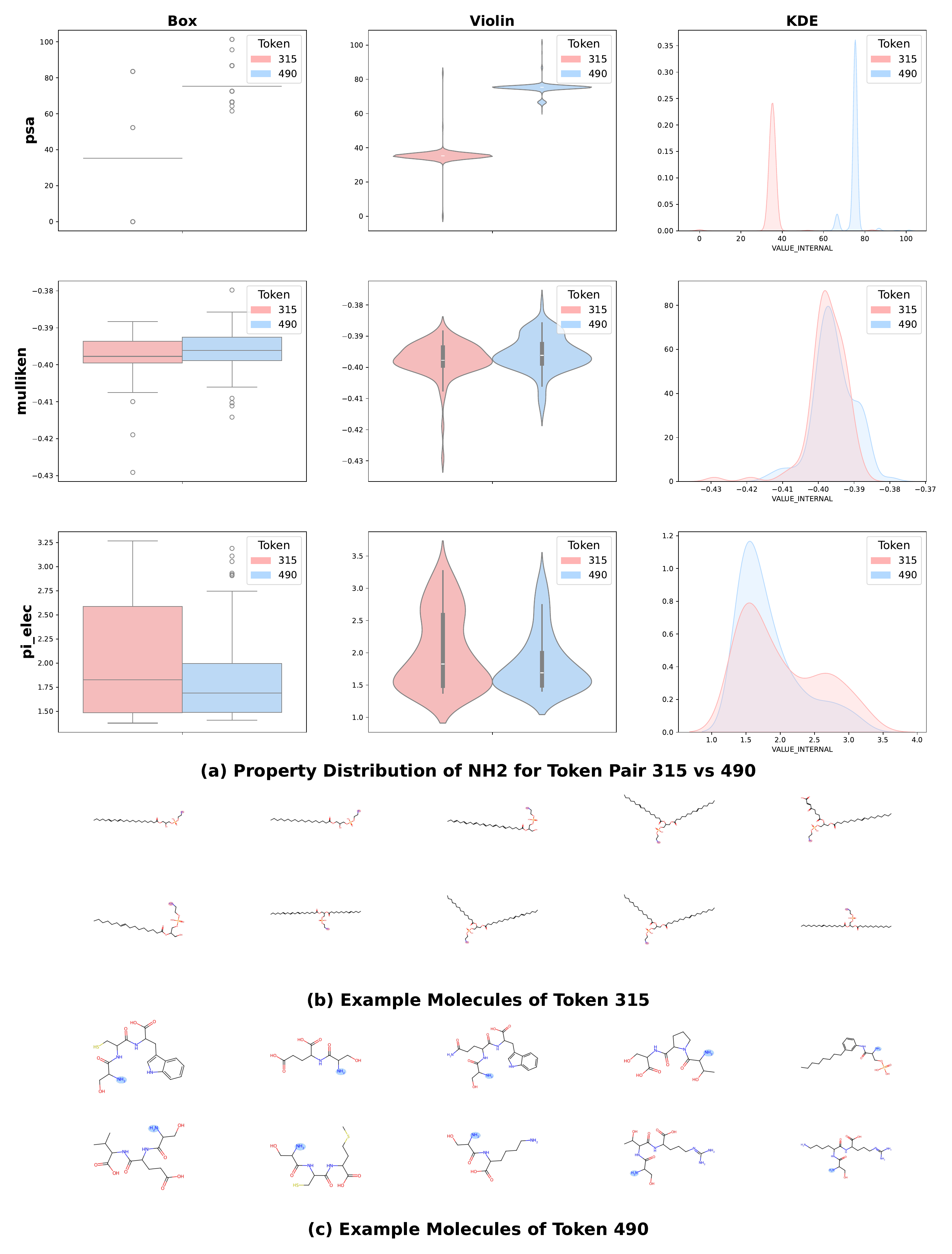}
    \caption{Distribution and molecular examples for amide assigned to token 315 and 490, 
        showing properties including PSA, Mulliken charge, and $\pi$-electron occupancy.}
        \label{fig:appendix_NH2_315_490}
\end{figure}

\begin{figure}[htbp]
    \centering
    \includegraphics[width=\linewidth]{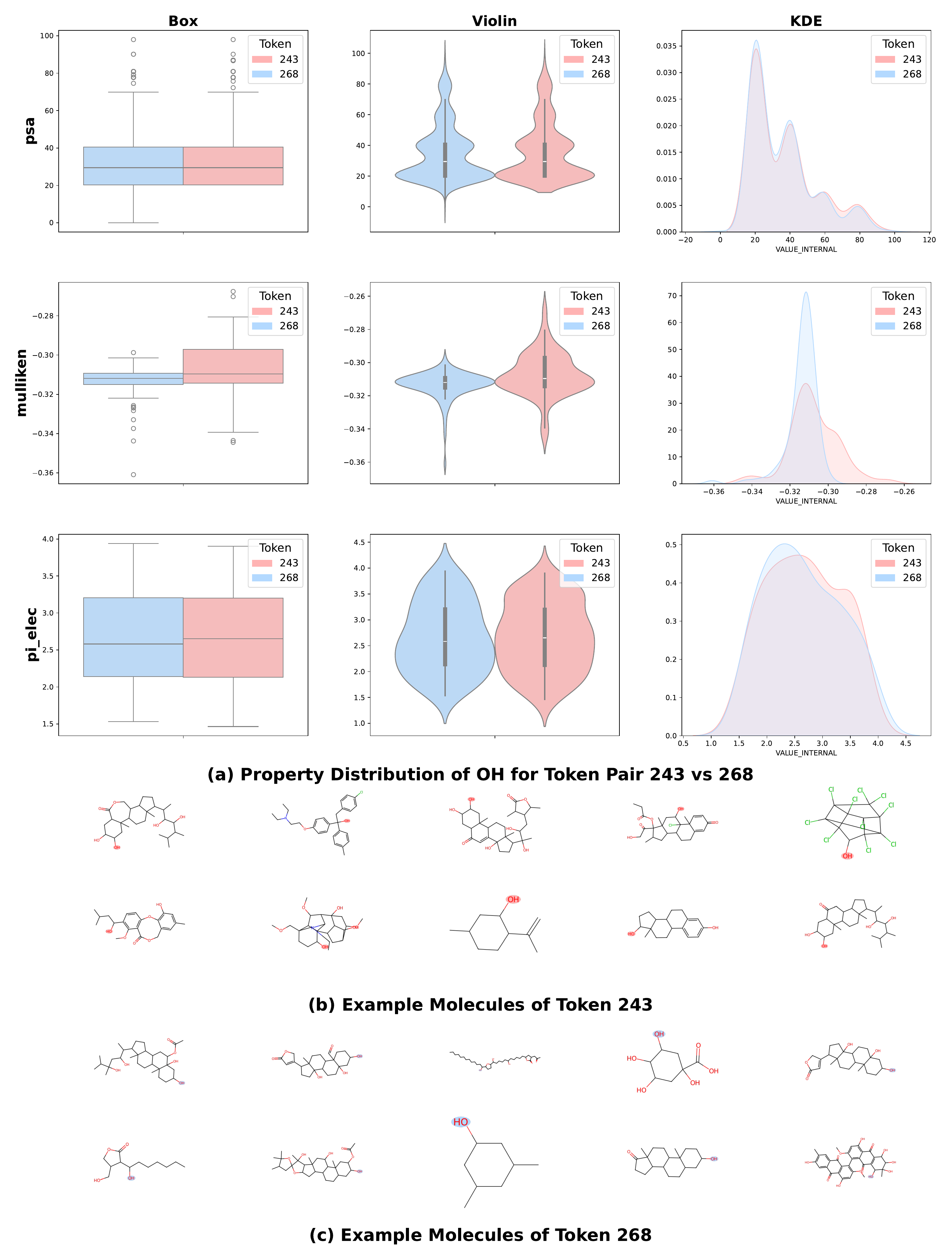}
    \caption{Distribution and molecular examples for OH assigned to token 243 and 268, 
        showing properties including PSA, Mulliken charge, and $\pi$-electron occupancy.}
    \label{fig:appendix_OH_243_468}
\end{figure}

\begin{figure}[htbp]
    \centering
    \includegraphics[width=\linewidth]{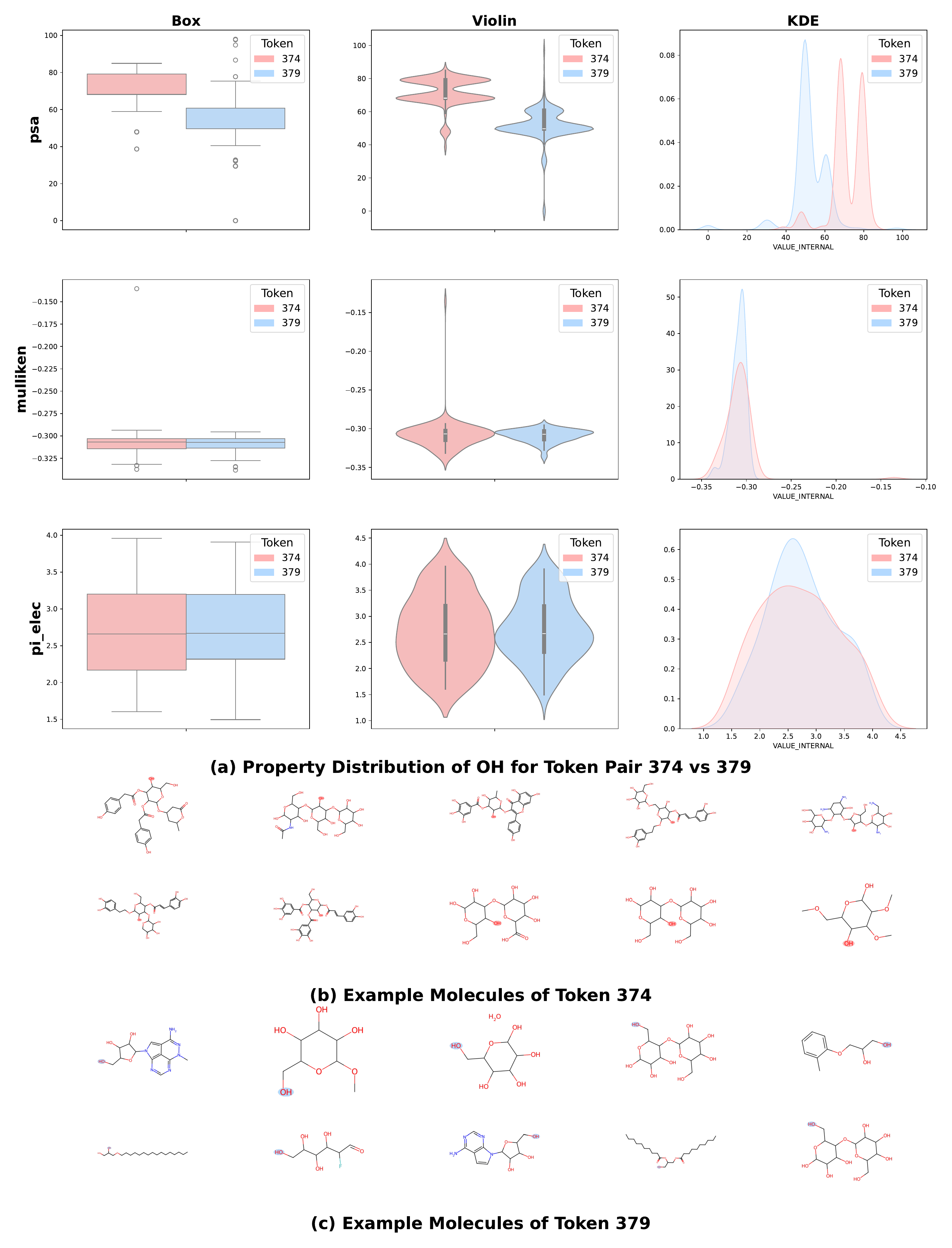}
    \caption{Distribution and molecular examples for OH assigned to token 374 and 379, 
        showing properties including PSA, Mulliken charge, and $\pi$-electron occupancy.}
    \label{fig:appendix_OH_374_379}
\end{figure}

\begin{figure}[htbp]
    \centering
    \includegraphics[width=\linewidth]{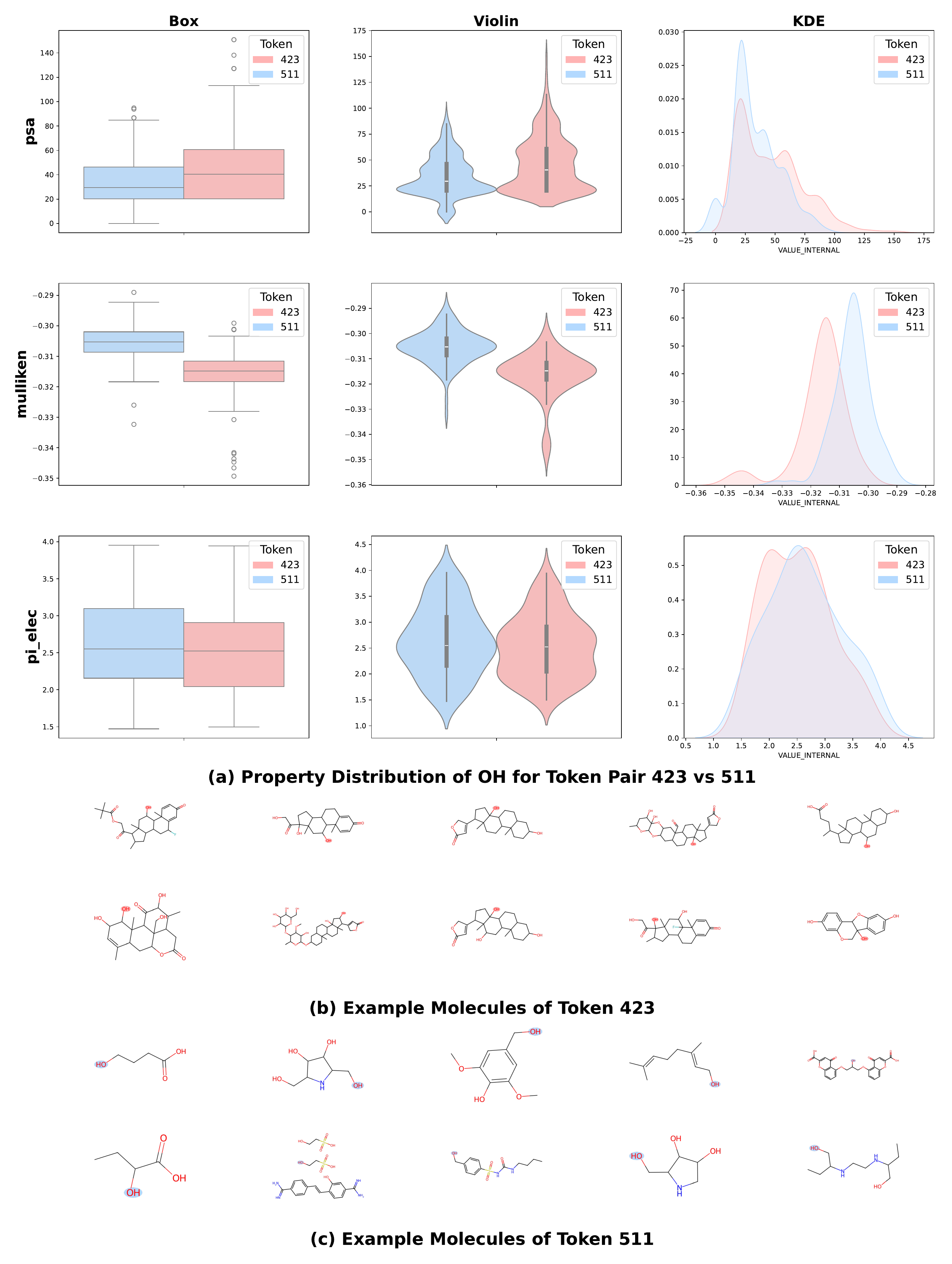}
    \caption{Distribution and molecular examples for OH assigned to token 423 and 511, 
        showing properties including PSA, Mulliken charge, and $\pi$-electron occupancy.}
    \label{fig:appendix_OH_423_511}
\end{figure}

\begin{figure}[htbp]
    \centering
    \includegraphics[width=\linewidth]{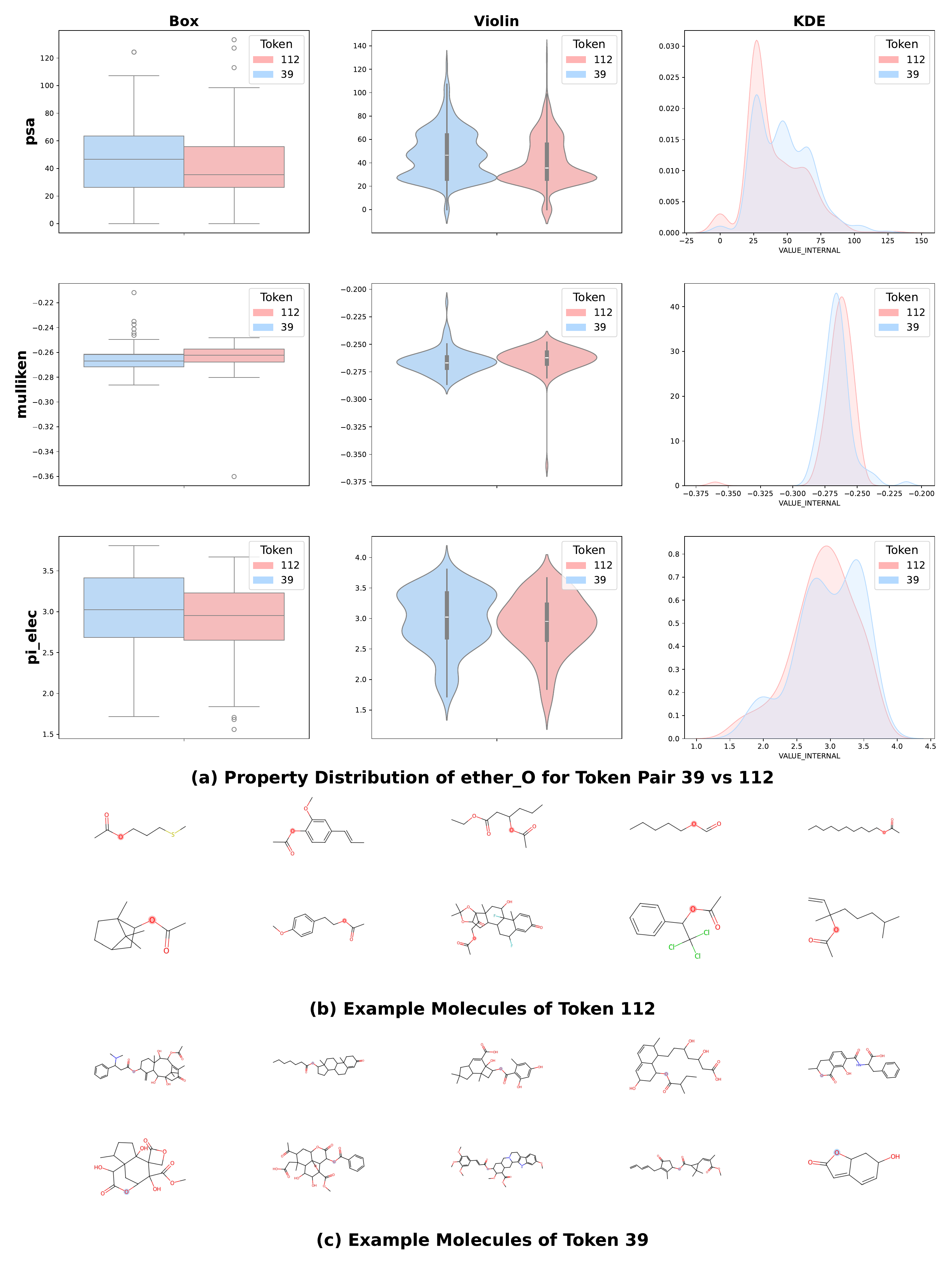}
    \caption{Distribution and molecular examples for ether assigned to token 39 and 112, 
        showing properties including PSA, Mulliken charge, and $\pi$-electron occupancy.}
    \label{fig:appendix_ether_39_112}
\end{figure}

\begin{figure}[htbp]
    \centering
    \includegraphics[width=\linewidth]{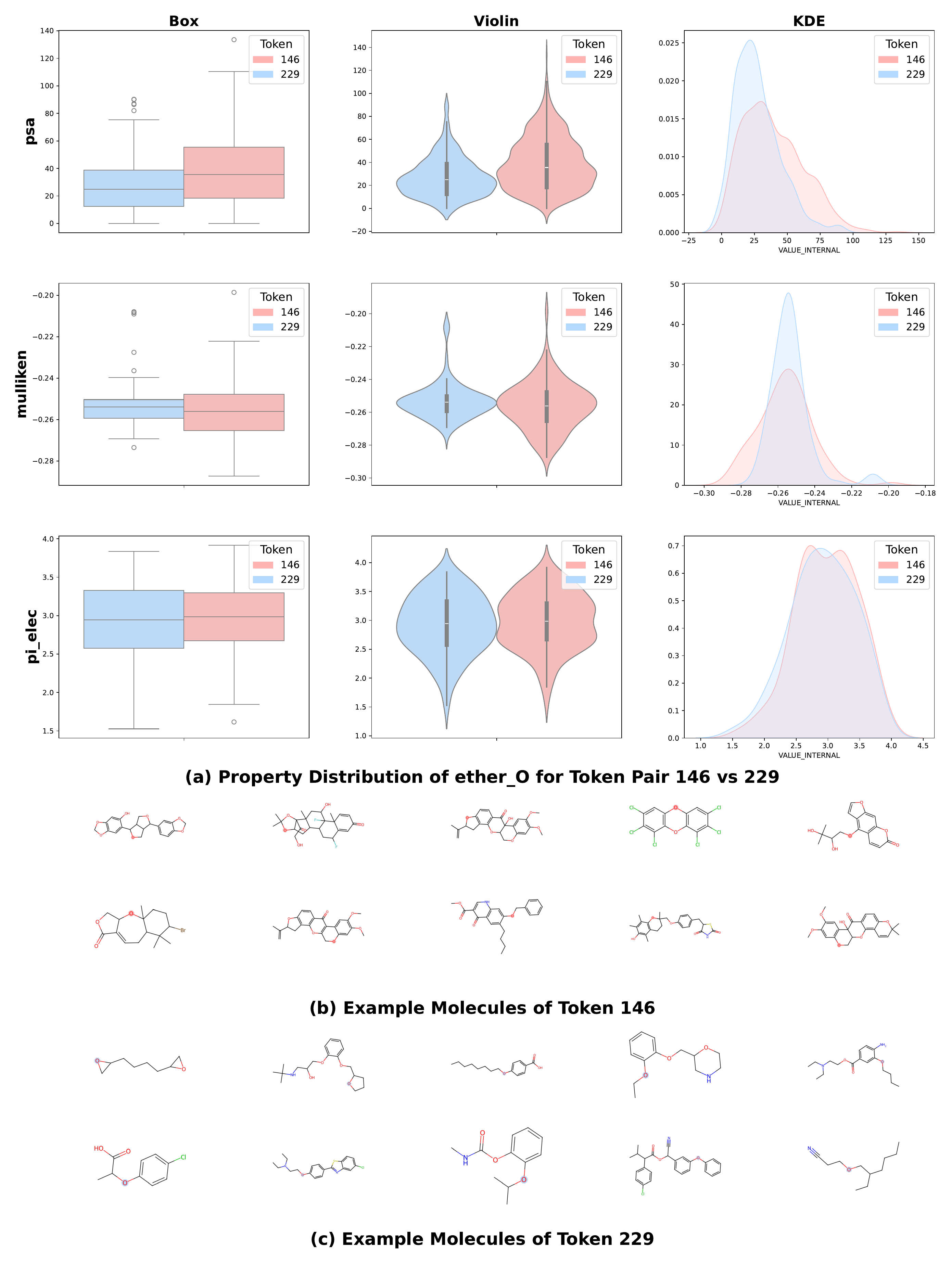}
    \caption{Distribution and molecular examples for ether assigned to token 146 and 229, 
        showing properties including PSA, Mulliken charge, and $\pi$-electron occupancy.}
    \label{fig:appendix_ether_146_229}
\end{figure}

\begin{figure}[htbp]
    \centering
    \includegraphics[width=\linewidth]{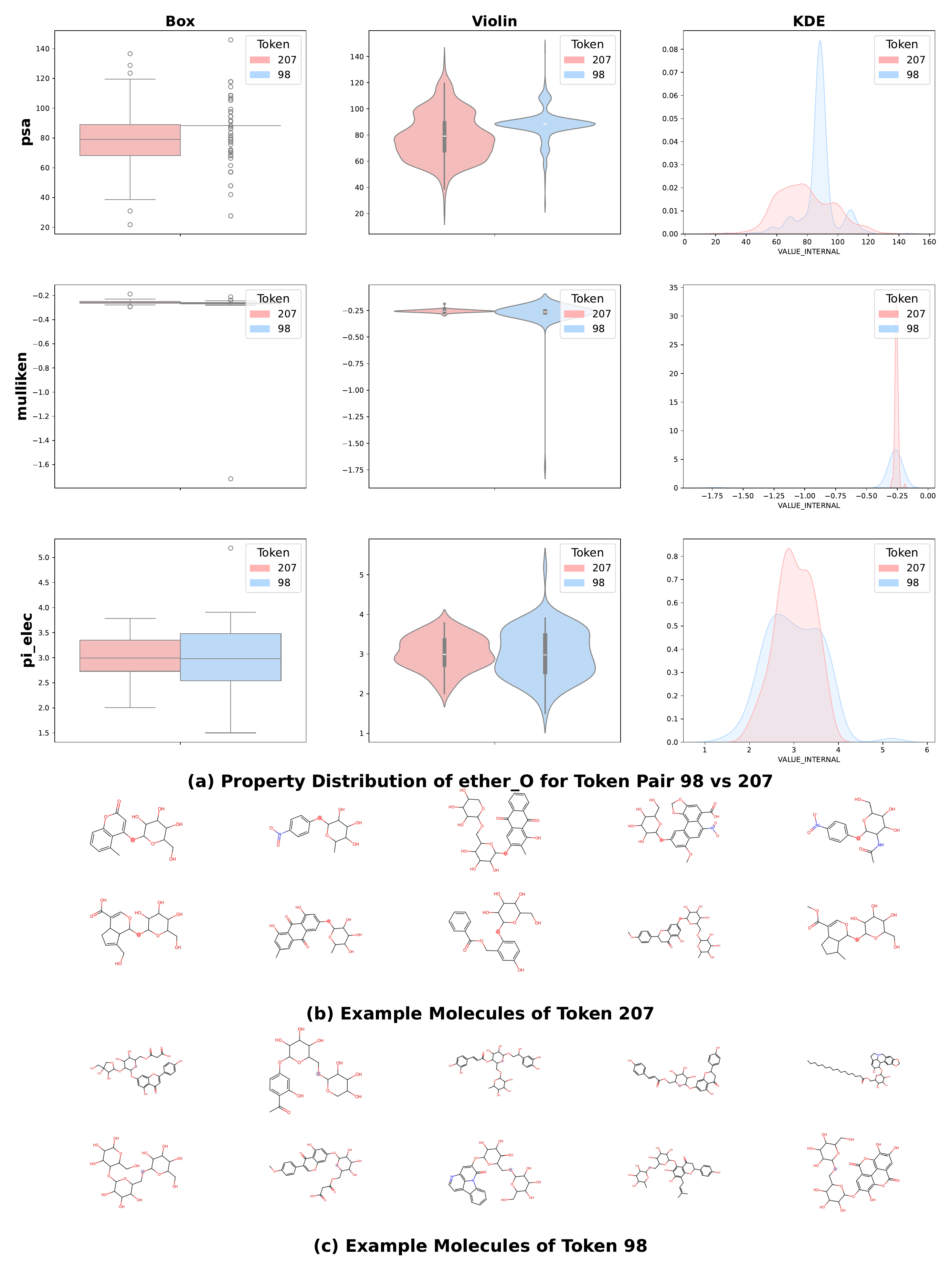}
    \caption{Distribution and molecular examples for ether assigned to token 98 and 207, 
        showing properties including PSA, Mulliken charge, and $\pi$-electron occupancy.}
    \label{fig:appendix_ether_98_207}
\end{figure}

In this section, we systematically analyze several key functional groups—including amines, hydroxyls, alkenes, ethers, and esters—focusing on cases where the same functional group is assigned to multiple distinct tokens by the AtomDisc tokenizer. For each functional group, we selected three representative token pairs and compared the local chemical environments and atomic property distributions associated with each token. This analysis demonstrates that AtomDisc is able to differentiate subtle context-dependent variations within the same functional group, providing further evidence for its fine-grained, interpretable encoding of molecular structure. The results for each functional group are summarized in Figures~\ref{fig:appendix_alkene_145_428} to~\ref{fig:appendix_ether_98_207} and Tables~\ref{tab:alkene_c_145_vs_428} to~\ref{tab:ether_o_207_vs_98}.

\begin{table}[htbp]
\centering
\caption{Statistical comparison of atomic properties for the \textbf{alkene\_C} functional group, comparing \textbf{Token 145} vs. \textbf{Token 428}. Metrics include mean $\pm$ std. dev., distribution distances (Wass., JS Div., Overlap), and p-values from the Mann-Whitney U (U p-val) and Kolmogorov-Smirnov (KS p-val) tests.}
\label{tab:alkene_c_145_vs_428}
\resizebox{\textwidth}{!}{%
\begin{tabular}{@{}lccccccccc@{}}
\toprule
& \multicolumn{2}{c}{\textbf{Distribution}} & \multicolumn{3}{c}{\textbf{Distribution Comparison}} & \multicolumn{4}{c}{\textbf{Significance Tests}} \\
\cmidrule(lr){2-3} \cmidrule(lr){4-6} \cmidrule(lr){7-10}
\textbf{Property} & \textbf{Token 145} & \textbf{Token 428} & \textbf{Wass.} & \textbf{JS Div.} & \textbf{Overlap} & \textbf{U-stat} & \textbf{U p-val} & \textbf{KS-stat} & \textbf{KS p-val} \\
\midrule
\textbf{psa} & 20.70 $\pm$ 17.82 & 8.76 $\pm$ 13.18 & 11.94 & 0.12 & 0.51 & 1.76e+05 & $<$1e-10 & 0.38 & $<$1e-10 \\
\textbf{mulliken} & -0.09 $\pm$ 0.06 & -0.01 $\pm$ 0.04 & 0.08 & 0.31 & 0.45 & 1.48e+3 & $<$1e-10 & 0.61 & $<$1e-10 \\
\textbf{pi\_elec} & 1.45 $\pm$ 0.15 & 1.39 $\pm$ 0.14 & 0.06 & 0.02 & 0.81 & 6.13e+3 & 3.8e-3 & 0.20 & 0.03 \\
\bottomrule
\end{tabular}%
}
\end{table}

\begin{table}[htbp]
\centering
\caption{Statistical comparison of atomic properties for the \textbf{alkene\_C} functional group, comparing \textbf{Token 167} vs. \textbf{Token 273}. Metrics include mean $\pm$ std. dev., distribution distances (Wass., JS Div., Overlap), and p-values from the Mann-Whitney U (U p-val) and Kolmogorov-Smirnov (KS p-val) tests.}
\label{tab:alkene_c_167_vs_273}
\resizebox{\textwidth}{!}{%
\begin{tabular}{@{}lccccccccc@{}}
\toprule
& \multicolumn{2}{c}{\textbf{Distribution}} & \multicolumn{3}{c}{\textbf{Distribution Comparison}} & \multicolumn{4}{c}{\textbf{Significance Tests}} \\
\cmidrule(lr){2-3} \cmidrule(lr){4-6} \cmidrule(lr){7-10}
\textbf{Property} & \textbf{Token 167} & \textbf{Token 273} & \textbf{Wass.} & \textbf{JS Div.} & \textbf{Overlap} & \textbf{U-stat} & \textbf{U p-val} & \textbf{KS-stat} & \textbf{KS p-val} \\
\midrule
\textbf{psa} & 22.09 $\pm$ 15.83 & 11.13 $\pm$ 15.27 & 11.38 & 0.16 & 0.48 & 1.73e+5 & $<$1e-10 & 0.44 & $<$1e-10 \\
\textbf{mulliken} & -0.02 $\pm$ 0.09 & -0.12 $\pm$ 0.03 & 0.11 & 0.52 & 0.29 & 9.15e+3 & $<$1e-10 & 0.80 & $<$1e-10 \\
\textbf{pi\_elec} & 1.43 $\pm$ 0.23 & 1.40 $\pm$ 0.14 & 0.04 & 0.02 & 0.83 & 5.19e+3 & 0.47 & 0.15 & 0.19 \\
\bottomrule
\end{tabular}%
}
\end{table}

\begin{table}[htbp]
\centering
\caption{Statistical comparison of atomic properties for the \textbf{alkene\_C} functional group, comparing \textbf{Token 240} vs. \textbf{Token 405}. Metrics include mean $\pm$ std. dev., distribution distances (Wass., JS Div., Overlap), and p-values from the Mann-Whitney U (U p-val) and Kolmogorov-Smirnov (KS p-val) tests.}
\label{tab:alkene_c_240_vs_405}
\resizebox{\textwidth}{!}{%
\begin{tabular}{@{}lccccccccc@{}}
\toprule
& \multicolumn{2}{c}{\textbf{Distribution}} & \multicolumn{3}{c}{\textbf{Distribution Comparison}} & \multicolumn{4}{c}{\textbf{Significance Tests}} \\
\cmidrule(lr){2-3} \cmidrule(lr){4-6} \cmidrule(lr){7-10}
\textbf{Property} & \textbf{Token 240} & \textbf{Token 405} & \textbf{Wass.} & \textbf{JS Div.} & \textbf{Overlap} & \textbf{U-stat} & \textbf{U p-val} & \textbf{KS-stat} & \textbf{KS p-val} \\
\midrule
\textbf{psa} & 11.66 $\pm$ 15.49 & 3.19 $\pm$ 9.42 & 8.47 & 0.21 & 0.36 & 1.68e+5 & $<$1e-10 & 0.35 & $<$1e-10 \\
\textbf{mulliken} & -0.01 $\pm$ 0.04 & -0.06 $\pm$ 0.01 & 0.05 & 0.69 & 0.17 & 8.20e+3 & $<$1e-10 & 0.80 & $<$1e-10 \\
\textbf{pi\_elec} & 1.40 $\pm$ 0.13 & 1.39 $\pm$ 0.13 & 0.02 & 0.01 & 0.83 & 5.21e+3 & 0.61 & 0.10 & 0.70 \\
\bottomrule
\end{tabular}%
}
\end{table}

\begin{table}[htbp]
\centering
\caption{Statistical comparison of atomic properties for the \textbf{ester\_O} functional group, comparing \textbf{Token 112} vs. \textbf{Token 223}. Metrics include mean $\pm$ std. dev., distribution distances (Wass., JS Div., Overlap), and p-values from the Mann-Whitney U (U p-val) and Kolmogorov-Smirnov (KS p-val) tests.}
\label{tab:ester_o_112_vs_223}
\resizebox{\textwidth}{!}{%
\begin{tabular}{@{}lccccccccc@{}}
\toprule
& \multicolumn{2}{c}{\textbf{Distribution}} & \multicolumn{3}{c}{\textbf{Distribution Comparison}} & \multicolumn{4}{c}{\textbf{Significance Tests}} \\
\cmidrule(lr){2-3} \cmidrule(lr){4-6} \cmidrule(lr){7-10}
\textbf{Property} & \textbf{Token 112} & \textbf{Token 223} & \textbf{Wass.} & \textbf{JS Div.} & \textbf{Overlap} & \textbf{U-stat} & \textbf{U p-val} & \textbf{KS-stat} & \textbf{KS p-val} \\
\midrule
\textbf{psa} & 41.85 $\pm$ 20.77 & 38.84 $\pm$ 15.43 & 4.60 & 0.07 & 0.75 & 1.15e+5 & 0.17 & 0.20 & 1.13e-8 \\
\textbf{mulliken} & -0.26 $\pm$ 0.01 & -0.26 $\pm$ 0.00 & 0.01 & 0.27 & 0.49 & 1.75e+3 & $<$1e-10 & 0.63 & $<$1e-10 \\
\textbf{pi\_elec} & 2.85 $\pm$ 0.43 & 2.89 $\pm$ 0.49 & 0.09 & 0.01 & 0.86 & 3.91e+3 & 0.49 & 0.12 & 0.49 \\
\bottomrule
\end{tabular}%
}
\end{table}

\begin{table}[htbp]
\centering
\caption{Statistical comparison of atomic properties for the \textbf{ester\_O} functional group, comparing \textbf{Token 171} vs. \textbf{Token 264}. Metrics include mean $\pm$ std. dev., distribution distances (Wass., JS Div., Overlap), and p-values from the Mann-Whitney U (U p-val) and Kolmogorov-Smirnov (KS p-val) tests.}
\label{tab:ester_o_171_vs_264}
\resizebox{\textwidth}{!}{%
\begin{tabular}{@{}lccccccccc@{}}
\toprule
& \multicolumn{2}{c}{\textbf{Distribution}} & \multicolumn{3}{c}{\textbf{Distribution Comparison}} & \multicolumn{4}{c}{\textbf{Significance Tests}} \\
\cmidrule(lr){2-3} \cmidrule(lr){4-6} \cmidrule(lr){7-10}
\textbf{Property} & \textbf{Token 171} & \textbf{Token 264} & \textbf{Wass.} & \textbf{JS Div.} & \textbf{Overlap} & \textbf{U-stat} & \textbf{U p-val} & \textbf{KS-stat} & \textbf{KS p-val} \\
\midrule
\textbf{psa} & 33.19 $\pm$ 16.15 & 61.00 $\pm$ 20.82 & 27.81 & 0.32 & 0.43 & 2.92e+4 & $<$1e-10 & 0.64 & $<$1e-10 \\
\textbf{mulliken} & -0.26 $\pm$ 0.01 & -0.32 $\pm$ 0.20 & 0.05 & 0.66 & 0.14 & 5.09e+3 & 3.71e-5 & 0.46 & 1.29e-8 \\
\textbf{pi\_elec} & 2.99 $\pm$ 0.43 & 2.95 $\pm$ 0.50 & 0.07 & 0.01 & 0.92 & 3.85e+3 & 0.63 & 0.10 & 0.73 \\
\bottomrule
\end{tabular}%
}
\end{table}

\begin{table}[htbp]
\centering
\caption{Statistical comparison of atomic properties for the \textbf{ester\_O} functional group, comparing \textbf{Token 318} vs. \textbf{Token 39}. Metrics include mean $\pm$ std. dev., distribution distances (Wass., JS Div., Overlap), and p-values from the Mann-Whitney U (U p-val) and Kolmogorov-Smirnov (KS p-val) tests.}
\label{tab:ester_o_318_vs_39}
\resizebox{\textwidth}{!}{%
\begin{tabular}{@{}lccccccccc@{}}
\toprule
& \multicolumn{2}{c}{\textbf{Distribution}} & \multicolumn{3}{c}{\textbf{Distribution Comparison}} & \multicolumn{4}{c}{\textbf{Significance Tests}} \\
\cmidrule(lr){2-3} \cmidrule(lr){4-6} \cmidrule(lr){7-10}
\textbf{Property} & \textbf{Token 318} & \textbf{Token 39} & \textbf{Wass.} & \textbf{JS Div.} & \textbf{Overlap} & \textbf{U-stat} & \textbf{U p-val} & \textbf{KS-stat} & \textbf{KS p-val} \\
\midrule
\textbf{psa} & 45.65 $\pm$ 21.54 & 46.90 $\pm$ 20.50 & 4.30 & 0.02 & 0.86 & 1.16e+5 & 0.11 & 0.12 & $<$1e-2 \\
\textbf{mulliken} & -0.27 $\pm$ 0.13 & -0.25 $\pm$ 0.08 & 0.04 & 0.10 & 0.75 & 6.54e+3 & 1.97e-4 & 0.37 & 1.68e-6 \\
\textbf{pi\_elec} & 2.95 $\pm$ 0.41 & 2.96 $\pm$ 0.41 & 0.06 & 0.01 & 0.93 & 4.38e+3 & 0.57 & 0.09 & 0.76 \\
\bottomrule
\end{tabular}%
}
\end{table}

\begin{table}[htbp]
\centering
\caption{Statistical comparison of atomic properties for the \textbf{NH2} functional group, comparing \textbf{Token 203} vs. \textbf{Token 442}. Metrics include mean $\pm$ std. dev., distribution distances (Wass., JS Div., Overlap), and p-values from the Mann-Whitney U (U p-val) and Kolmogorov-Smirnov (KS p-val) tests.}
\label{tab:nh2_203_vs_442}
\resizebox{\textwidth}{!}{%
\begin{tabular}{@{}lccccccccc@{}}
\toprule
& \multicolumn{2}{c}{\textbf{Distribution}} & \multicolumn{3}{c}{\textbf{Distribution Comparison}} & \multicolumn{4}{c}{\textbf{Significance Tests}} \\
\cmidrule(lr){2-3} \cmidrule(lr){4-6} \cmidrule(lr){7-10}
\textbf{Property} & \textbf{Token 203} & \textbf{Token 442} & \textbf{Wass.} & \textbf{JS Div.} & \textbf{Overlap} & \textbf{U-stat} & \textbf{U p-val} & \textbf{KS-stat} & \textbf{KS p-val} \\
\midrule
\textbf{psa} & 63.37 $\pm$ 20.59 & 68.33 $\pm$ 24.33 & 15.18 & 0.26 & 0.46 & 9.62e+4 & 1.15e-10 & 0.42 & $<$1e-10 \\
\textbf{mulliken} & -0.39 $\pm$ 0.01 & -0.40 $\pm$ 0.01 & 0.01 & 0.19 & 0.57 & 7.75e+3 & $<$1e-10 & 0.53 & $<$1e-10 \\
\textbf{pi\_elec} & 2.01 $\pm$ 0.53 & 2.05 $\pm$ 0.58 & 0.07 & 0.01 & 0.78 & 4.70e+3 & 0.54 & 0.13 & 0.33 \\
\bottomrule
\end{tabular}%
}
\end{table}

\begin{table}[htbp]
\centering
\caption{Statistical comparison of atomic properties for the \textbf{NH2} functional group, comparing \textbf{Token 23} vs. \textbf{Token 267}. Metrics include mean $\pm$ std. dev., distribution distances (Wass., JS Div., Overlap), and p-values from the Mann-Whitney U (U p-val) and Kolmogorov-Smirnov (KS p-val) tests.}
\label{tab:nh2_23_vs_267}
\resizebox{\textwidth}{!}{%
\begin{tabular}{@{}lccccccccc@{}}
\toprule
& \multicolumn{2}{c}{\textbf{Distribution}} & \multicolumn{3}{c}{\textbf{Distribution Comparison}} & \multicolumn{4}{c}{\textbf{Significance Tests}} \\
\cmidrule(lr){2-3} \cmidrule(lr){4-6} \cmidrule(lr){7-10}
\textbf{Property} & \textbf{Token 23} & \textbf{Token 267} & \textbf{Wass.} & \textbf{JS Div.} & \textbf{Overlap} & \textbf{U-stat} & \textbf{U p-val} & \textbf{KS-stat} & \textbf{KS p-val} \\
\midrule
\textbf{psa} & 33.16 $\pm$ 15.92 & 58.56 $\pm$ 11.76 & 25.40 & 0.56 & 0.26 & 1.93e+4 & $<$1e-10 & 0.85 & $<$1e-10 \\
\textbf{mulliken} & -0.40 $\pm$ 0.01 & -0.40 $\pm$ 0.01 & $<$1e-2 & 0.12 & 0.66 & 3.51e+3 & 5.78e-4 & 0.34 & 1.44e-5 \\
\textbf{pi\_elec} & 2.18 $\pm$ 0.64 & 1.95 $\pm$ 0.73 & 0.33 & 0.05 & 0.67 & 6.07e+3 & 0.01 & 0.24 & 0.01 \\
\bottomrule
\end{tabular}%
}
\end{table}

\begin{table}[htbp]
\centering
\caption{Statistical comparison of atomic properties for the \textbf{NH2} functional group, comparing \textbf{Token 315} vs. \textbf{Token 490}. Metrics include mean $\pm$ std. dev., distribution distances (Wass., JS Div., Overlap), and p-values from the Mann-Whitney U (U p-val) and Kolmogorov-Smirnov (KS p-val) tests.}
\label{tab:nh2_315_vs_490}
\resizebox{\textwidth}{!}{%
\begin{tabular}{@{}lccccccccc@{}}
\toprule
& \multicolumn{2}{c}{\textbf{Distribution}} & \multicolumn{3}{c}{\textbf{Distribution Comparison}} & \multicolumn{4}{c}{\textbf{Significance Tests}} \\
\cmidrule(lr){2-3} \cmidrule(lr){4-6} \cmidrule(lr){7-10}
\textbf{Property} & \textbf{Token 315} & \textbf{Token 490} & \textbf{Wass.} & \textbf{JS Div.} & \textbf{Overlap} & \textbf{U-stat} & \textbf{U p-val} & \textbf{KS-stat} & \textbf{KS p-val} \\
\midrule
\textbf{psa} & 35.34 $\pm$ 5.40 & 74.87 $\pm$ 3.41 & 39.53 & 1.00 & 1.73e-3 & 1.47e+3 & $<$1e-10 & 0.99 & $<$1e-10 \\
\textbf{mulliken} & -0.40 $\pm$ 0.01 & -0.40 $\pm$ 0.01 & 1.84e-3 & 0.04 & 0.85 & 3.84e+3 & 0.05 & 0.18 & 0.09 \\
\textbf{pi\_elec} & 2.02 $\pm$ 0.58 & 1.84 $\pm$ 0.46 & 0.18 & 0.03 & 0.69 & 5.34e+3 & 0.14 & 0.17 & 0.10 \\
\bottomrule
\end{tabular}%
}
\end{table}

\begin{table}[htbp]
\centering
\caption{Statistical comparison of atomic properties for the \textbf{OH} functional group, comparing \textbf{Token 243} vs. \textbf{Token 268}. Metrics include mean $\pm$ std. dev., distribution distances (Wass., JS Div., Overlap), and p-values from the Mann-Whitney U (U p-val) and Kolmogorov-Smirnov (KS p-val) tests.}
\label{tab:oh_243_vs_268}
\resizebox{\textwidth}{!}{%
\begin{tabular}{@{}lccccccccc@{}}
\toprule
& \multicolumn{2}{c}{\textbf{Distribution}} & \multicolumn{3}{c}{\textbf{Distribution Comparison}} & \multicolumn{4}{c}{\textbf{Significance Tests}} \\
\cmidrule(lr){2-3} \cmidrule(lr){4-6} \cmidrule(lr){7-10}
\textbf{Property} & \textbf{Token 243} & \textbf{Token 268} & \textbf{Wass.} & \textbf{JS Div.} & \textbf{Overlap} & \textbf{U-stat} & \textbf{U p-val} & \textbf{KS-stat} & \textbf{KS p-val} \\
\midrule
\textbf{psa} & 36.30 $\pm$ 18.91 & 35.23 $\pm$ 18.03 & 1.13 & 1.15e-3 & 0.97 & 1.28e+5 & 0.48 & 0.04 & 0.90 \\
\textbf{mulliken} & -0.31 $\pm$ 0.01 & -0.31 $\pm$ 0.01 & 0.01 & 0.14 & 0.69 & 6.25e+3 & 8.34e-4 & 0.33 & 2.86e-5 \\
\textbf{pi\_elec} & 2.67 $\pm$ 0.66 & 2.65 $\pm$ 0.67 & 0.06 & 7.42e-3 & 0.90 & 5.00e+3 & 0.80 & 0.09 & 0.81 \\
\bottomrule
\end{tabular}%
}
\end{table}

\begin{table}[htbp]
\centering
\caption{Statistical comparison of atomic properties for the \textbf{OH} functional group, comparing \textbf{Token 374} vs. \textbf{Token 379}. Metrics include mean $\pm$ std. dev., distribution distances (Wass., JS Div., Overlap), and p-values from the Mann-Whitney U (U p-val) and Kolmogorov-Smirnov (KS p-val) tests.}
\label{tab:oh_374_vs_379}
\resizebox{\textwidth}{!}{%
\begin{tabular}{@{}lccccccccc@{}}
\toprule
& \multicolumn{2}{c}{\textbf{Distribution}} & \multicolumn{3}{c}{\textbf{Distribution Comparison}} & \multicolumn{4}{c}{\textbf{Significance Tests}} \\
\cmidrule(lr){2-3} \cmidrule(lr){4-6} \cmidrule(lr){7-10}
\textbf{Property} & \textbf{Token 374} & \textbf{Token 379} & \textbf{Wass.} & \textbf{JS Div.} & \textbf{Overlap} & \textbf{U-stat} & \textbf{U p-val} & \textbf{KS-stat} & \textbf{KS p-val} \\
\midrule
\textbf{psa} & 71.85 $\pm$ 8.45 & 51.97 $\pm$ 10.04 & 20.03 & 0.68 & 0.14 & 2.32e+5 & $<$1e-10 & 0.91 & $<$1e-10 \\
\textbf{mulliken} & -0.31 $\pm$ 0.02 & -0.31 $\pm$ 0.01 & 3.29e-3 & 0.06 & 0.78 & 5.00e+3 & 0.99 & 0.11 & 0.58 \\
\textbf{pi\_elec} & 2.71 $\pm$ 0.68 & 2.75 $\pm$ 0.60 & 0.09 & 0.01 & 0.86 & 4.79e+3 & 0.61 & 0.13 & 0.37 \\
\bottomrule
\end{tabular}%
}
\end{table}

\begin{table}[htbp]
\centering
\caption{Statistical comparison of atomic properties for the \textbf{OH} functional group, comparing \textbf{Token 423} vs. \textbf{Token 511}. Metrics include mean $\pm$ std. dev., distribution distances (Wass., JS Div., Overlap), and p-values from the Mann-Whitney U (U p-val) and Kolmogorov-Smirnov (KS p-val) tests.}
\label{tab:oh_423_vs_511}
\resizebox{\textwidth}{!}{%
\begin{tabular}{@{}lccccccccc@{}}
\toprule
& \multicolumn{2}{c}{\textbf{Distribution}} & \multicolumn{3}{c}{\textbf{Distribution Comparison}} & \multicolumn{4}{c}{\textbf{Significance Tests}} \\
\cmidrule(lr){2-3} \cmidrule(lr){4-6} \cmidrule(lr){7-10}
\textbf{Property} & \textbf{Token 423} & \textbf{Token 511} & \textbf{Wass.} & \textbf{JS Div.} & \textbf{Overlap} & \textbf{U-stat} & \textbf{U p-val} & \textbf{KS-stat} & \textbf{KS p-val} \\
\midrule
\textbf{psa} & 46.40 $\pm$ 26.61 & 34.25 $\pm$ 19.58 & 12.15 & 0.05 & 0.79 & 1.52e+5 & 1.56e-9 & 0.21 & 1.95e-10 \\
\textbf{mulliken} & -0.32 $\pm$ 0.01 & -0.31 $\pm$ 0.01 & 0.01 & 0.30 & 0.45 & 1.26e+3 & $<$1e-10 & 0.63 & $<$1e-10 \\
\textbf{pi\_elec} & 2.52 $\pm$ 0.60 & 2.64 $\pm$ 0.67 & 0.13 & 0.01 & 0.86 & 4.46e+3 & 0.23 & 0.13 & 0.34 \\
\bottomrule
\end{tabular}%
}
\end{table}

\begin{table}[htbp]
\centering
\caption{Statistical comparison of atomic properties for the \textbf{ether\_O} functional group, comparing \textbf{Token 112} vs. \textbf{Token 39}. Metrics include mean $\pm$ std. dev., distribution distances (Wass., JS Div., Overlap), and p-values from the Mann-Whitney U (U p-val) and Kolmogorov-Smirnov (KS p-val) tests.}
\label{tab:ether_o_112_vs_39}
\resizebox{\textwidth}{!}{%
\begin{tabular}{@{}lccccccccc@{}}
\toprule
& \multicolumn{2}{c}{\textbf{Distribution}} & \multicolumn{3}{c}{\textbf{Distribution Comparison}} & \multicolumn{4}{c}{\textbf{Significance Tests}} \\
\cmidrule(lr){2-3} \cmidrule(lr){4-6} \cmidrule(lr){7-10}
\textbf{Property} & \textbf{Token 112} & \textbf{Token 39} & \textbf{Wass.} & \textbf{JS Div.} & \textbf{Overlap} & \textbf{U-stat} & \textbf{U p-val} & \textbf{KS-stat} & \textbf{KS p-val} \\
\midrule
\textbf{psa} & 40.63 $\pm$ 20.86 & 46.96 $\pm$ 20.55 & 6.41 & 0.02 & 0.83 & 1.02e+5 & 2.45e-7 & 0.19 & 1.22e-8 \\
\textbf{mulliken} & -0.26 $\pm$ 0.01 & -0.27 $\pm$ 0.01 & 0.01 & 0.06 & 0.79 & 5.98e+3 & 2.98e-3 & 0.25 & 3.62e-3 \\
\textbf{pi\_elec} & 2.90 $\pm$ 0.47 & 2.97 $\pm$ 0.49 & 0.09 & 0.01 & 0.88 & 4.37e+3 & 0.28 & 0.17 & 0.09 \\
\bottomrule
\end{tabular}%
}
\end{table}

\begin{table}[htbp]
\centering
\caption{Statistical comparison of atomic properties for the \textbf{ether\_O} functional group, comparing \textbf{Token 146} vs. \textbf{Token 229}. Metrics include mean $\pm$ std. dev., distribution distances (Wass., JS Div., Overlap), and p-values from the Mann-Whitney U (U p-val) and Kolmogorov-Smirnov (KS p-val) tests.}
\label{tab:ether_o_146_vs_229}
\resizebox{\textwidth}{!}{%
\begin{tabular}{@{}lccccccccc@{}}
\toprule
& \multicolumn{2}{c}{\textbf{Distribution}} & \multicolumn{3}{c}{\textbf{Distribution Comparison}} & \multicolumn{4}{c}{\textbf{Significance Tests}} \\
\cmidrule(lr){2-3} \cmidrule(lr){4-6} \cmidrule(lr){7-10}
\textbf{Property} & \textbf{Token 146} & \textbf{Token 229} & \textbf{Wass.} & \textbf{JS Div.} & \textbf{Overlap} & \textbf{U-stat} & \textbf{U p-val} & \textbf{KS-stat} & \textbf{KS p-val} \\
\midrule
\textbf{psa} & 39.31 $\pm$ 22.80 & 27.22 $\pm$ 17.18 & 12.08 & 0.06 & 0.75 & 1.64e+5 & $<$1e-10 & 0.32 & $<$1e-10 \\
\textbf{mulliken} & -0.26 $\pm$ 0.01 & -0.25 $\pm$ 0.01 & 4.90e-3 & 0.08 & 0.76 & 4.29e+3 & 0.29 & 0.20 & 0.03 \\
\textbf{pi\_elec} & 2.98 $\pm$ 0.48 & 2.92 $\pm$ 0.51 & 0.06 & 2.98e-3 & 0.93 & 4.96e+3 & 0.51 & 0.10 & 0.72 \\
\bottomrule
\end{tabular}%
}
\end{table}

\begin{table}[htbp]
\centering
\caption{Statistical comparison of atomic properties for the \textbf{ether\_O} functional group, comparing \textbf{Token 207} vs. \textbf{Token 98}. Metrics include mean $\pm$ std. dev., distribution distances (Wass., JS Div., Overlap), and p-values from the Mann-Whitney U (U p-val) and Kolmogorov-Smirnov (KS p-val) tests.}
\label{tab:ether_o_207_vs_98}
\resizebox{\textwidth}{!}{%
\begin{tabular}{@{}lccccccccc@{}}
\toprule
& \multicolumn{2}{c}{\textbf{Distribution}} & \multicolumn{3}{c}{\textbf{Distribution Comparison}} & \multicolumn{4}{c}{\textbf{Significance Tests}} \\
\cmidrule(lr){2-3} \cmidrule(lr){4-6} \cmidrule(lr){7-10}
\textbf{Property} & \textbf{Token 207} & \textbf{Token 98} & \textbf{Wass.} & \textbf{JS Div.} & \textbf{Overlap} & \textbf{U-stat} & \textbf{U p-val} & \textbf{KS-stat} & \textbf{KS p-val} \\
\midrule
\textbf{psa} & 78.78 $\pm$ 17.78 & 87.65 $\pm$ 11.69 & 11.66 & 0.25 & 0.47 & 8.09e+4 & $<$1e-10 & 0.50 & $<$1e-10 \\
\textbf{mulliken} & -0.26 $\pm$ 0.01 & -0.28 $\pm$ 0.15 & 0.02 & 0.38 & 0.36 & 6.98e+3 & 1.27e-6 & 0.35 & 7.85e-6 \\
\textbf{pi\_elec} & 3.02 $\pm$ 0.42 & 3.00 $\pm$ 0.61 & 0.16 & 0.05 & 0.78 & 5.16e+3 & 0.70 & 0.16 & 0.15 \\
\bottomrule
\end{tabular}%
}
\end{table}

\newpage
\subsubsection{Systematic Significance of Token–Property Differences}
\label{appendix:sys_token_diff}

\begin{figure}[htbp]
    \centering
    \includegraphics[width=\textwidth]{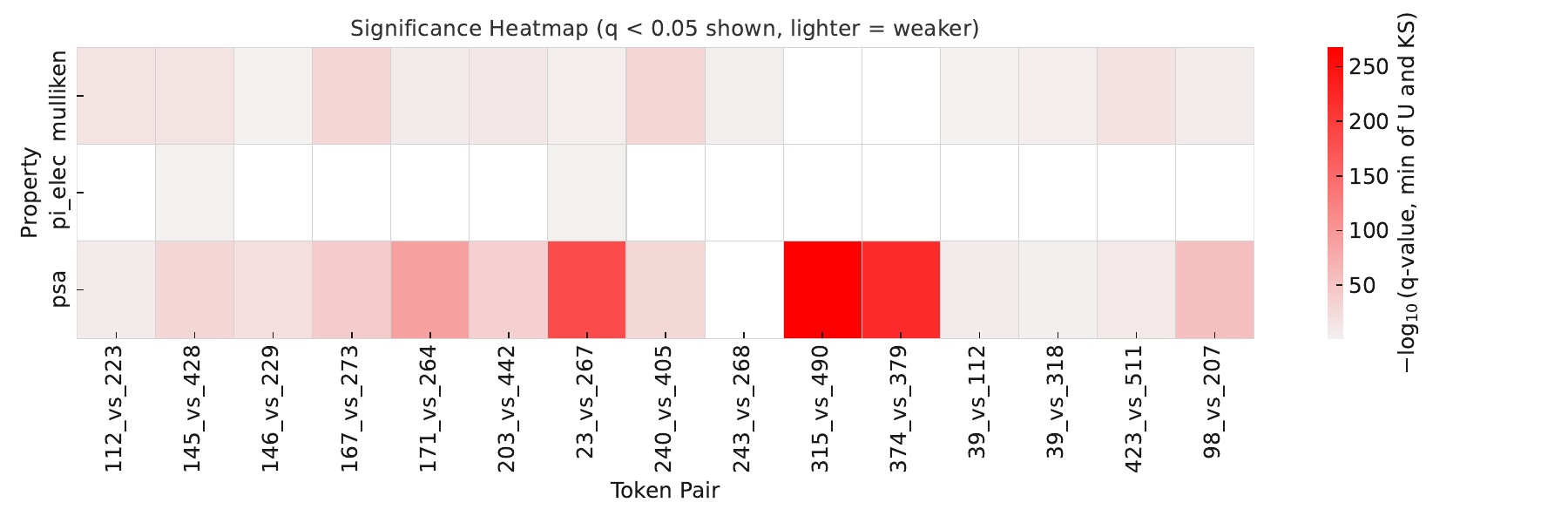}
    \caption{Significance heatmap for 45 functional group--token--property comparisons. White cells denote non-significant results ($q \ge 0.05$); red shading intensity indicates stronger significance ($q < 0.05$).}
    \label{fig:sig_heatmap}
\end{figure}

To rigorously evaluate whether the observed property differences between token pairs are robust rather than artifacts of multiple comparisons, we performed Benjamini–Hochberg correction (FDR = 0.05) across all 45 functional group–token–property combinations. For each pair, we considered the minimum $q$-value from both the Mann–Whitney U and Kolmogorov–Smirnov tests as the measure of significance. After correction, 29 out of 45 comparisons (64.4\%) remained statistically significant ($q<0.05$), demonstrating that the majority of property distinctions between token pairs are highly robust and unlikely to be due to chance. Importantly, for every functional group examined ($15/15$ token pairs), at least one property exhibited a significant difference after FDR adjustment, indicating that fine-grained chemical context is systematically reflected in the tokenizer assignments. Figure~\ref{fig:sig_heatmap} provides an overview of these results, where colored cells indicate significant comparisons and color intensity reflects $-\log_{10}(q)$ significance.

\newpage
\subsubsection{Verification with heavier quantum-chemistry settings}

To assess the robustness of our statistical conclusions against the choice of quantum-chemistry settings, we recomputed $\pi$-electron occupancy for all five functional group–token pairs using a heavier level of theory (B3LYP/6-31G*) in place of the lighter HF/STO-3G setup. For each pair, we performed a two-sided Kolmogorov–Smirnov (KS) test to compare the corresponding distributions, and quantified effect sizes using KDE-based distribution overlap and Jensen–Shannon divergence (JSD).

As summarized in Table~\ref{tab:pi_elec_light_vs_heavy_redux}, the KS test produced identical significance calls ($p<0.05$ or not) between the light and heavy settings for all 5/5 cases, indicating complete agreement in detecting distributional differences. Effect-size changes between the two settings were small in magnitude: the shift in distribution overlap ($\Delta Overlap$) was within $\pm$ 0.081 and the shift in JSD ($\Delta JSD$) within $\pm$ 0.020, with the direction of separation preserved across all comparisons. These results confirm that the qualitative distributional differences observed for $\pi$-electron occupancy are stable with respect to the choice of basis set and functional, and that heavier DFT calculations yield conclusions consistent with those obtained using the lighter HF/STO-3G setting.

Overall, the heavier B3LYP/6-31G* calculations confirm the qualitative distributional differences obtained with HF/STO-3G, 
supporting the robustness of our conclusions.

\begin{table}[htbp]
\centering
\caption{Comparison of $\pi$-electron occupancy significance (KS test) between light (HF/STO-3G) and heavy (B3LYP/6-31G*) settings, along with effect-size shifts. 
Columns show KS $p$-values under each setting (and whether the significant/non-significant decision agrees at $\alpha=0.05$), plus the change in distribution overlap and Jensen--Shannon divergence (heavy $-$ light).}
\label{tab:pi_elec_light_vs_heavy_redux}
\begin{tabular}{@{}lcccccc@{}}
\toprule
\textbf{FG} & \textbf{Token Pair} & \textbf{KS $p$ (light)} & \textbf{KS $p$ (heavy)} & \textbf{Consistency} & \textbf{$\Delta$Overlap} & \textbf{$\Delta$JSD} \\
\midrule
hydroxyl        & 374 vs 379 & 0.368 & 0.295 & Yes & 0.008 & 0.001 \\
amide    & 23 vs 267  & 0.006 & $9.8\times10^{-5}$ & Yes & 0.011 & 0.002 \\
ether  & 39 vs 112  & 0.093 & 0.096 & Yes & 0.001 & 0.001 \\
ester     & 39 vs 318  & 0.760 & 0.065 & Yes & 0.081 & 0.019 \\
alkene & 145 vs 428 & 0.028 & 0.016 & Yes & 0.003 & 0.002 \\
\bottomrule
\end{tabular}
\end{table}

\newpage
\putbib[ref]   
\end{bibunit}
\end{document}